\RequirePackage[right]{lineno}
\documentclass[aps,prl,preprint,nopacs,superscriptaddress]{revtex4}
\usepackage{graphicx}
\usepackage{verbatim}
\usepackage{relsize}
\usepackage{mathrsfs}
\usepackage{color}
\usepackage{ulem}
\usepackage{amsmath,amsfonts,amssymb}
\usepackage{graphicx}
\usepackage{braket}
\usepackage{bbding}
\usepackage{pifont}
\usepackage{wasysym}
\usepackage{amssymb}
\usepackage{titlesec}

\usepackage{multirow}
\usepackage{tikz} 
\usetikzlibrary{arrows,scopes}

\newcommand{\rc}{%
\resizebox{!}{1.25ex}{%
    \begin{tikzpicture}[>=round cap]
        \clip (0.09em,-0.05ex) rectangle (0.61em,0.81ex);
        \draw [line width=.11ex, <->, rounded corners=0.13ex] (0.1em,0.1ex) .. controls (0.24em,0.4ex) .. (0.35em,0.8ex) .. controls (0.29em,0.725ex) .. (0.25em,0.6ex) .. controls (0.7em,0.8ex) and (0.08em,-0.4ex) .. (0.55em,0.25ex);
    \end{tikzpicture}%
}%
}

\newcommand{\brc}{%
\resizebox{!}{1.3ex}{%
    \begin{tikzpicture}[>=round cap]
        \clip (0.085em,-0.1ex) rectangle (0.61em,0.875ex);
        \draw [line width=.2ex, <->, rounded corners=0.13ex] (0.1em,0.1ex) .. controls (0.24em,0.4ex) .. (0.35em,0.8ex) .. controls (0.29em,0.725ex) .. (0.25em,0.6ex) .. controls (0.7em,0.8ex) and (0.08em,-0.4ex) .. (0.55em,0.25ex);
    \end{tikzpicture}%
}%
}

\def \beq {\begin{equation}}
\def \eeq {\end{equation}}

\begin{document}

\title{
Quantum metric nonlinear Hall effect in a topological antiferromagnetic heterostructure

}

\author{Anyuan Gao}\affiliation{\footnotesize Department of Chemistry and Chemical Biology, Harvard University, Massachusetts 02138, USA}

\author{Yu-Fei Liu}\affiliation{\footnotesize Department of Chemistry and Chemical Biology, Harvard University, Massachusetts 02138, USA}\affiliation{\footnotesize Department of Physics, Harvard University, Cambridge, MA 02138, USA}
\author{Jian-Xiang Qiu}\affiliation{\footnotesize Department of Chemistry and Chemical Biology, Harvard University, Massachusetts 02138, USA}

\author{Barun Ghosh}\affiliation{\footnotesize Department of Physics, Northeastern University, Boston, MA 02115, USA}
\author{Tha\'is V. Trevisan}\affiliation{\footnotesize Department of Physics and Astronomy, Iowa State University, Ames, Iowa 50011, USA}\affiliation{\footnotesize Ames National Laboratory, Ames, Iowa 50011, USA}

\author{Yugo Onishi}\affiliation{\footnotesize Department of Physics, Massachusetts Institute of Technology, Cambridge, MA 02139, USA}

\author{Chaowei Hu}\affiliation{\footnotesize Department of Physics and Astronomy and California NanoSystems Institute, University of California, Los Angeles, Los Angeles, CA 90095, USA.}
\author{Tiema Qian}\affiliation{\footnotesize Department of Physics and Astronomy and California NanoSystems Institute, University of California, Los Angeles, Los Angeles, CA 90095, USA.}

\author{Hung-Ju Tien}\affiliation{\footnotesize Department of Physics, National Cheng Kung University, Tainan 701, Taiwan}
\author{Shao-Wen Chen}\affiliation{\footnotesize Department of Physics, Harvard University, Cambridge, MA 02138, USA}

\author{Mengqi Huang}\affiliation{\footnotesize Department of Physics, University of California, San Diego, La Jolla, CA, 92093, USA}

\author{Damien B\'erub\'e}\affiliation{\footnotesize Department of Chemistry and Chemical Biology, Harvard University, Massachusetts 02138, USA}

\author{Houchen Li}\affiliation{\footnotesize Department of Chemistry and Chemical Biology, Harvard University, Massachusetts 02138, USA}

\author{Christian Tzschaschel}\affiliation{\footnotesize Department of Chemistry and Chemical Biology, Harvard University, Massachusetts 02138, USA}

\author{Thao Dinh}\affiliation{\footnotesize Department of Chemistry and Chemical Biology, Harvard University, Massachusetts 02138, USA}\affiliation{\footnotesize Department of Physics, Harvard University, Cambridge, MA 02138, USA}

\author{Zhe Sun}\affiliation{\footnotesize Department of Chemistry and Chemical Biology, Harvard University, Massachusetts 02138, USA}\affiliation{\footnotesize Department of Physics, Boston College, Chestnut Hill, MA, USA}
\author{Sheng-Chin Ho}\affiliation{\footnotesize Department of Chemistry and Chemical Biology, Harvard University, Massachusetts 02138, USA}

\author{Shang-Wei Lien}\affiliation{\footnotesize Department of Physics, National Cheng Kung University, Tainan 701, Taiwan}

\author{Bahadur Singh}\affiliation{\footnotesize Department of Condensed Matter Physics and Materials Science, Tata Institute of Fundamental Research, Colaba, Mumbai, India}

\author{Kenji Watanabe}\affiliation{\footnotesize International Center for Materials Nanoarchitectonics, National Institute for Materials Science,  1-1 Namiki, Tsukuba 305-0044, Japan}

\author{Takashi Taniguchi}\affiliation{\footnotesize International Center for Materials Nanoarchitectonics, National Institute for Materials Science,  1-1 Namiki, Tsukuba 305-0044, Japan}

\author{David C. Bell}\affiliation{\footnotesize Harvard John A. Paulson School of Engineering and Applied Sciences,Harvard University, Cambridge, Massachusetts 02138, USA}\affiliation{\footnotesize Center for Nanoscale Systems, Harvard University, Cambridge, Massachusetts 02138, USA}

\author{Hsin Lin}
\affiliation{\footnotesize Institute of Physics, Academia Sinica, Taipei 11529, Taiwan}

\author{Tay-Rong Chang}\affiliation{\footnotesize Department of Physics, National Cheng Kung University, Tainan 701, Taiwan}

\author{Chunhui Rita Du}
\affiliation{\footnotesize Department of Physics, University of California, San Diego, La Jolla, CA, 92093, USA}

\author{Arun Bansil}
\affiliation{\footnotesize Department of Physics, Northeastern University, Boston, MA 02115, USA}

\author{Liang Fu}\affiliation{\footnotesize Department of Physics, Massachusetts Institute of Technology, Cambridge, MA 02139, USA}

\author{Ni Ni}\affiliation{\footnotesize Department of Physics and Astronomy and California NanoSystems Institute, University of California, Los Angeles, Los Angeles, CA 90095, USA.}

\author{Peter P. Orth}\affiliation{\footnotesize Department of Physics and Astronomy, Iowa State University, Ames, Iowa 50011, USA}\affiliation{\footnotesize Ames National Laboratory, Ames, Iowa 50011, USA}

\author{Qiong Ma}\affiliation{\footnotesize Department of Physics, Boston College, Chestnut Hill, MA, USA}

\author{Su-Yang Xu\footnote{Corresponding author (email): suyangxu@fas.harvard.edu}}\affiliation{\footnotesize Department of Chemistry and Chemical Biology, Harvard University, Massachusetts 02138, USA}

\pacs{}
\maketitle

\textbf{Quantum geometry - the geometry of electron Bloch wavefunctions - is central to modern condensed matter physics. Due to the quantum nature, quantum geometry has two parts, the real part quantum metric and the imaginary part Berry curvature. The studies of Berry curvature have led to countless breakthroughs, ranging from the quantum Hall effect in 2DEGs to the anomalous Hall effect (AHE) in ferromagnets. However, in contrast to Berry curvature, the quantum metric has rarely been explored. Here, we report a new nonlinear Hall effect induced by quantum metric by interfacing even-layered MnBi$_2$Te$_4$ (a $\mathcal{PT}$-symmetric antiferromagnet (AFM)) with black phosphorus. This novel nonlinear Hall effect switches direction upon reversing the AFM spins and exhibits distinct scaling that suggests a non-dissipative nature. Like the AHE brought Berry curvature under the spotlight, our results open the door to discovering quantum metric responses. Moreover, we demonstrate that the AFM can harvest wireless electromagnetic energy via the new nonlinear Hall effect, therefore enabling intriguing applications that bridges nonlinear electronics with AFM spintronics.}

\vspace{0.5cm}
\textbf{Introduction}

\begin{figure*}[t]
\includegraphics[width=16cm]{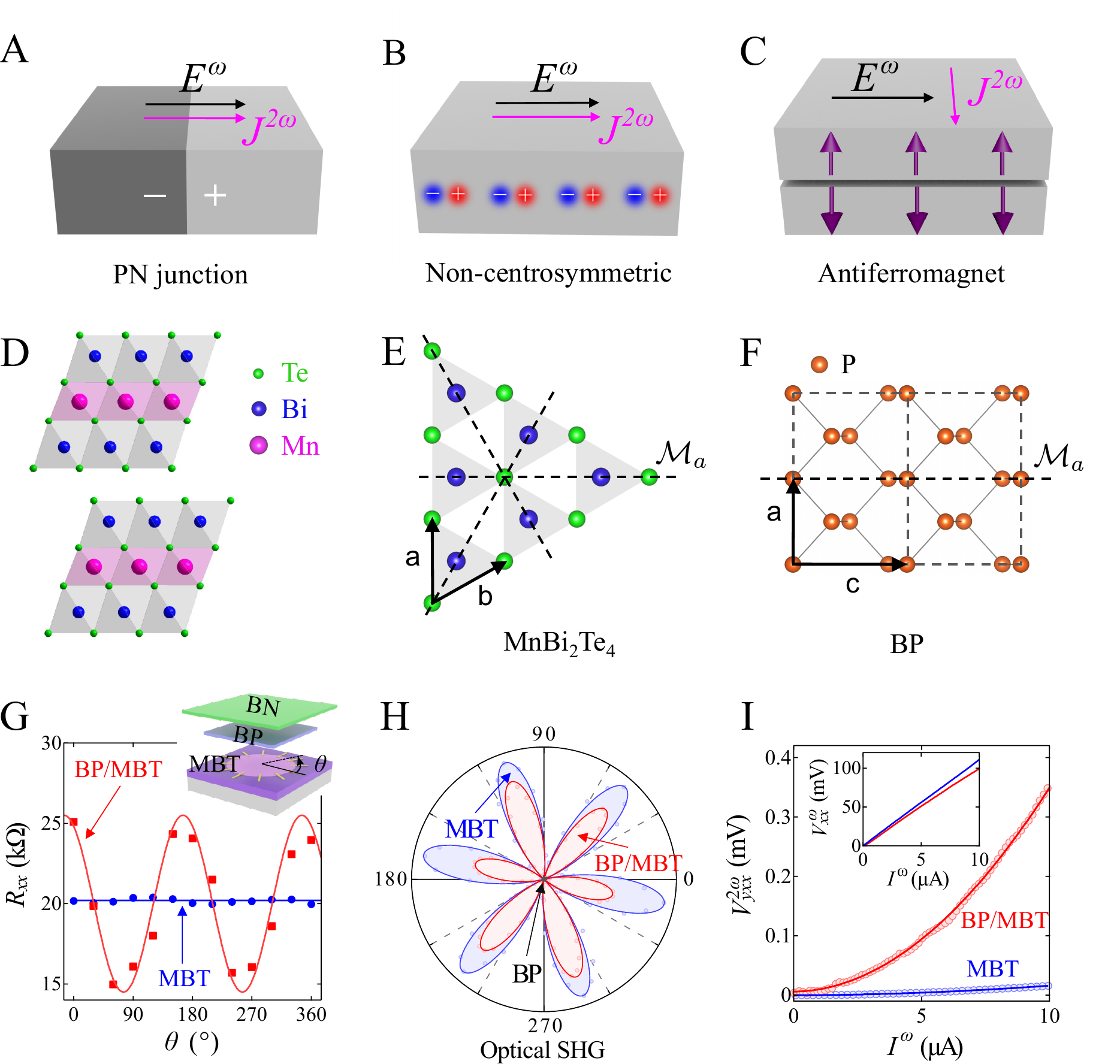}
\vspace{0cm}
\caption{{\bf Spin-induced electrical nonlinearity in $\mathcal{PT}$-symmetric antiferromagnets and introduction to our sample.} (\textbf{A} and \textbf{B}) Nonlinear electrical transport in PN junctions and noncentrosymmetric conductors (charge-induced electrical nonlinearity). (\textbf{C}) Nonlinear electrical transport in $\mathcal{PT}$-symmetric AFMs (spin-induced electrical nonlinearity).   (\textbf{D} to \textbf{F}) Lattice structures of the MnBi$_2$Te$_4$ and BP. (\textbf{G} and \textbf{H}) Angle-resolved resistance and optical second-harmonic generation (SHG) measurements of a 6SL MnBi$_2$Te$_4$ before and after interfaced with BP (Device-BM1). (\textbf{I}) The nonlinear Hall signal $V^{2\omega}_{yxx}$ of Device-BM1 before and after interfaced with BP at $B$ = 0 T.}
\label{Fig1}
\end{figure*}

Nonlinearities are crucial in many branches of physics, ranging from atomic physics to condensed matter and complex dynamical systems. Nonlinear electrical transport is the foundation of applications such as rectification and wave mixing. Classically, the most well-known nonlinear device is a PN diode (Fig.~\ref{Fig1}A). Noncentrosymmetric polar materials (Fig.~\ref{Fig1}B) are similar to PN diodes as they both possess an electric dipole. They have recently been discovered to show intrinsic nonlinear electrical transport, which not only suggests novel nonlinear applications but also provides a powerful probe of the quantum geometry of the conduction electrons \cite{provost1980riemannian,xiao2010berry, tokura2018nonreciprocal,ma2021topology,orenstein2021topology,moore2010confinement,sodemann2015quantum,rectification, Kang2019nonlinear,ma2019observation,dzsaber2021giant, kumar2021room,yasuda2020large, zhao2020magnetic, isobe2020high,Lai2021third,he2022graphene,zhang2022non, sinha2022berry}. Broadly, the nonlinear transport in both diodes (Fig.~\ref{Fig1}A) and noncentrosymmetric conductors (Fig.~\ref{Fig1}B) arise from an inversion asymmetric charge distributions (e.g. an electric dipole). Since the electron has another fundamental degree of freedom, spin, an interesting question is whether spin can also lead to an electrical nonlinearity even in a centrosymmetric lattice.  One ideal platform is the $\mathcal{PT}$-symmetric AFMs \cite{Zhang2022Diodic}, where only the spins feature a noncentrosymmetric distribution (Fig.~\ref{Fig1}C).

Important clues can be drawn from previous optical experiments, where optical second-harmonic generation (SHG) has been observed in the $\mathcal{PT}$-symmetric AFMs including Cr$_2$O$_3$ \cite{fiebig1994second} and CrI$_3$ \cite{sun2019giant}. Nevertheless, nonlinear transport is distinct because it directly probes the Fermi surface electrons and in many cases their geometrical properties \cite{tokura2018nonreciprocal,ma2021topology,orenstein2021topology, provost1980riemannian,xiao2010berry}. As such, it enables a probe of the quantum geometry \cite{tokura2018nonreciprocal,ma2021topology,orenstein2021topology, provost1980riemannian,xiao2010berry} of the topological bands at the Fermi level of novel conductors.

The quantum geometry has two parts, $T=g-\frac{i}{2}\Omega$ \cite{provost1980riemannian,xiao2010berry} ($T$ is the quantum geometrical tensor). The imaginary part is the well-known Berry curvature $\Omega_{\alpha\beta}=-2\text{Im}\sum_{m\neq n}[\langle u_n\arrowvert i\partial_{k_\alpha}u_m\rangle\langle u_m\arrowvert i\partial_{k_\beta}u_n\rangle]$, which describes the curvature of wavefunction in Hilbert space ($n,m$ are band indices and $\alpha, \beta$ are spatial directions). Berry curvature has been identified as the source of many novel electronic and optical responses. By contrast, the real part is the quantum metric, $g_{\alpha\beta}=\text{Re}\sum_{m\neq n}[\langle u_n\arrowvert i\partial_{k_\alpha}u_m\rangle\langle u_m\arrowvert i\partial_{k_\beta}u_n\rangle]$, which measures the distance between neighboring Bloch wavefunctions in Hilbert space (i.e., the distance when Bloch wavefunctions are mapped onto a Bloch sphere, see SM. IV.1). Although being equally important, the quantum metric is much less explored. There have been a few examples related to the quantum metric, including prediction for the electrical and orbital magnetic susceptibilities \cite{gao2015geometrical}, observation of a third order Hall effect \cite{Lai2021third} and the quantum metric in atomic physics \cite{Gianfrate2020measurement}. However, examples have remained limited and how quantum metric regulates the electronic motion remains largely unknown. Recently, theory has started to predict a wide range of exotic quantum metric responses \cite{gao2014field,Wang2021Intrinsic,liu2021intrinsic, lahiri2022intrinsic, Smith2022momentum, Arora2022quantuma,Mitscherling2021, Rhim2020quantum, ledwith2020fractional, Holder2020consequences,  Watanabe2021chiral,Huhtinen2022revisting, Hofmann2022superconductivity,Hu2020quantum}. One particularly intriguing platform is the $\mathcal{PT}$-symmetric AFM \cite{gao2014field,Wang2021Intrinsic,liu2021intrinsic, lahiri2022intrinsic, Smith2022momentum}, because $\mathcal{PT}$ forces the Berry curvature to vanish identically, hence isolating novel phenomena related to quantum metric.

Here, we focus on the recent proposal of a nondissipative, intrinsic second-order Hall effect induced by the quantum metric dipole \cite{Wang2021Intrinsic,gao2014field,Smith2022momentum}. We design and fabricate a feasible material platform and demonstrate the first realization. To conceptualize this new nonlinear Hall effect, we draw comparison with the well-known AHE in ferromagnetic metals \cite{nagaosa2010anomalous}, where Berry curvature leads to the anomalous velocity and therefore the AHE, $v_{\textrm{anomalous}}\propto \int_{\textbf{k}} \mathbf{E}_{\|} \times \mathbf{\Omega}$, ($\mathbf{E}_{\|}$ is the in-plane source-drain electric field). By contrast, in a $\mathcal{PT}$-symmetric AFM, Berry curvature is zero due to $\mathcal{PT}$. However, a nonzero quantum metric $g$ in the two-band limit can induce an anomalous velocity to the second-order of $\mathbf{E}_{\|}$, $v_{\textrm{anomalous}}\propto \int_{\textbf{k}}  \mathbf{E}_{\|} \times [\nabla_{\mathbf{k}}\times(g\mathbf{E}_{\|})]$, as proposed in \cite{gao2014field}. This leads to the intrinsic second-order Hall effect. From the expression above, one can show that this effect is nonzero only when the system breaks both $\mathcal{P}$ and $\mathcal{T}$. Therefore, we need $\mathcal{PT}$-symmetric AFM conductors with a large quantum metric on the Fermi surface.  We have carefully considered possible materials, and identified 2D even-layered MnBi$_2$Te$_4$ \cite{Otrokov2019a, Zhang2019a, Deng2020, Liu2020a, deng2021high, Ovchinnikov2020,gao2021layer, Lee2019a, cai2021electric, tai2021polarity, bac2022topological, zhang2022non} as an ideal platform. Even-layered MnBi$_2$Te$_4$ is a $\mathcal{PT}$-symmetric AFM. Moreover, its topological bands support gate-tunable transport and a giant quantum metric. However, its lattice has $\mathcal{C}_{3z}$ rotational symmetry (Figs.~\ref{Fig1}D,E), which forces the effect to vanish \cite{Wang2021Intrinsic}. To break $C_{3z}$, we interface it with black phosphorus (BP) \cite{akamatsu2021van}.

\vspace{0.5cm}
\textbf{Demonstration of rotational symmetry breaking}

We start by showing that interfacing MnBi$_2$Te$_4$ with BP indeed breaks its $\mathcal{C}_{3z}$ rotational symmetry. To this end, we study the directional dependence of the resistance \cite{xia2014rediscovering, Kang2019nonlinear} of MnBi$_2$Te$_4$ without and with BP. We fabricated a 6-septuple-layer (6SL) MnBi$_2$Te$_4$ device with radially distributed electrical contacts (Device-BM1). As shown by the blue curve in Fig.~\ref{Fig1}G, the four-probe resistance ($T=1.8$ K) is found to be fully isotropic, consistent with the presence of the  $\mathcal{C}_{3z}$ symmetry. We then stacked a BP layer ($\sim10$ nm) onto this MnBi$_2$Te$_4$ sample and performed the measurements again. As shown by the red curve in Fig.~\ref{Fig1}G, the resistance develops a clear anisotropy with a $180^{\circ}$ periodicity, providing a clear signature of the breaking of $\mathcal{C}_{3z}$ symmetry (In SM. I.3, we present additional experiments to show that the transport signal is dominated by the MnBi$_2$Te$_4$ layer of the heterostructure). The transverse resistance and two-probe resistance also show the breaking of $\mathcal{C}_{3z}$ (fig. S6). We further substantiate the breaking of $\mathcal{C}_{3z}$ symmetry by an independent method, the optical second harmonic generation (SHG) at room temperature. As shown in Fig.~\ref{Fig1}H, our SHG data also shows the clear breaking of $\mathcal{C}_{3z}$ symmetry (see detailed discussions in SM. I.5 and fig. S7). Our demonstration of $\mathcal{C}_{3z}$ breaking establishes the BP/MnBi$_2$Te$_4$ heterostructure as an ideal platform to search for this effect.

\vspace{0.5cm}
\textbf{Observation of the nonlinear Hall effect}

In order to measure the linear and nonlinear electrical transport, we pass a current at frequency $\omega$ ($I^{\omega}$) and use the lock-in technique to detect linear voltage $V^{\omega}$ and nonlinear voltage $V^{2\omega}$. We describe the nonlinear voltage as $V^{2\omega}_{ijk}$, where $i$ is the direction of the nonlinear voltage $V^{2\omega}$ and $j,k$ are the directions of the injected current $I^{\omega}$.  All measurements are performed at $B=0$. 

Figure~\ref{Fig1}I shows the nonlinear Hall voltage $V^{2\omega}_{yxx}$ of the Device-BM1 before and after interfaced with BP. Remarkably, a prominent nonlinear Hall signal only emerges after BP is introduced. This is in sharp contrast to the linear voltage (inset of Fig.~\ref{Fig1}I), which becomes even slightly smaller upon the introduction of BP. Such observation agrees well with the theoretical expectation of the intrinsic nonlinear Hall effect induced by a quantum metric dipole. To exclude that the effect is caused by a Berry curvature dipole \cite{sodemann2015quantum,ma2019observation,Kang2019nonlinear,kumar2021room}, which leads to a second-order Hall effect in nonmagnetic, noncentrosymmetric conductors, we study the relationship between the second-order nonlinear Hall effect and the AFM order in MnBi$_2$Te$_4$.

\vspace{0.5cm}
\textbf{The AFM spin-induced nonlinearity}

\begin{figure*}[t]
\includegraphics[width=12cm]{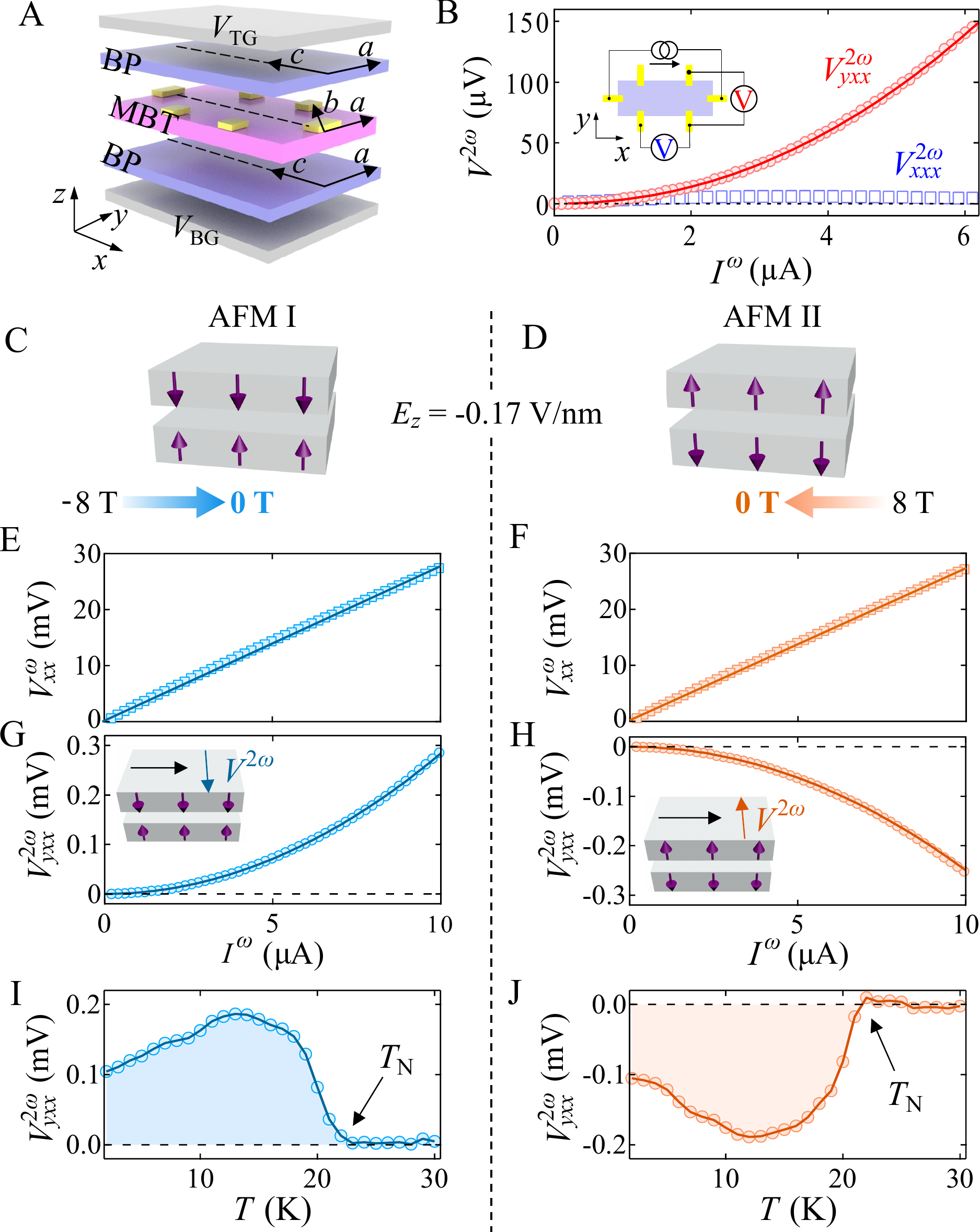}
\vspace{-0.5cm}
\caption{{\bf Observation of the antiferromagnetic nonlinear Hall effect.} (\textbf{A}) Schematic illustration of 2L BP/6SL MnBi$_2$Te$_4$/2L BP device (Device-BMB1). The crystalline $a$ axes of the BPs and the MnBi$_2$Te$_4$ were all aligned (see fig. S10).  (\textbf{B}) The longitudinal ($V^{2\omega}_{xxx}$) and Hall ($V^{2\omega}_{yxx}$) components of the nonlinear voltage. (\textbf{C} and \textbf{D}) We follow the procedures established by previous works \cite{gao2021layer}: under a fixed $E_z$ ($-0.17$ V/nm), we sweep $B_z$ from $-8$ T to 0 T or from $+8$ T to 0 T to prepare the two AFM states. (\textbf{E} and \textbf{F}) Linear longitudinal voltage as a function of incidence current for AFM I and AFM II. (\textbf{G} and \textbf{I}) Nonlinear Hall voltage as a function of incident current and temperature of AFM I. (\textbf{H} and \textbf{J}) The same as panels (G) and (I) but for AFM II. }
\label{Fig2}
\end{figure*}

Overall, we have fabricated 26 BP/MnBi$_2$Te$_4$ heterostructure devices. In all of the 26 devices, we have observed the nonlinear Hall effect with consistent behaviors as a function of AFM order, spatial direction, scattering time, vertical electric field and doping (see fig. S15 and table S1 for a summary of all 26 devices). Here, we focus on the Device-BMB1 (Fig.~\ref{Fig2}A), which has 2L BP on both sides of 6SL MnBi$_2$Te$_4$. Moreover, we have made sure that the crystalline $a$ axes of the BPs and the MnBi$_2$Te$_4$ are all aligned (Fig.~\ref{Fig2}A). Such a carefully controlled configuration is important to preserve MnBi$_2$Te$_4$'s $\mathcal{PT}$ symmetry, which enforces the Berry curvature and Berry curvature dipole to vanish. Figure~\ref{Fig2}B shows the basic nonlinear transport responses. A large transverse nonlinear response $V^{2\omega}_{yxx}$ is found, showing the nonlinear Hall effect in Device-BMB1. We have also measured the longitudinal nonlinear response $V^{2\omega}_{xxx}$, which shows no observable signal. Therefore, our data reveals an interesting ``Hall dominance'' in the nonlinear transport.

We now focus on exploring how the nonlinear Hall signal depends on opposite AFM states. In ferromagnets, the opposite FM states can be controlled by sweeping $B$ field. In $\mathcal{PT}$-symmetric AFMs including Cr$_2$O$_3$, even-layered CrI$_3$ and even-layered MnBi$_2$Te$_4$ \cite{iyama2013magnetoelectric,jiang2018electric,gao2021layer}, previous works have shown that the opposite AFM states can be controlled by sweeping vertical $B_z$ field under a fixed vertical $E_z$ field. Hence, we follow the procedures established by previous works \cite{gao2021layer}:  under a fixed $E_z$ ($E_z=-0.17$ V/nm), we sweep $B_z$ from $-8$ T to 0 T or from $+8$ T to 0 T to prepare the two AFM states (Fig.~\ref{Fig2}, C and D). We first study the AFM-I. The linear voltage $V^{\omega}_{xx}$ (Fig.~\ref{Fig2}E) exhibits a typical Ohm's law behavior. The nonlinear voltage $V^{2\omega}_{yxx}$ (Fig.~\ref{Fig2}G) is prominent and its sign is positive. We then prepare AFM-II. The linear voltage $V^{\omega}_{xx}$ (Fig.~\ref{Fig2}F) remains unchanged. In sharp contrast, the nonlinear voltage $V^{2\omega}_{yxx}$ (Fig.~\ref{Fig2}H) flips sign. For both AFM-I and II, if we measure $V^{2\omega}_{yxx}$ while warming up, we found that the nonlinear Hall effect is only present in the AFM phase but is absent in the nonmagnetic phase (Fig.~\ref{Fig2}, I and J). Therefore, we demonstrate that our nonlinear Hall effect arises from a spin-induced nonlinearity in the Fermi surface electrons.

\begin{figure*}[t]
\includegraphics[width=11.5cm]{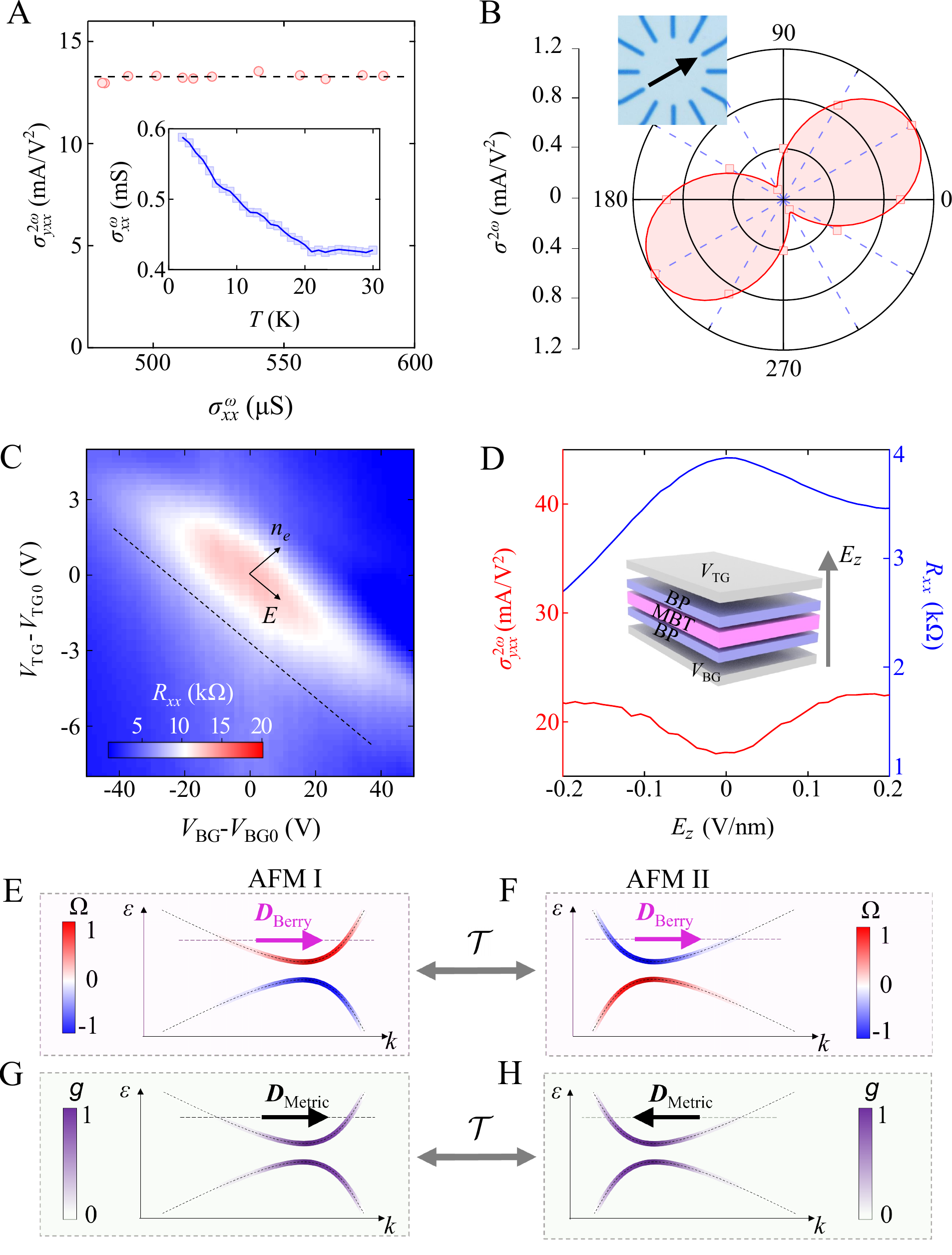}
\vspace{-0.5cm}
\caption{{\bf Systematic investigations of the nonlinear Hall effect.} (\textbf{A}) The scaling between the nonlinear Hall conductivity and the Drude conductivity $\sigma_{xx}=1/R_{xx}$. The nonlinear Hall conductivity can be directly extracted from the data as $\sigma^{2\omega}_{yxx}=\frac{V^{2\omega}_{yxx}}{{I^{\omega}_{x}}^{2}R_{xx}^3}\frac{l^3}{w^2d}$. (\textbf{B}) Angular dependence of the nonlinear Hall conductivity $\sigma^{2\omega}$ in Device-BM1. (\textbf{C}) Dual gated resistance map of the 2L BP/6SL MnBi$_2$Te$_4$/2L BP heterostructure (Device-BMB1). The vertical electric field $E_z$ and carrier density dependence can be independently tuned by combining the top and bottom gate voltages. (\textbf{D}) $E_z$ dependence of the nonlinear Hall conductivity $\sigma^{2\omega}_{yxx}$ and linear longitudinal resistance $R_{xx}$. $E_z$ follows the dashed line in (C). (\textbf{E} to \textbf{H}) Schematic illustration of the Berry curvature dipole ($D_{\textrm{Berry}}$) and quantum metric dipole ($D_{\textrm{Metric}}$) for the AFM I and AFM II. Although we aligned the crystalline axes of BP and MnBi$_2$Te$_4$ in our Device-BMB1 (Fig.~\ref{Fig2}A), realistically it is difficult to make alignment perfect. If the alignment is imperfect and $\mathcal{PT}$ symmetry is broken, a Berry curvature dipole is allowed. }
\label{Fig3}
\end{figure*}

We now perform further systematic studies. Because the nonlinear Hall current flips sign upon reversing the AFM order, all the nonlinear Hall data (apart from Fig. 2) are obtained by taking the difference between the two AFM domains. First, the intrinsic nonlinear Hall effect is expected to be dissipationless. Interestingly, this represents the first known dissipationless nonlinear transport effect. Here, ``dissipationless'' means that the intrinsic nonlinear Hall conductivity is independent of the scattering time $\tau$ \cite{gao2014field, Wang2021Intrinsic,liu2021intrinsic}, just like the intrinsic AHE in ferromagnetic metals was referred as a dissipationless effect \cite{nagaosa2010anomalous} when the anomalous Hall conductivity is independent of $\tau$. In both cases, there is still dissipation through the linear Drude conductivity $\sigma_{xx}$. So they are different from the QAHE that has no dissipation channel at all.  The nonlinear Hall conductivity can be directly extracted from our data by $\sigma^{2\omega}_{yxx}=J^{2\omega}_{yxx}/{E^{\omega}_{x}}^{2}=\frac{V^{2\omega}_{yxx}}{{I^{\omega}_{x}}^{2}R_{xx}^3}\frac{l^3}{w^2d}$, where $l,w,d$ are the length, width and thickness of the sample. Previous experiments have studied the scattering time $\tau$ dependence of various Hall effects \cite{nagaosa2010anomalous,Kang2019nonlinear,kumar2021room,he2022graphene} by investigating the scaling between the corresponding Hall conductivity and the Drude conductivity. Therefore, following the established method, we study the scaling between $\sigma^{2\omega}_{yxx}$ and $\sigma_{xx}$. Our data (Fig.~\ref{Fig3}A) show that $\sigma^{2\omega}_{yxx}$ is independent of $\sigma_{xx}$, consistent with being non-dissipative. Second, the intrinsic nonlinear Hall effect does not require a noncentrosymmetric lattice or any explicit breaking of $\mathcal{PT}$ symmetry. To test this, we explicitly break $\mathcal{PT}$ by applying a vertical $E_z$ field via dual gating. As shown in Fig.~\ref{Fig3}D, the nonlinear Hall signal is already prominent even at $E_z=0$, confirming that it does not require any $\mathcal{PT}$ breaking. Moreover, the nonlinear Hall signal is symmetric for $\pm E_z$, also consistent with the expectation (see SM. IV.2). Third, the nonlinear Hall effect is expected to be sensitive to the direction of the incident current $I^{\omega}$. In Fig.~\ref{Fig3}B, we measure the nonlinear Hall conductivity as a function of the direction of $I^{\omega}$. Indeed, we found that the signal is most prominent when $I^{\omega}$ is along a particular in-plane direction. In this way, we managed to experimentally map out the direction of the relevant geometrical dipole (in our case it is the quantum metric dipole as we demonstrate next). 
 
\vspace{0.5cm}
\textbf{Demonstrating the quantum metric mechanism by excluding competing mechanisms}

Although we tried to eliminate Berry curvature dipole by aligning the crystalline $a$ axes between BPs and MnBi$_2$Te$_4$ to preserve $\mathcal{PT}$ symmetry (Fig.~\ref{Fig2}A). Let us assume that the alignment is imperfect, so Berry curvature dipole is allowed. We now show that the observed relationship between the nonlinear Hall signal and AFM order can discern Berry curvature dipole $D_{\mathrm{Berry}}$ and quantum metric dipole $D_{\mathrm{Metric}}$ \cite{Wang2021Intrinsic}. $D_{\mathrm{Berry}}$ can be understood as a distribution of the Berry curvature around the Fermi surface such that it is larger on one side of the Fermi surface than on the opposite side. A similar picture holds for $D_{\mathrm{Metric}}$ (Fig.~\ref{Fig3}). As we observe that the nonlinear Hall signal changes sign upon the reversal of AFM order, the dipole that causes our observed nonlinear Hall signal must also flip. Let us assume that the AFM-I has $D_{\mathrm{Berry}}>0$ and $D_{\mathrm{Metric}}>0$, which is visualized in a tilted gapped Dirac band structure in Figs.~\ref{Fig3}E and G. We now flip the AFM order to the AFM-II by performing time reversal $\mathcal{T}$. Under $\mathcal{T}$, the bands are flipped between $\pm \mathbf{k}$ (Figs.~\ref{Fig3}F-H), the Berry curvature flips sign ($\Omega(k) \xrightarrow{\mathcal{T}} -\Omega(-k)$), but the quantum metric keeps the same sign ($g(k) \xrightarrow{\mathcal{T}} g(-k)$). Hence, from Figs.~\ref{Fig3}F-H, one can see that, $D_{\mathrm{Berry}}(\textrm{AFM-II})=D_{\mathrm{Berry}}(\textrm{AFM-I})$, but $D_{\mathrm{Metric}}(\textrm{AFM-II})=-D_{\mathrm{Metric}}(\textrm{AFM-I})$. Therefore, our observation that the nonlinear Hall signal flips sign upon reversing the AFM order excludes the Berry curvature dipole mechanism.

Within the nonlinear effects that flip sign upon reversing the AFM order, there is another possibility, the second-order Drude effect \cite{Wang2021Intrinsic, rectification, isobe2020high, Zhang2022Diodic}. This effect can be ruled out based on our scaling data in Fig.~\ref{Fig3}A, because it is expected to be proportional to $\tau^2$ \cite{Wang2021Intrinsic}. Moreover, the nonlinear Hall effect  (NHE) is antisymmetric (upon exchanging the first two indices) $\sigma^{\textrm{NHE}}_{\alpha\beta\gamma}=-\sigma^{\textrm{NHE}}_{\beta\alpha\gamma}$  but the second-order Drude effect (SODE) is symmetric $\sigma^{\textrm{SODE}}_{\alpha\beta\gamma}=\sigma^{\textrm{SODE}}_{\beta\alpha\gamma}$ \cite{Wang2021Intrinsic}. Using a novel electrical sum-frequency generation method (SM. II.2), we showed that our signal is indeed antisymmetric, i.e., $\sigma^{2\omega}_{yxx}=-\sigma^{2\omega}_{xyx}$, which demonstrates that the SODE is insignificant in our signal (SM II.2). Finally, we also carefully addressed other competing origins such as thermal and accidental diode junctions (SM. II.3). By excluding competing mechanisms, we establish the quantum metric dipole as the underlying interpretation.

\vspace{0.5cm}
\textbf{Energy-resolved probe of quantum metric in $\mathcal{PT}$-symmetric AFM}

\begin{figure*}[t]
\includegraphics[width=14.5cm]{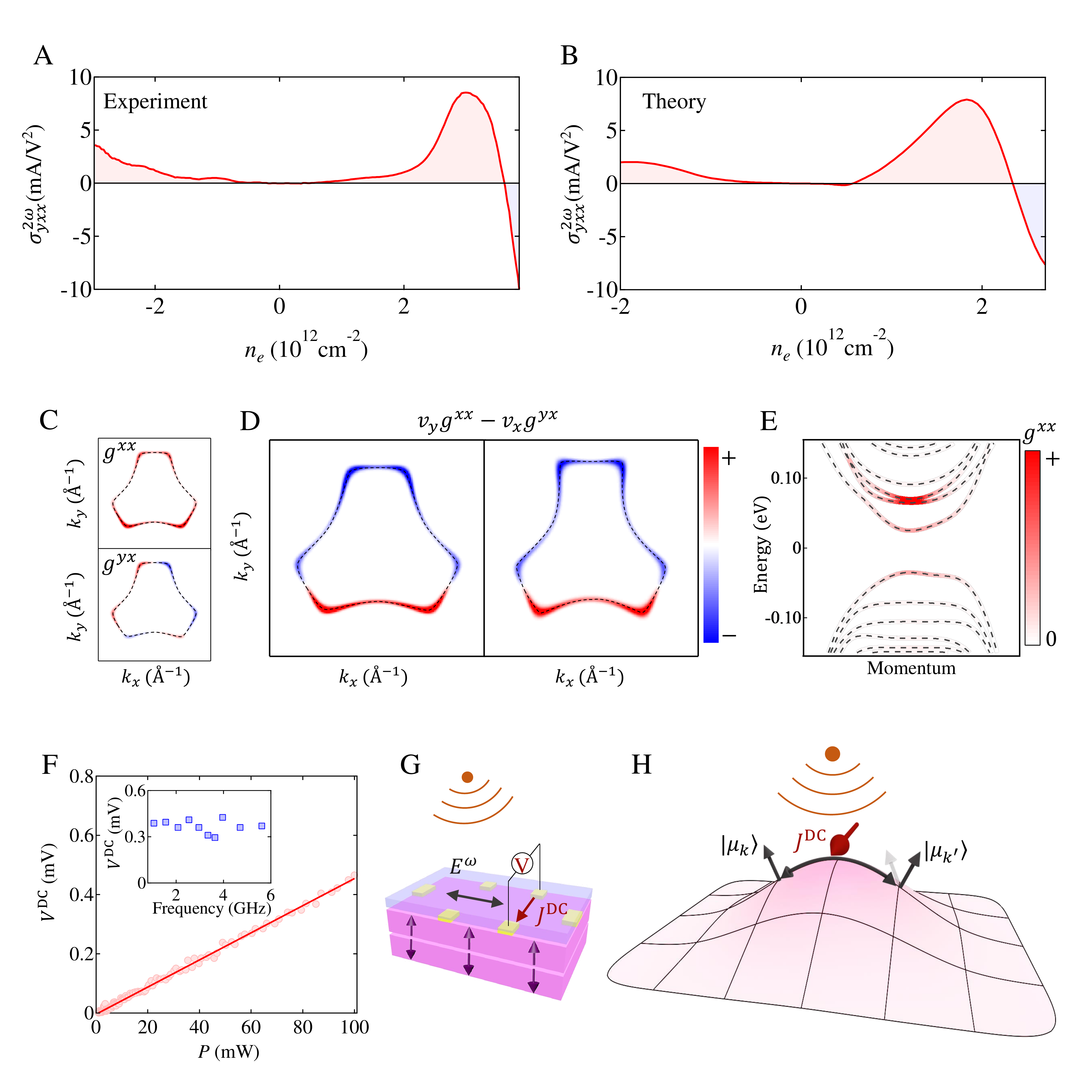}
\vspace{-0.5cm}
\caption{{\bf The quantum metric dipole as the microscopic geometrical origin.} (\textbf{A}) Experimentally measured nonlinear Hall conductivity $\sigma^{2\omega}_{yxx}$ as a function of carrier density $n$. (\textbf{B}) Theoretically calculated $\sigma^{2\omega}_{yxx}$ as a function of $n$ based on the BP/6SL MnBi$_2$Te$_4$/BP band structure (see text). (\textbf{C} to \textbf{E}) The electronic structure of the BP/6SL MnBi$_2$Te$_4$/BP heterostructure calculated with an effective model (see text). (\textbf{C}) Fermi surface at $-50$ meV (the lower part of the surface Dirac cone). The coupling between MnBi$_2$Te$_4$ and BP is turned off, so that contour respects $C_{3z}$ symmetry. The quantum metric $g_{xx}$ and $g_{yx}$ plotted around the Fermi surface. (\textbf{D}) The nonlinear Hall conductivity $\sigma^{2\omega}_{yxx}$ is given by the integral of $(v_yg_{xx}-v_xg_{yx})$ (the difference between the two quantum metric dipoles) around the Fermi surface. With $C_{3z}$ symmetry, the integral goes to zero. After turning on the coupling between MnBi$_2$Te$_4$ and BP, $C_{3z}$ is broken, making the integral of ($v_yg_{xx}-v_xg_{yx}$) around the Fermi surface nonzero. (\textbf{E}) Band structure of BP/6SL MnBi$_2$Te$_4$/BP heterostructure. Color represents the quantum metric $g_{xx}$ of the bands. (\textbf{F}) Measured microwave rectification based on the intrinsic nonlinear Hall effect. Inset is the DC signal $V^{\textrm{DC}}$ as a function of microwave frequency. (\textbf{G}) Schematic illustration of microwave rectification. (\textbf{H}) Schematic illustration of quantum metric induced nonlinear responses.}
\label{Fig4}
\end{figure*}

We also study the evolution of the nonlinear conductivity $\sigma_{yxx}^{2\omega}$ with the charge density $n$. As shown in Fig.~\ref{Fig4}A, the nonlinear Hall signal is zero inside the charge neutrality gap. This is consistent with the expectation that the nonlinear Hall effect is a Fermi surface property. As we tune the Fermi energy away from the charge neutrality, the nonlinear Hall signal emerges. Importantly, the conductivity in electron and hole regimes have the same sign. As we go deeper into the electron-doped regime, the signal reverses sign again.

We now provide an intuitive physical picture to understand the large quantum metric dipole and its Fermi level dependence. MnBi$_2$Te$_4$ features Dirac surface states, which are gapped due to the AFM, leading to large quantum metric near the gap edge. Moreover, because the AFM order breaks both $\mathcal{T}$ and $\mathcal{P}$, the Dirac bands are asymmetric about $\mathbf{k}=0$, as shown in Fig.~\ref{Fig3}G. Hence, at a fixed energy, positive and negative momenta have different quantum metric, leading to a nonzero quantum metric dipole. Intuitively, we can understand the sign of the nonlinear Hall signal by which momentum side has a larger quantum metric. We see from Fig.~\ref{Fig3}G that both upper and lower parts of the Dirac cone have $g(+k_{\mathrm{F}})>g(-k_{\mathrm{F}}$), suggesting that the nonlinear Hall signals should show the same sign in electron and hole regimes, consistent with our data (Fig.~\ref{Fig4}A). The additional sign change in the electron-doped regime is beyond this simple picture.

\newenvironment{psmallmatrix}

To achieve a more comprehensive understanding, we built an effective model of the BP/6SL MnBi$_2$Te$_4$/BP  heterostructure.
Due to the incommensurability of the BP and MnBi$_2$Te$_4$ lattices, we need to derive the coupling between the Bloch states of the two materials in the real-space continuum (i.e. within the extended Brillouin zone BZ). The low-energy bands are located in the BZ center $\Gamma$, so only Bloch bands with the same momentum hybridize. The coupling amplitude depends only on the characteristic decay length of the atomic orbitals as any discrete lattice structure is averaged out \cite{akamatsu2021van}. The Hamiltonian reads $\hat{h}(k_x, k_y)=\left(\begin{smallmatrix}
        \hat{h}_{\text{MBT}} & \hat{U}_t & \hat{U}_b\\
        \hat{U}_t^{\dag} & \hat{h}_{\text{BP},t} & 0\\
        \hat{U}_b^{\dag} & 0 & \hat{h}_{\text{BP},b}
    \end{smallmatrix}\right)$ where  $\hat{h}_{\text{MBT}}$ and $\hat{h}_{\text{BP},t(b)}$ are Hamiltonians for 6SL MnBi$_2$Te$_4$ and top (bottom) BP, respectively. $\hat{U}_t$ and $\hat{U}_b$ denote the nearest-neighbors coupling between MnBi$_2$Te$_4$ and BP, which is crucial for breaking the $C_{3z}$. 
    
We first turn off the coupling between the MnBi$_2$Te$_4$ and BP ($\hat{U}_t=\hat{U}_b=0$). The Fermi surface shown in Fig.~\ref{Fig4}C ($-50$ meV) is $C_{3z}$ symmetric and there are already large quantum metric ($g_{xx}$ and $g_{yx}$) around it. According to Ref. \cite{Wang2021Intrinsic}, the $D_{\mathrm{Metric}}$ responsible for the nonlinear Hall is given by $D_{\mathrm{Metric}}=\int_\mathbf{k} (v_yg_{xx}-v_xg_{yx}) \delta(\varepsilon-\varepsilon_{\mathrm{F}})$ ($v$ is the Fermi velocity). We plot the integral kernel ($v_yg_{xx}-v_xg_{yx}$) as color in Fig.~\ref{Fig4}D. Positive and negative contributions around the contour exactly cancel because of $C_{3z}$ symmetry. So the integral goes to zero (the left panel in Fig.~\ref{Fig4}D). We then turn on the MnBi$_2$Te$_4$-BP  couplings, which breaks $C_{3z}$. For the $C_{3z}$-breaking contour, we observe unequal contributions from the two colors, leading to a nonzero $D_{\textrm{Metric}}$ (the right panel in Fig.~\ref{Fig4}D). Figure~\ref{Fig4}E shows the band structure of the BP/6SL MnBi$_2$Te$_4$/BP heterostructure, based on which we can compute the intrinsic nonlinear Hall conductivity $\sigma^{2\omega}_{yxx}$ as a function of chemical potential. In particular, near the charge neutrality gap, we found that $\sigma^{2\omega}_{yxx}$ indeed mainly comes from the quantum metric of the Dirac surface states, consistent with the intuitive picture above. The sign inversion in the electron-doped regime mainly comes from the quantum metric of the avoided crossing inside conduction bands. Note that due to the multiband nature of our model,  the $\sigma_{yxx}^{2\omega}$ was calculated by the general expression $\sigma_{yxx}^{2\omega}= -2 e^3 \sum^{\varepsilon_n \neq \varepsilon_m}_{n,m} \text{Re} \int_\mathbf{k} \bigl( \frac{v_y^n \langle u_n\arrowvert i\partial_{k_x}u_m\rangle \langle u_m\arrowvert i\partial_{k_x}u_n\rangle}{\varepsilon_n - \varepsilon_m} -  \frac{v_x^n \langle u_n\arrowvert i\partial_{k_y}u_m\rangle \langle u_m\arrowvert i\partial_{k_x}u_n\rangle}{\varepsilon_n - \varepsilon_m} \bigr)\delta(\varepsilon_n-\varepsilon_{\mathrm{F}})$~\cite{Wang2021Intrinsic}. This general expression can be decomposed into the quantum metric dipole $D_{\mathrm{Metric}}$ contribution plus additional inter-band contributions (AIC), 
\begin{equation}
\sigma_{yxx}^{2\omega}= -2 e^3 \sum_{n} \int_\mathbf{k}\frac{v_y^n g^n_{xx}-v_x^n g^n_{yx}}{\varepsilon_n - \varepsilon_{\bar{n}}}\delta(\varepsilon_n-\varepsilon_{\mathrm{F}}) + \textrm{AIC},
\end{equation}
where the first term is the quantum metric dipole contribution, and the second term is $\textrm{AIC}=-2 e^3 \sum_{n,m}^{\varepsilon_m \neq \varepsilon_n, \varepsilon_{\bar{n}}} \text{Re} \int_\mathbf{k} \bigl( \frac{v_y^n \langle u_n\arrowvert i\partial_{k_x}u_m\rangle \langle u_m\arrowvert i\partial_{k_x}u_n\rangle}{\varepsilon_n - \varepsilon_m} -  \frac{v_x^n \langle u_n\arrowvert i\partial_{k_y}u_m\rangle \langle u_m\arrowvert i\partial_{k_x}u_n\rangle}{\varepsilon_n - \varepsilon_m} \bigr)\frac{\varepsilon_m-\varepsilon_{\bar{n}}}{\varepsilon_n - \varepsilon_{\bar{n}}}\delta(\varepsilon_n-\varepsilon_{\mathrm{F}})$ ($\bar{n}$ is the band whose energy is closest to $n$). In our BP/6SL MnBi$_2$Te$_4$/BP system, we found that the quantum metric dipole contribution strongly dominates, whereas the AIC is small (see details in SM. IV.3).

By comparing the calculated and measured $\sigma^{2\omega}_{yxx}$ (Fig.~\ref{Fig4}, A and B), we found a good agreement. Therefore, our nonlinear Hall measurement is a powerful, energy-resolved probe of the quantum metric.

\vspace{0.5cm}
\textbf{AFM spin-based wireless rectification and outlook}

The second-order nonlinear effect enables not only frequency doubling ($\omega\rightarrow2\omega$) but also rectification ($\omega\rightarrow\textrm{DC}$). The rectification is crucial for harvesting electromagnetic radiation energy \cite{kumar2021room,isobe2020high} because we can convert the electromagnetic radiation into DC electricity. We use the intrinsic AFM nonlinear Hall effect to demonstrate wireless rectification with zero external bias (battery-free) and without magnetic field. We inject microwave radiation and measure the DC signal. As shown in Fig.~\ref{Fig4}F, we observe clear rectification DC voltage in response to the microwave radiation, which shows a broad band response, including the WiFi frequencies (2.4 and 5 GHz) and even higher frequencies (see fig. S21).

In summary, we have presented the first experimental realization of the intrinsic second-order Hall effect. This effect realizes an electrical nonlinearity induced by the AFM spins and provides a rare example of a quantum metric response. Both aspects are of fundamental interest. Just like the AHE about a decade ago inspired the discoveries of a variety of Berry curvature responses, we hope that our work opens the door to experimentally search for quantum metric responses. As highlighted by recent theoretical studies, the influence of the quantum metric is expected to span many different areas, ranging from nonlinear responses in $\mathcal{PT}$-symmetric AFMs to flat band conductivity, superconductivity and charge orders in moir\'e systems, the fractional Chern insulator, and $\mathbf{k}$-space dual of gravity \cite{gao2014field,Wang2021Intrinsic,liu2021intrinsic, lahiri2022intrinsic, Smith2022momentum, Arora2022quantuma,Mitscherling2021, Rhim2020quantum, ledwith2020fractional, Holder2020consequences,  Watanabe2021chiral,Huhtinen2022revisting, Hofmann2022superconductivity,Hu2020quantum}. Another interesting future direction is to explore the nonlinear responses in canted AFM materials, where nonzero Berry curvature of higher order in magnetization have recently been observed (\cite{Lee2019a, bac2022topological, kipp2021chiral}).
 In terms of materials, the vdW interface engineering has been widely applied to engineer band structure, such as the band alignment in semiconductors. We show that, beyond ``band structure engineering'', the vdW interfaces can be used to engineer the properties of the wavefunction i.e., ``quantum geometry engineering'' \cite{akamatsu2021van}. We demonstrate that, the topological Dirac surface state on the interface of a TI can be the source of a wide range of novel topological and geometrical phenomena beyond the Berry curvature upon proper engineering. In terms of spin-induced electrical nonlinearity, our observation enables the possibility to use AFM spins to harvest electromagnetic energy and to realize self-powered AFM spintronic devices.

\vspace{0.5cm}

\bibliographystyle{Science}
\bibliography{Quantum_Metric}

\begin{thebibliography}{10}

\bibitem{provost1980riemannian}
J.~Provost, G.~Vallee, Riemannian structure on manifolds of quantum states,
  {\it Commun. Math. Phys.\/} {\bf 76}, 289 (1980).

\bibitem{xiao2010berry}
D.~Xiao, M.-C. Chang, Q.~Niu, Berry phase effects on electronic properties,
  {\it Rev. Mod. Phys.\/} {\bf 82}, 1959 (2010).

\bibitem{tokura2018nonreciprocal}
Y.~Tokura, N.~Nagaosa, Nonreciprocal responses from non-centrosymmetric quantum
  materials, {\it Nature Commun.\/} {\bf 9}, 1 (2018).

\bibitem{ma2021topology}
Q.~Ma, A.~G. Grushin, K.~S. Burch, Topology and geometry under the nonlinear
  electromagnetic spotlight, {\it Nature Mater.\/} {\bf 20}, 1601 (2021).

\bibitem{orenstein2021topology}
J.~Orenstein, {\it et~al.\/}, Topology and symmetry of quantum materials via
  nonlinear optical responses, {\it Annu. Rev. Condens. Matter Phys.\/} {\bf
  12}, 247 (2021).

\bibitem{moore2010confinement}
J.~E. Moore, J.~Orenstein, Confinement-induced {B}erry phase and
  helicity-dependent photocurrents, {\it Phys. Rev. Lett.\/} {\bf 105}, 026805
  (2010).

\bibitem{sodemann2015quantum}
I.~Sodemann, L.~Fu, Quantum nonlinear {H}all effect induced by {B}erry
  curvature dipole in time-reversal invariant materials, {\it Phys. Rev.
  Lett.\/} {\bf 115}, 216806 (2015).

\bibitem{rectification}
T.~Ideue, {\it et~al.\/}, Bulk rectification effect in a polar semiconductor,
  {\it Nature Phys.\/} {\bf 13}, 578 (2017).

\bibitem{Kang2019nonlinear}
K.~Kang, T.~Li, E.~Sohn, J.~Shan, K.~F. Mak, Nonlinear anomalous {H}all effect
  in few-layer {W}{T}e$_2$, {\it Nature Mater.\/} {\bf 18}, 324 (2019).

\bibitem{ma2019observation}
Q.~Ma, {\it et~al.\/}, Observation of the nonlinear {H}all effect under
  time-reversal-symmetric conditions, {\it Nature\/} {\bf 565}, 337 (2019).

\bibitem{dzsaber2021giant}
S.~Dzsaber, {\it et~al.\/}, Giant spontaneous {H}all effect in a nonmagnetic
  {W}eyl-{K}ondo semimetal, {\it PNAS\/} {\bf 118}, e2013386118 (2021).

\bibitem{kumar2021room}
D.~Kumar, {\it et~al.\/}, Room-temperature nonlinear {H}all effect and wireless
  radiofrequency rectification in {W}eyl semimetal {T}a{I}r{T}e$_4$, {\it
  Nature Nanotech.\/} {\bf 16}, 421 (2021).

\bibitem{yasuda2020large}
K.~Yasuda, {\it et~al.\/}, Large non-reciprocal charge transport mediated by
  quantum anomalous Hall edge states, {\it Nature Nanotech.\/} {\bf 15}, 831
  (2020).

\bibitem{zhao2020magnetic}
W.~Zhao, {\it et~al.\/}, Magnetic proximity and nonreciprocal current switching
  in a monolayer {WT}e$_2$ helical edge, {\it Nature Mater.\/} {\bf 19}, 503
  (2020).

\bibitem{isobe2020high}
H.~Isobe, S.-Y. Xu, L.~Fu, High-frequency rectification via chiral Bloch
  electrons, {\it Science Advances\/} {\bf 6}, eaay2497 (2020).

\bibitem{Lai2021third}
S.~Lai, {\it et~al.\/}, Third-order nonlinear {H}all effect induced by the
  {B}erry-connection polarizability tensor, {\it Nature Nanotech.\/} {\bf 16},
  869 (2021).

\bibitem{he2022graphene}
P.~He, {\it et~al.\/}, Graphene moir\'e superlattices with giant quantum
  nonlinearity of chiral Bloch electrons, {\it Nature Nanotech.\/} {\bf 17},
  378 (2022).

\bibitem{zhang2022non}
Z.~Zhang, {\it et~al.\/}, Non-reciprocal charge transport in an intrinsic
  magnetic topological insulator {M}n{B}i$_2${T}e$_4$  Preprint at
  https://arxiv.org/abs/2203.09350 (2022).

\bibitem{sinha2022berry}
S.~Sinha, {\it et~al.\/}, Berry curvature dipole senses topological transition
  in a moir\'e superlattice, {\it Nature Phys.\/}  1--6 (2022).

\bibitem{Zhang2022Diodic}
N.~J. Zhang, {\it et~al.\/}, Diodic transport response and the loop current
  state in twisted trilayer graphene  Preprint at
  https://arxiv.org/abs/2209.12964 (2022).

\bibitem{fiebig1994second}
M.~Fiebig, D.~Fr{\"o}hlich, B.~Krichevtsov, R.~V. Pisarev, Second harmonic
  generation and magnetic-dipole-electric-dipole interference in
  antiferromagnetic {C}r$_2${O}$_3$, {\it Phys. Rev. Lett.\/} {\bf 73}, 2127
  (1994).

\bibitem{sun2019giant}
Z.~Sun, {\it et~al.\/}, Giant nonreciprocal second-harmonic generation from
  antiferromagnetic bilayer {C}r{I}$_3$, {\it Nature\/} {\bf 572}, 497 (2019).

\bibitem{gao2015geometrical}
Y.~Gao, S.~A. Yang, Q.~Niu, Geometrical effects in orbital magnetic
  susceptibility, {\it Phys. Rev. B\/} {\bf 91}, 214405 (2015).

\bibitem{Gianfrate2020measurement}
A.~Gianfrate, {\it et~al.\/}, Measurement of the quantum geometric tensor and
  of the anomalous {H}all drift, {\it Nature\/} {\bf 578}, 381 (2020).

\bibitem{gao2014field}
Y.~Gao, S.~A. Yang, Q.~Niu, Field induced positional shift of {B}loch electrons
  and its dynamical implications, {\it Phys. Rev. Lett.\/} {\bf 112}, 166601
  (2014).

\bibitem{Wang2021Intrinsic}
C.~Wang, Y.~Gao, D.~Xiao, Intrinsic Nonlinear {H}all Effect in
  Antiferromagnetic Tetragonal {C}u{M}n{A}s, {\it Phys. Rev. Lett.\/} {\bf
  127}, 277201 (2021).

\bibitem{liu2021intrinsic}
H.~Liu, {\it et~al.\/}, Intrinsic Second-Order Anomalous {H}all Effect and Its
  Application in Compensated Antiferromagnets, {\it Phys. Rev. Lett.\/} {\bf
  127}, 277202 (2021).

\bibitem{lahiri2022intrinsic}
S.~Lahiri, K.~Das, D.~Culcer, A.~Agarwal, Intrinsic nonlinear conductivity
  induced by the quantum metric dipole  Preprint at
  https://arxiv.org/abs/2207.02178 (2022).

\bibitem{Smith2022momentum}
T.~B. Smith, L.~Pullasseri, A.~Srivastava, Momentum-space gravity from the
  quantum geometry and entropy of {B}loch electrons, {\it Phys. Rev.
  Research\/} {\bf 4}, 013217 (2022).

\bibitem{Arora2022quantuma}
A.~Arora, M.~S. Rudner, J.~C.~W. Song, Quantum metric dipole and non-reciprocal
  bulk plasmons in parity-violating magnets  Preprint at
  https://arxiv.org/abs/2202.08284 (2022).

\bibitem{Mitscherling2021}
J.~Mitscherling, T.~Holder, Bound on resistivity in flat-band materials due to
  the quantum metric, {\it Phys. Rev. B\/} {\bf 105}, 085154 (2021).

\bibitem{Rhim2020quantum}
J.-W. Rhim, K.~Kim, B.-J. Yang, Quantum distance and anomalous {L}andau levels
  of flat bands, {\it Nature\/} {\bf 584}, 59 (2020).

\bibitem{ledwith2020fractional}
P.~J. Ledwith, G.~Tarnopolsky, E.~Khalaf, A.~Vishwanath, Fractional {C}hern
  insulator states in twisted bilayer graphene: An analytical approach, {\it
  Phys. Rev. Research\/} {\bf 2}, 023237 (2020).

\bibitem{Holder2020consequences}
T.~Holder, D.~Kaplan, B.~Yan, Consequences of time-reversal-symmetry breaking
  in the light-matter interaction: {B}erry curvature, quantum metric, and
  diabatic motion, {\it Phys. Rev. Research\/} {\bf 2}, 033100 (2020).

\bibitem{Watanabe2021chiral}
H.~Watanabe, Y.~Yanase, Chiral Photocurrent in Parity-Violating Magnet and
  Enhanced Response in Topological Antiferromagnet, {\it Phys. Rev. X\/} {\bf
  11}, 011001 (2021).

\bibitem{Huhtinen2022revisting}
K.-E. Huhtinen, J.~Herzog-Arbeitman, A.~Chew, B.~A. Bernevig, P.~T{\"o}rm{\"a},
  Revisiting flat band superconductivity: dependence on minimal quantum metric
  and band touchings  Preprint at https://arxiv.org/abs/2203.11133 (2022).

\bibitem{Hofmann2022superconductivity}
J.~S. Hofmann, E.~Berg, D.~Chowdhury, Superconductivity, charge density wave,
  and supersolidity in flat bands with tunable quantum metric  Preprint at
  https://arxiv.org/abs/2204.02994 (2022).

\bibitem{Hu2020quantum}
X.~Hu, T.~Hyart, D.~I. Pikulin, E.~Rossi, Quantum-metric-enabled exciton
  condensate in double twisted bilayer graphene, {\it Phys. Rev. B\/} {\bf
  105}, L140506 (2022).

\bibitem{nagaosa2010anomalous}
N.~Nagaosa, J.~Sinova, S.~Onoda, A.~MacDonald, N.~P. Ong, Anomalous {H}all
  effect, {\it Rev. Mod. Phys.\/} {\bf 82}, 1539 (2010).

\bibitem{Otrokov2019a}
M.~M. Otrokov, {\it et~al.\/}, Prediction and observation of an
  antiferromagnetic topological insulator, {\it Nature\/} {\bf 576}, 416
  (2019).

\bibitem{Zhang2019a}
D.~Zhang, {\it et~al.\/}, Topological axion states in the magnetic insulator
  {M}n{B}i$_2${T}e$_4$ with the quantized magnetoelectric effect, {\it Phys.
  Rev. Lett.\/} {\bf 122}, 206401 (2019).

\bibitem{Deng2020}
Y.~Deng, {\it et~al.\/}, Quantum anomalous {H}all effect in intrinsic magnetic
  topological insulator {M}n{B}i$_2${T}e$_4$, {\it Science\/} {\bf 367}, 895
  (2020).

\bibitem{Liu2020a}
C.~Liu, {\it et~al.\/}, Robust axion insulator and {C}hern insulator phases in
  a two-dimensional antiferromagnetic topological insulator, {\it Nature
  Mater.\/} {\bf 19}, 522 (2020).

\bibitem{deng2021high}
H.~Deng, {\it et~al.\/}, High-temperature quantum anomalous Hall regime in a
  {MnBi$_2$Te$_4$/Bi$_2$Te$_3$} superlattice, {\it Nature Phys.\/} {\bf 17}, 36
  (2021).

\bibitem{Ovchinnikov2020}
D.~Ovchinnikov, {\it et~al.\/}, Intertwined Topological and Magnetic Orders in
  Atomically Thin {C}hern Insulator {M}n{B}i$_2${T}e$_4$, {\it Nano Lett.\/}
  {\bf 21}, 2544 (2021).

\bibitem{gao2021layer}
A.~Gao, {\it et~al.\/}, Layer {H}all effect in a 2{D} topological axion
  antiferromagnet, {\it Nature\/} {\bf 595}, 521 (2021).

\bibitem{Lee2019a}
S.H. ~Lee, {\it et~al.\/}, Spin scattering and noncollinear spin structure-induced intrinsic anomalous {H}all effect in antiferromagnetic topological insulator {M}n{B}i$_2${T}e$_4$,
 {\it Phys. Rev. Res.\/} {\bf 1}, 012011 (2019).

\bibitem{cai2021electric}
J.~Cai, {\it et~al.\/}, Electric control of a canted-antiferromagnetic {C}hern
  insulator, {\it Nature Commun.\/} {\bf 13}, 1668 (2022).

\bibitem{tai2021polarity}
L.~Tai, {\it et~al.\/}, Distinguishing two-component anomalous {H}all effect
  from topological {H}all effect in magnetic topological insulator
  {M}n{B}i$_2${T}e$_4$.  Preprint at https://arxiv.org/abs/2103.09878 (2021).

\bibitem{bac2022topological}
S.-K.~Bac, {\it et~al.\/}, Topological response of the anomalous {H}all effect in {M}n{B}i$_2${T}e$_4$ due to magnetic canting,  {\it npj Quantum Materials.\/} {\bf 7}, 1-7 (2022).

\bibitem{akamatsu2021van}
T.~Akamatsu, {\it et~al.\/}, A van der {W}aals interface that creates in-plane
  polarization and a spontaneous photovoltaic effect, {\it Science\/} {\bf
  372}, 68 (2021).

\bibitem{xia2014rediscovering}
F.~Xia, H.~Wang, Y.~Jia, Rediscovering black phosphorus as an anisotropic
  layered material for optoelectronics and electronics, {\it Nature Commun.\/}
  {\bf 5}, 1 (2014).

\bibitem{iyama2013magnetoelectric}
A.~Iyama, T.~Kimura, Magnetoelectric hysteresis loops in {C}r$_2${O}$_3$ at
  room temperature, {\it Phys. Rev. B\/} {\bf 87}, 180408 (2013).

\bibitem{jiang2018electric}
S.~Jiang, J.~Shan, K.~F. Mak, Electric-field switching of two-dimensional van
  der {W}aals magnets, {\it Nature Mater.\/} {\bf 17}, 406 (2018).
  
 \bibitem{kipp2021chiral}
J.~Kipp, {\it et~al.\/}, The chiral Hall effect in canted ferromagnets and
  antiferromagnets, {\it Commun. Phys.\/} {\bf 4}, 99 (2021).

\bibitem{yan2019}
J.-Q. Yan, {\it et~al.\/}, Crystal growth and magnetic structure of
  {M}n{B}i$_2${T}e$_4$, {\it Phys. Rev. Materials\/} {\bf 3}, 064202 (2019).

\bibitem{zhao2021Emergent}
S.~Zhao, {\it et~al.\/}, Emergent interfacial superconductivity between twisted
  cuprate superconductors.  Preprint at https://arxiv.org/abs/2108.13455
  (2021).

\bibitem{DFT2}
G.~Kresse, J.~Furthm\"uller, Efficient iterative schemes for ab initio
  total-energy calculations using a plane-wave basis set, {\it Phys. Rev. B\/}
  {\bf 54}, 11169 (1996).

\bibitem{steiner2016calculation}
S.~Steiner, S.~Khmelevskyi, M.~Marsmann, G.~Kresse, Calculation of the magnetic
  anisotropy with projected-augmented-wave methodology and the case study of
  disordered {F}e$_{1- x}${C}o$_x$ alloys, {\it Phys. Rev. B\/} {\bf 93},
  224425 (2016).

\bibitem{lian2020flat}
B.~Lian, Z.~Liu, Y.~Zhang, J.~Wang, Flat {C}hern Band from Twisted Bilayer
  {M}n{B}i$_2${T}e$_4$, {\it Phys. Rev. Lett.\/} {\bf 124}, 126402 (2020).

\bibitem{rudenko2015toward}
A.~Rudenko, S.~Yuan, M.~Katsnelson, Toward a realistic description of
  multilayer black phosphorus: From GW approximation to large-scale
  tight-binding simulations, {\it Phys. Rev. B\/} {\bf 92}, 085419 (2015).

\bibitem{hsieh2011Selective}
D.~Hsieh, {\it et~al.\/}, Selective probing of photoinduced charge and spin
  dynamics in the bulk and surface of a topological insulator, {\it Phys. Rev.
  Lett.\/} {\bf 107}, 077401 (2011).

\bibitem{Ghosh2016Electric}
B.~Ghosh, B.~Singh, R.~Prasad, A.~Agarwal, Electric-field tunable Dirac
  semimetal state in phosphorene thin films, {\it Phys. Rev. B\/} {\bf 94},
  205426 (2016).

\bibitem{walmsley2017determination}
P.~Walmsley, I.~Fisher, Determination of the resistivity anisotropy of
  orthorhombic materials via transverse resistivity measurements, {\it Rev.
  Sci. Ins.\/} {\bf 88}, 043901 (2017).

\bibitem{Zheng2022symmetry}
X.~Zheng, {\it et~al.\/}, Symmetry Engineering Induced In-Plane Polarization in
  {M}o{S}$_2$ through Van der {W}aals Interlayer Coupling, {\it Adv. Funct.
  Mater.\/}  2202658 (2022).

\bibitem{Watanabe2020Nonlinear}
H.~Watanabe, Y.~Yanase, Nonlinear electric transport in odd-parity magnetic
  multipole systems: Application to {M}n-based compounds, {\it Phys. Rev.
  Mater.\/} {\bf 2}, 043081 (2020).

\bibitem{tian2009proper}
Y.~Tian, L.~Ye, X.~Jin, Proper scaling of the anomalous {H}all effect, {\it
  Phys. Rev. Lett.\/} {\bf 103}, 087206 (2009).

\bibitem{Liu2020}
Z.~Liu, J.~Wang, Anisotropic topological magnetoelectric effect in axion
  insulators, {\it Phys. Rev. B\/} {\bf 101}, 205130 (2020).

\bibitem{varjas2018qsymm}
D.~Varjas, T.~{\"O}. Rosdahl, A.~R. Akhmerov, Qsymm: {A}lgorithmic symmetry
  finding and symmetric {H}amiltonian generation, {\it New J. of Phys.\/} {\bf
  20}, 093026 (2018).

\bibitem{ezawa2014topological}
M.~Ezawa, Topological origin of quasi-flat edge band in phosphorene, {\it New
  J. of Phys.\/} {\bf 16}, 115004 (2014).

\bibitem{slater1954simplified}
J.~C. Slater, G.~F. Koster, Simplified {LCAO} method for the periodic potential
  problem, {\it Phys. Rev.\/} {\bf 94}, 1498 (1954).

\bibitem{hemour2014towards}
S.~Hemour, {\it et~al.\/}, Towards low-power high-efficiency {RF} and microwave
  energy harvesting, {\it IEEE Trans. Microw. Theory Tech.\/} {\bf 62}, 965
  (2014).
 
\bibitem{Varjas_2018}
D.~Varjas, T.~. Rosdahl, A.~R. Akhmerov, Qsymm: algorithmic symmetry finding
  and symmetric Hamiltonian generation, {\it New Journal of Physics\/} {\bf
  20}, 093026 (2018).

\bibitem{thiel2019probing}
L.~Thiel, {\it et~al.\/}, Probing magnetism in 2{D} materials at the nanoscale
  with single-spin microscopy, {\it Science\/} {\bf 364}, 973 (2019).

\bibitem{hemour2014radio}
S.~Hemour, K.~Wu, Radio-frequency rectifier for electromagnetic energy
  harvesting: {D}evelopment path and future outlook, {\it Proceedings of the
  IEEE\/} {\bf 102}, 1667 (2014).

\end{thebibliography}

\vspace{0.5cm}

\textbf{Author contributions:}  SYX conceived the experiments and supervised the project. AG fabricated the devices, performed the measurements and analyzed data with help from YFL, DB, JXQ, HCL, CT, TD, ZS, SCH, DCB and QM. AG and SWC performed the microwave rectification experiments. CH, TQ and NN grew the bulk MnBi$_2$Te$_4$ single crystals. BG made the theoretical studies including first-principles calculations and effective modeling with the help from TVT, YO, SWL, BS, HL, AB, TRC, LF and PPO. TVT developed the effective model with help from BG under the guidance of PPO. KW and TT grew the bulk hBN single crystals. SYX, AG and QM wrote the manuscript with input from all authors. 

\textbf{Acknowledgement:} We gratefully thank Amir Yacoby for his generous help for the microwave measurements. We also thank Marie Wesson and Nick Poniatowski for technical support during the microwave measurements. We thank Yang Gao and Junyeong Ahn, Philip Kim for helpful discussions. We also gratefully thank Linda Ye, Masataka Mogi, Yukako Fujishiro, and Takashi Kurumaji for extensive discussions on the scaling of AHE. 
Work in the SYX group was partly supported through the Center for the Advancement of Topological Semimetals (CATS), an Energy Frontier Research Center (EFRC) funded by the U.S. Department of Energy (DOE) Office of Science (fabrication and measurements), through the Ames National Laboratory under contract DE-AC0207CH11358. and partly through Air Force Office of Scientific Research (AFOSR) grant FA9550-23-1-0040 (data analysis and manuscript writing). SYX acknowledges the Corning Fund for Faculty Development. QM and LF acknowledge support from the NSF Convergence program (NSF ITE-2235945) and the CIFAR program. SYX and DB were supported by the NSF Career DMR-2143177. CT and ZS acknowledge support from the Swiss National Science Foundation under project P2EZP2\_191801 and P500PT\_206914, respectively. YFL, SYX, DCB, YO and LF were supported by the STC Center for Integrated Quantum Materials (CIQM), NSF Grant No. DMR-1231319. This work was performed in part at the Center for Nanoscale Systems (CNS) Harvard University, a member of the National Nanotechnology Coordinated Infrastructure Network (NNCI), which is supported by the National Science Foundation under NSF award no.1541959. Bulk single crystal growth and characterization of MnBi2Te4 were performed at UCLA, which were supported by the DOE, office of Science, under Award Number DE-SC0021117. The work at Northeastern University was supported by the Air Force Office of Scientific Research under award number FA9550-20-1-0322, and it benefited from the computational resources of Northeastern University's Advanced Scientific Computation Center (ASCC) and the Discovery Cluster. The work in the QM group was partly supported through the CATS, an EFRC funded by the DOE Office of Science, through the Ames National Laboratory under contract DE-AC0207CH11358 (fabrication and measurements) and partly through NSF DMR-2143426 (data analysis and manuscript writing). TVT and PPO were supported from the CATS, an EFRC funded by the DOE Office of Science, through the Ames National Laboratory under contract DE-AC0207CH11358. TRC was supported by the 2030 Cross-Generation Young Scholars Program from the National Science and Technology Council (NSTC) in Taiwan (program no. MOST111-2628-M- 006-003-MY3), National Cheng Kung University (NCKU), Taiwan, and the National Center for Theoretical Sciences (NCTS), Taiwan. This research was supported, in part, by the Higher Education Sprout Project, Ministry of Education to the Headquarters of University Advancement at NCKU. HL acknowledges the support by the National Science and Technology Council (NSTC) in Taiwan under grant number MOST 111-2112-M-001-057-MY3. The work at TIFR Mumbai was supported by the Department of Atomic Energy of the Government of India under Project No. 12-R\&D-TFR-5.10-0100 and benefited from the computational resources of TIFR Mumbai. KW and TT acknowledge support from the JSPS KAKENHI (Grant Numbers 20H00354, 21H05233 and 23H02052) and World Premier International Research Center Initiative (WPI), MEXT, Japan. MH and CRD were supported by the AFOSR under award no. FA9550-20-1-0319. SWC acknowledges partial support from the Harvard Quantum Initiative in Science and Engineering.

\textbf{Competing financial interests:} The authors declare no competing financial interests.

\clearpage

\clearpage


\clearpage

\clearpage

\renewcommand{\thesubsection}{ IV.\arabic{subsection}}
\renewcommand{\thesubsubsection}{ IV.\arabic{subsection}.\arabic{subsubsection}}

\vspace{3cm}
\Large
\begin{center}
Supplementary Materials for\\
\vspace{1cm}
\textbf{Quantum metric nonlinear Hall effect in a topological antiferromagnetic heterostructure}

\vspace{0.5cm}

\vspace{0.5cm}
Correspondence to: suyangxu$@$fas.harvard.edu
\end{center}

\clearpage
\vspace{3cm}
\textbf{
This file includes:
}

\vspace{3mm}
\normalsize
\textbf{Materials and Methods}

\hspace{5mm}		Bulk crystal growth 

\hspace{5mm}		Sample fabrication

\hspace{5mm}		Nonlinear electrical transport measurements

\hspace{5mm}		Optical second harmonic generation and polarized Raman measurements

\hspace{5mm}		Wireless radio frequency (RF) rectification measurements

\hspace{5mm}		First-principles calculations 

\hspace{5mm}		Theoretical modeling

\textbf{Supplementary Text}

\hspace{5mm}    I. Basic characterization of BP/MnBi$_2$Te$_4$ heterostructure

\hspace{10mm}   I.1. The symmetry of MnBi$_2$Te$_4$, BP/MnBi$_2$Te$_4$ and BP/MnBi$_2$Te$_4$/BP 

\hspace{10mm}   I.2. Determining the crystalline directions for MnBi$_2$Te$_4$ and BP

\hspace{10mm}   I.3. Basic transport characterizations

\hspace{10mm}   I.4. Angular-resolved transport

\hspace{10mm} I.5. Optical second-harmonic generation 

\hspace{5mm}    II. Addressing alternative mechanisms for the nonlinear Hall signals

\hspace{10mm}	II.1. Berry curvature dipole 
 
\hspace{10mm}   II.2. Second-order Drude conductivity

\hspace{10mm}   II.3. Joule heating induced Anomalous Nernst effect

\hspace{10mm}   II.4. Other extrinsic effects

\hspace{10mm}   II.5. Addressing the nonlinear Hall signals induced by skew scattering

\hspace{5mm}    III. Additional data

\hspace{5mm}    IV. Theoretical studies

\hspace{10mm}   IV.1 An intuitive picture of quantum metric

\hspace{10mm}   IV.2 Symmetry analysis

\hspace{15mm}   IV.2.1 Symmetric nonlinear Hall conductivity as a function of $E_z$

\hspace{15mm}   IV.2.2 Symmetry comparison of second-order Drude and quantum metric Hall nonlinear conductivities

\hspace{10mm}   IV.3 Quantum metric dipole contribution dominated nonlinear Hall signal

\hspace{10mm}   IV.4 Low-energy model for BP/MnBi$_2$Te$_4$/BP heterostructure

\hspace{15mm}   IV.4.1 MnBi$_2$Te$_4$ septuple Hamiltonian

\hspace{15mm}   IV.4.2 BP Hamiltonian

\hspace{15mm}   IV.4.3 MnBi$_2$Te$_4$ and BP coupling Hamiltonian

\hspace{10mm}   IV.5 Modeling MnBi$_2$Te$_4$ septuple layers

\hspace{10mm}   IV.6 Effects of strain on low-energy model of MnBi$_2$Te$_4$

\hspace{10mm}   IV.7 Modeling the BP monolayers

\hspace{10mm}   IV.8 Modeling the coupling between MnBi$_2$Te$_4$ and BP

\hspace{15mm}   IV.8.1 Construction of the MnBi$_2$Te$_4$ and BP coupling Hamiltonian

\hspace{15mm}   IV.8.2 MnBi$_2$Te$_4$-BP hopping integrals using Slater-Koster approach 

\hspace{5mm}   IV.9 Parameter set used in the main text

\textbf{Figs. S1 to S28}

\textbf{Tables S1 to S3}

\normalsize
\clearpage
\section*{\underline{Materials and Methods}}
\subsection*{\underline{Bulk crystal growth}}
\hspace{4mm} Our MnBi$_2$Te$_4$ bulk crystals were grown by two methods: the Bi$_2$Te$_3$ flux method \cite{yan2019} and solid-state reaction method with extra Mn and I$_2$. In the Bi$_2$Te$_3$ flux method, elemental Mn, Bi and Te were mixed at a molar ratio of $15:170:270$, loaded in a crucible, and sealed in a quartz tube under one-third atmospheric pressure of Ar. The ampule was first heated to $900^{\circ}$C for $5$ hours. It was then moved to another furnace where it slowly cooled from $597^{\circ}$C to $587^{\circ}$C and stayed at $587^{\circ}$C for one day. Finally, MnBi$_2$Te$_4$ were obtained by centrifuging the ampule to separate the crystals from Bi$_2$Te$_3$ flux. In the solid-state reaction method, elemental form of Mn, Bi, Te and I$_2$ were first mixed at a stoichiometric ratio of $1.5:2:4:0.5$ and sealed in a quartz ampoule under vacuum. The sample was heated to $900^{\circ}$C in 24 hours in a box furnace and stayed at the temperature for over 5 hours to ensure a good mixture. The ampoule was then air quenched and moved to another furnace preheated at $597^{\circ}$C, where it slowly cooled to $587^{\circ}$C in 72 hours and stayed at the final temperature for two weeks. The high purity bulk BP crystals were bought from Smart-elements GmbH company.

\vspace{3mm}
\subsection*{\underline{Sample fabrication}}
\hspace{4mm} To address the sensitive chemical nature of 2D MnBi$_2$Te$_4$ flakes, all fabrication processes were completed in an argon environment without exposure to air, chemicals, or heat. Specifically, the argon-filled glovebox maintained O$_2$ and H$_2$O level below $0.01$ ppm and a dew point below $-96^{\circ}$C. The glovebox was attached to an e-beam evaporator, allowing us to make metal deposition without exposure to air. For the BP/MnBi$_2$Te$_4$ devices, MnBi$_2$Te$_4$ was mechanically exfoliated onto a $300$-nm SiO$_2/$Si wafer using Scotch-tape. Once identifying a proper thin MnBi$_2$Te$_4$ flake \cite{gao2021layer}, a tip was used to scratch the flake to a rectangular/circular shape. After that a stencil mask technique \cite{gao2021layer} was used to make Cr/Au contacts on top of MnBi$_2$Te$_4$ without exposure to air or chemical. BP flakes were then exfoliated onto a polydimethylsiloxane (PDMS) film, and a BP flake with proper shape and thickness was then identified and transferred onto the MnBi$_2$Te$_4$ flake. Next, a 20-50 nm BN flake was transferred onto the BP/MnBi$_2$Te$_4$ heterostructure as the top gate dielectric layer. A metal gate was evaporated onto the BP/MnBi$_2$Te$_4$ heterostructure. For the BP/MnBi$_2$Te$_4$/BP devices, both the bottom layer BP and MnBi$_2$Te$_4$ were mechanically exfoliated onto $300$-nm SiO$_2/$Si wafers. After a proper BP and MnBi$_2$Te$_4$ flakes were identified, the MnBi$_2$Te$_4$ flake was transferred onto the BP flake using the cryogenic pickup method developed in Ref. \cite{zhao2021Emergent}, where a thin piece of PDMS was cooled to $-110^{\circ}$C by liquid nitrogen to achieve the pickup. The rest of the procedures were the same as the BP/MnBi$_2$Te$_4$ heterostructure, which included making the contacts by shadow masks, transferring the top BP flake, and making the top gate.

BP and MnBi$_2$Te$_4$ flakes with long, straight edges were deliberately chosen (those straight edges are likely to be along the crystalline direction). When making the stack, the flakes were aligned along their straight edges. After the transport measurements were done, the samples were taken out for both Raman and SHG measurements to check the crystalline direction. The advantage of this approach is that the flakes were kept inside the glovebox throughout the fabrication process. The disadvantage is that the straight edges may turn out to be not along the crystalline direction. As a result, a few devices were made at once to make sure at least one was aligned.

\vspace{3mm}

\subsection*{\underline{Nonlinear electrical transport measurements}}
\hspace{4mm} Electrical transport measurements were carried out in a PPMS (Quantum Design DynaCool). The base temperature is 1.65 K and maximum magnetic field is 9 T. The magnetic field was applied along the out-of-plane direction. The gate voltages were applied by Keithley 2400 source meters. Longitudinal and Hall voltages were measured simultaneously. Both first- and second-harmonic signals were collected by standard Lock-in techniques (Stanford Research Systems Model SR830) with excitation frequencies between 1-600 Hz. We have also performed the electrical sum frequency generation measurements, which will be discussed below in Supplementary Materials (SM) II.2.

\vspace{3mm}

\subsection*{\underline{Optical second harmonic generation and polarized Raman measurements}}
\hspace{4mm} All SHG experiments were performed using a near-infrared femtosecond laser at room temperatures. The light source is an amplified Yb:KGW laser (Pharos, LightConversion) emitting $168$ fs pulses at $1.2$ eV with a pulse energy of $100$ $\mu$J and a default repetition rate of $100$ kHz. All measurements were performed at normal incidence. The polarization of the incident laser was controlled using an achromatic half-wave plate, while a Glan-Laser (GL) polarizer prism was used as an analyzer to select the polarization of the outgoing SHG signal. Both half-wave plate and GL prism were mounted on motorized rotation stages. The SHG data and corresponding symmetry analysis will be presented in SM I.5.

The polarized Raman was performed on a Horiba LabRam HR Evolution Raman spectrometer using a 532-nm laser in a backscattering configuration at room temperature. The polarization of the incident laser beam was controlled using a rotating achromatic half-wave plate and the scattering data with all the polarized directions were received and detected by the spectrometer. An approximately 1-$\mathrm{\mu}$m laser beam was focused on the sample by a 100$\times$ objective. A 1800 l/mm grating and 2 s exposure time were chosen to characterize the crystalline orientation of BP. The samples were protected by BN during the measurements of polarized Raman.

\vspace{3mm}

\subsection*{\underline{Wireless radio frequency (RF) rectification measurements}}
\hspace{4mm} We set up a simple experiment to harvest wireless RF signals and recorded the generated DC signals. The RF signal generator was Hittite HMC-T2220 with a frequency range of 10 MHz to 20 GHz. The samples were connected to a low temperature probe. The RF signals went into the low temperature probe through a coaxial cable and the other end of the coaxial cable was connected to an antenna. The antenna was made by a 50-mm long conducting wire with a $\sim$0.2-mm diameter. The end of the antenna was parallel to the sample and the spacing was $\sim$10 mm. The electrical field direction and power of the RF signals shone on the sample were not well defined. BP/MnBi$_2$Te$_4$ samples were made on a highly doped silicon wafer covered by a 300-nm SiO$_2$. The DC voltage signals were first passed through a voltage amplifier SR560 and then recorded by an Agilent 34401A Digital Multimeter.

\vspace{3mm}

\subsection*{\underline{First-principles calculations}}
\hspace{4mm} First-principles calculations were performed using the projector augmented wave (PAW) method as implemented in the VASP suite of codes \cite{DFT2}. The exchange-correlation part of the potential was treated within the generalized gradient approximation (GGA) scheme developed by Perdew-Burke-Ernzerhof (PBE). A $9\times5\times1$ Monkhorst-Pack k-grid was adapted for the Brillouin zone integration. The kinetic energy cutoff for the plane wave basis was set to 270 eV. The heterostructure was created by placing 2L-MBT in between monolayer BP on top and bottom, with the armchair direction of the MBT aligned along the zigzag direction of the BP. Atomic positions were relaxed until the force on each atom became less than 0.001 eV/$\mathrm{\AA}$. In order to treat the localized Mn 3d orbitals, we used an onsite U = 5.0 eV \cite{Otrokov2019a}.

\vspace{3mm}

\subsection*{\underline{Theoretical modeling}}
\hspace{4mm} We use a $40$-band continuum $k \cdot p$ model to describe the low-energy electronic properties of the BP/MnBi$_2$Te$_4$/BP heterostructure. It is defined around the $\Gamma$ point in the Brillouin zone (BZ) and contains terms of up $\mathcal{O}(k^3)$:
    $\hat{H}(\mathbf{k})=\left(\begin{smallmatrix}
        \hat{h}_{\text{MBT}}(\mathbf{k}) & \hat{U}_t(\mathbf{k}) & \hat{U}_b(\mathbf{k})\\ 
        \hat{U}^{\dag}_t(\mathbf{k}) & \hat{h}_{\text{BP},t}(\mathbf{k}) & 0\\
        \hat{U}^{\dag}_b(\mathbf{k}) & 0 & \hat{h}_{\text{BP},b}(\mathbf{k})
    \end{smallmatrix}\right)$
where $\mathbf{k}=(k_x,k_y)$ is an in-plane momentum. $\hat{h}_{\text{MBT}}(\mathbf{k})$, $\hat{h}_{\text{BP},t}(\mathbf{k})$ and $\hat{h}_{\text{BP},b}(\mathbf{k})$ are the Hamiltonians of MnBi$_2$Te$_4$, top BP and bottom BP, respectively. $\hat{h}_{\text{MBT}}(\mathbf{k})$, $\hat{h}_{\text{BP},t}(\mathbf{k})$ and $\hat{h}_{\text{BP},b}(\mathbf{k})$ share similar work function, which is consistent with the DFT calculation of the BP/MnBi$_2$Te$_4$/BP heterostructure (Fig.~\ref{DFT}). The $24 \times 24$ matrix $\hat{h}_{\text{MBT}}(\mathbf{k})$ describes the low-energy band structure of 6 septuple-layer (SL) MnBi$_2$Te$_4$, where each SL $i$ is modeled by a four-band Hamiltonian~\cite{lian2020flat}:
$\hat{h}_{\text{MBT},ii}(\mathbf{k}) \equiv \hat{h}_{\text{SL},i}(\mathbf{k})=\hat{h}_{\text{N}}(\mathbf{k})-\gamma_{af}\, \hat{h}_{\text{AFM},i}(\mathbf{k})$.
The low-energy basis is formed by the states $\left\{\left|p_{z,\text{Bi}}^{+},\uparrow\right\rangle, \left|p_{z,\text{Te}}^{-},\downarrow\right\rangle, \left|p_{z,\text{Te}}^{-},\uparrow\right\rangle,\left|p_{z,\text{Bi}}^{+},\downarrow\right\rangle\right\}$, which are the symmetric ($+$) superposition of Bi $p_z$ orbitals on the two Bi layers and the antisymmetric ($-$) superposition of Te $p_z$ orbitals on the top and bottom layers in the SL. The part $\hat{h}_\text{N}$ accounts for the normal state, where we include a cubic $\mathcal{O}(k^3)$ warping term to obtain the correct threefold symmetric Fermi surface shape. The parameter $\gamma_{af}$ controls the strength of the magnetic contribution $\hat{h}_{\text{AFM}}$ in the presence of A-type AFM magnetic order. Nearest neighbor SLs are coupled via a hopping term $\hat{h}_{\text{MBT},i, i\pm 1}(\mathbf{k}) = \hat{T}_0(\mathbf{k})$ that is derived from the $k_z$ dispersion of bulk MnBi$_2$Te$_4$. To account for lattice strain (with strength $\gamma_s$) induced by the encapsulation with BP, we add symmetry allowed terms $\gamma_s\left[\hat{h}_{\text{N},s}(\mathbf{k})-\gamma_{af}\hat{h}_{\text{AFM},s}(\mathbf{k})\right]$ and $\gamma_{s}\hat{T}_s(\mathbf{k})$. 
The two $8 \times 8$ blocks $\hat{h}_{\text{BP,t}}$ and $\hat{h}_{\text{BP,b}}$ describe the low-energy band structure of top (t) and bottom (b) BP monolayer~\cite{rudenko2015toward}

\begin{equation}
\hat{h}_{\text{BP,t}}=\hat{h}_{\text{BP,b}}=\sum\limits_{\sigma}\left[f_1(\mathbf{k})\left(\hat{\phi}_{k,1\sigma}^{\dag}\hat{\phi}_{k,2\sigma}+\hat{\phi}_{k,3\sigma}^{\dag}\hat{\phi}_{k,4\sigma}\right)+f_2(\mathbf{k})\left(\hat{\phi}_{k,2\sigma}^{\dag}\hat{\phi}_{k,3\sigma}+\hat{\phi}_{k,1\sigma}^{\dag}\hat{\phi}_{k,4\sigma}\right)+\text{h.c}\right] \text{ ,}
\end{equation}

\noindent where $f_{1}(k_x,k_y)=2\tilde{t}_1e^{i\frac{ak_x}{2\sqrt{3}}}\cos\left(\frac{bk_y}{2}\right)$ and $f_{2}(k_x,k_y)=\tilde{t}_2e^{-i\frac{ak_x}{\sqrt{3}}}$. Here, $\phi_{\mathbf{k}\nu,\sigma}^{\dag}$ creates an electron at one of the four P $p_z$ orbitals in the unit cell and  $\tilde{t}_1$ and $\tilde{t}_2$ are set to obtain the experimentally observed BP band gap of $0.3$~eV.
The off-diagonal $24 \times 8$ blocks $\hat{U}_t$ and $\hat{U}_b$ describe the electronic coupling between MnBi$_2$Te$_4$ and BP monolayer on top and bottom surfaces.  Due to the mismatch of lattice geometries, the hybridization of MnBi$_2$Te$_4$ and BP bands leads to a breaking of threefold rotation and translation symmetry.
The coupling between the Bloch states of the two layers thus needs to be derived in the real-space continuum leading to 
$\hat{H}_{\text{int}}^{(b)}=\sum\limits_{\mu,\nu,\sigma}\sum\limits_{\mathbf{k}}\left[t_{\mu\nu}^{(b)}(\mathbf{k})\psi_{\mathbf{k}\mu,\sigma}^{\dag}\phi_{b,\mathbf{k}\nu,\sigma}+\text{h.c.}\right]$. Here, $\psi_{\mathbf{k}\mu,\sigma}^{\dag}$ creates an electron with momentum $\mathbf{k}$ and spin $\sigma$ in one of the seven orbitals $\mu$ in the MnBi$_2$Te$_4$ unit cell and the MnBi$_2$Te$_4$-BP interface hopping amplitude is given by
    $t_{\mu\nu}^{(b)}(\mathbf{k})\equiv\int d^2y\, t_b\left(\mathbf{y}+s_{\mu\nu}\hat{\mathbf{z}}\right)e^{-i\mathbf{k}\cdot \mathbf{y}}$.
The size of the hopping is controlled by the orbital distance with 
out-of-plane component $s_{\mu\nu}=\left(\boldsymbol{\tau}_{\mu}-\boldsymbol{\rho}_{\nu}\right)\cdot\hat{\mathbf{z}}$. The integration over the in-plane distance $\mathbf{y}$ is due to the lattice incommensurability. The real space hopping elements are parameterized using the Slater-Koster approach as
    $t_{b}(\mathbf{r})=A_{\sigma}e^{-r/a_\sigma}\cos^{2}\theta+A_{\pi}e^{-r/a_\pi}\sin^{2}\theta$,
where $A_\sigma, A_\pi$ denote characteristic energy and $a_\sigma, a_\pi$ denote characteristic length scales of the hopping integrals for $\sigma$ and $\pi$ bonding. Projection onto the relevant MnBi$_2$Te$_4$ low-energy manifold yields the coupling matrix $\hat{U}_b$. The coupling block $\hat{U}_t$ is obtained from inversion symmetry.
More details of these theoretical models are shown in Secs. IV.3-IV.8.

\section*{\underline{Supplementary Text}}
\subsection*{\underline{I. Basic characterization of BP/MnBi$_2$Te$_4$ heterostructure}}

\subsection*{\underline{I.1. The symmetry of MnBi$_2$Te$_4$, BP/MnBi$_2$Te$_4$ and BP/MnBi$_2$Te$_4$/BP}}
\hspace{4mm} The even-layered MnBi$_2$Te$_4$ is a fully compensated antiferromagnet at low temperature. Although the lattice of MnBi$_2$Te$_4$ is centrosymmetric, the spin breaks the inversion and time reversal symmetry. Therefore, even-layered MnBi$_2$Te$_4$ breaks $\mathcal{P}$ and $\mathcal{T}$ but preserves the $\mathcal{PT}$ symmetry  (Fig.~\ref{MBT_Symmetry}A). After stacking BP on MnBi$_2$Te$_4$, the inversion symmetry of the lattice is also broken.  Therefore, the BP/MnBi$_2$Te$_4$ breaks $\mathcal{P}$, $\mathcal{T}$ and also $\mathcal{PT}$ symmetry (Fig.~\ref{MBT_Symmetry}B). Sandwiching MnBi$_2$Te$_4$ between two 2L BP, the lattice of the heterostructure is still centrosymmetric when the crystallographic $a$ axes of the BP layers and the MnBi$_2$Te$_4$ layer are all aligned. Therefore, the aligned BP/MnBi$_2$Te$_4$/BP heterostructure breaks $\mathcal{P}$ and $\mathcal{T}$ but still preserves the $\mathcal{PT}$ symmetry (Fig.~\ref{MBT_Symmetry}C).

\subsection*{\underline{I.2. Determining the crystalline directions for MnBi$_2$Te$_4$ and BP}}

\hspace{4mm} The crystalline directions of the MnBi$_2$Te$_4$ flakes were determined by optical SHG measurements at room temperature. At room temperature, the interior of thin MnBi$_2$Te$_4$  flake's crystal structure is centrosymmetric (group $D_{3d}$). Therefore, the SHG signals are only expected to originate from the surface (surface group $C_{3v}$). The largest SHG signals are along the mirror plane of the MnBi$_2$Te$_4$ surface according to the symmetry. Figure~\ref{SHG_And_Raman}A and ~\ref{SHG_And_Raman}B show how our SHG data correspond to the crystalline axes of MnBi$_2$Te$_4$. These results are consistent with previous SHG results on Bi$_2$Se$_3$ (same symmetry as MnBi$_2$Te$_4$ above Neel temperature) \cite{hsieh2011Selective}. 

The crystalline directions of the BP were determined by polarized Raman scattering measurements. Figure ~\ref{SHG_And_Raman}C and ~\ref{SHG_And_Raman}D show how our Raman data correspond to the crystalline axes of BP. These results are consistent with previous Raman measurements on BP \cite{hsieh2011Selective}.

\subsection*{\underline{I.3. Basic transport characterizations}}

\hspace{4mm} Figure~\ref{Magnetotransport_MBT_vs_BP_MBT} shows the magneto-transport data of MnBi$_2$Te$_4$ before and after covering BP of Device-BM19. At charge neutrality, both MnBi$_2$Te$_4$ and BP/MnBi$_2$Te$_4$ heterostructure show that the Hall resistivity ($R_{xy}$) is nearly zero in the AFM phase ($-3$ T $\lesssim B \lesssim +3$ T) and is nearly quantized in the FM phase ($|B| \gtrsim 6$ T). These data suggest that interfacing MnBi$_2$Te$_4$ with BP does not change the topological phases of even-layered MnBi$_2$Te$_4$: it is a Chern insulator in the FM phase and an Axion insulator in the AFM phase. The magneto-transport data is consistent with the DFT calculated band structure of BP/MnBi$_2$Te$_4$/BP (Fig.~\ref{DFT}), in which the lowest conducting band and highest valance band are both derived from MnBi$_2$Te$_4$ band.

Figure~\ref{Gate_BP_MBT} shows the gate-dependent four-point resistance of bare BP and 6SL MnBi$_2$Te$_4$. BP was found to be significantly more resistive than 6SL MnBi$_2$Te$_4$. The resistance of BP is two orders of magnitude lager than that of MnBi$_2$Te$_4$ for all of the gate voltages. The resistivity data can be explained by the electronic structures. MnBi$_2$Te$_4$ is a Dirac material while BP is a semiconductor (band gap $\sim0.3$ eV for thick flake and $\sim0.7$ eV for bilayer \cite{Ghosh2016Electric}). Therefore, it is reasonable to assume that the current flowing in BP layer is small.\\

To further confirm that, we designed and fabricated another kind of device. As shown by the schematic in Fig.~\ref{Two_pair_contacts}B, we injected current through the drain electrode, then collected the current from two source electrodes (S$_{\textrm{MBT}}$ and S$_{\textrm{BP}}$). Figures~\ref{Two_pair_contacts}C-F show the currents collected from S$_{\textrm{MBT}}$ and S$_{\textrm{BP}}$ and their ratio as a function of $V_{\textrm{BG}}$ and $V_{\textrm{TG}}$. These data show that the current flowing through BP layer is small. Therefore, in the BP/MnBi$_2$Te$_4$ heterostructure, the current mainly flows in MnBi$_2$Te$_4$ layer. 

\color{black}
\subsection*{\underline{I.4. Angular-resolved transport}}

\hspace{4mm} Here we use transport method to determine the resistivity anisotropy of the sample. Assuming the resistivity of the sample is $\rho_a$ and $\rho_b$ along $a$ and $b$ axis ($a$, $b$ are in the in-plane crystallographic axes of the lattice, $a \bot b$). The measurements were performed in Cartesian basis and the angle between $x$ and $a$ is $\theta$. The resistivity in longitudinal ($\rho_{xx}$) and transverse ($\rho_{yx}$) direction are expressed as\cite{walmsley2017determination}:

\begin{equation}
\rho_{xx}=\rho_a \textrm{cos}^2(\theta)+\rho_b \textrm{sin}^2(\theta)=\frac{1}{2}(\rho_a + \rho_b)+\frac{1}{2}(\rho_a - \rho_b)\textrm{cos}(2 \theta)
\label{eq:Rxx}
\end{equation}
\begin{equation}
\rho_{yx}=(\rho_a-\rho_b)\textrm{cos}(\theta)\textrm{sin}(\theta)=\frac{1}{2}(\rho_a-\rho_b)\textrm{sin}(2 \theta)
\label{eq:Ryx}
\end{equation}

Bare MnBi$_2$Te$_4$ has $C_{3z}$ rotational symmetry. Therefore, $\rho_a=\rho_b$, and then $\rho_{xx}=\rho_a$, $\rho_{yx}=0$. Hence there is no anisotropy for bare MnBi$_2$Te$_4$, consistent with our results showing both $R_{xx}$ and $R_{yx}$ are independent of the measurement directions and the $R_{yx}$ is always around zero (Fig.~\ref{Angular_Resolved_Transport}). After interfacing MnBi$_2$Te$_4$ with BP, the $C_{3z}$ of the sample is broken. so, $\rho_a \neq \rho_b$. According to the Eqs.~\ref{eq:Rxx} and \ref{eq:Ryx}, both $R_{xx}$ and $R_{yx}$ depend on measurement direction and have a 180$^{\circ}$ periodicity, which is consistent with our measurement (Fig.~\ref{Angular_Resolved_Transport}). Moreover, for the angular-dependent of $R_{xx}$ and $R_{yx}$, $R_{xx}$ has maximum/minimum value when the $R_{yx}$ crosses zero. This further confirms that MnBi$_2$Te$_4$ has anisotropy after interfacing with BP, i.e. $\rho_a \neq \rho_b$.

To further exclude the effects of nonlocal transport, we also performed two-probe measurement (Fig.~\ref{Angular_Resolved_Transport}C). The two-probe resistance showed the same angular dependence as four-probe measurements, and the overall resistance value is larger due to the additional contact resistance. Hence, two-probe measurements again confirm that interfacing MnBi$_2$Te$_4$ with BP breaks the $C_{3z}$ rotational symmetry. \color{black}

\subsection*{\underline{I.5. Optical second-harmonic generation}}

\hspace{4mm} Optical SHG is an effective way for probing sample's symmetry because it is sensitive to crystal symmetry \cite{Zheng2022symmetry}. Here, we use optical SHG to demonstrate that interfacing MnBi$_2$Te$_4$ with BP can break the $C_{3z}$ rotational symmetry of MnBi$_2$Te$_4$. The optical SHG measurements were performed at room temperature. At room temperature, the interior of MnBi$_2$Te$_4$ is centrosymmetric which prohibits any SHG signal generation. However, the surfaces of MnBi$_2$Te$_4$ break the inversion symmetry and have the $C_{3z}$ rotational symmetry. Therefore, for the normal incidence measurement, the SHG signals of MnBi$_2$Te$_4$ are mainly from the top surface. For BP, because both its interior and surfaces have the two-fold rotational symmetry, BP does not generate any SHG signal, which is confirmed by our measurements (Fig. 1H). Therefore, for the BP/MnBi$_2$Te$_4$ heterostructure, the SHG signals come from the surface of MnBi$_2$Te$_4$. Hence, optical SHG is an effective way for probing the surface symmetry of MnBi$_2$Te$_4$.

Figure~\ref{Angular_Resolved_SHG}B shows the SHG signals of a BP/MnBi$_2$Te$_4$ heterostructure. The SHG pattern is asymmetric which demonstrates that the $C_{3z}$ symmetry of MnBi$_2$Te$_4$ is broken. To double check the $C_{3z}$ symmetry breaking was induced by BP, we removed the BP layer of the heterostructure using the Scotch tape (Fig.~\ref{Angular_Resolved_SHG}C). We probed SHG from the same area of the MnBi$_2$Te$_4$. As shown in Fig.~\ref{Angular_Resolved_SHG}D, after removing the BP, the SHG signals of MnBi$_2$Te$_4$ recovers the symmetric pattern. We repeated these measurements on three devices, all of them showed consistent results. Therefore, the optical SHG measurements demonstrate that the $C_{3z}$ symmetry is broken after interfacing MnBi$_2$Te$_4$ with BP.

\section*{\underline{II. Addressing alternative mechanisms for the nonlinear Hall signals}}

\subsection*{\underline{II.1. Berry curvature dipole}}

\hspace{4mm} The most important competing mechanism is the Berry curvature dipole induced nonlinear Hall effect \cite{sodemann2015quantum}. The key here is the $\mathcal{PT}$ symmetry: $\mathcal{PT}$ prohibits Berry curvature dipole but allows quantum metric dipole. Because we have thoroughly considered this in the main text, we will summarize the crucial points.

The aligned BP/MnBi$_2$Te$_4$/BP device is expected to respect $\mathcal{PT}$ symmetry. $\mathcal{PT}$ enforces Berry curvature and Berry curvature dipole to vanish. Admittedly, it is difficult to achieve  a perfect $\mathcal{PT}$ symmetry. Therefore, we assume $\mathcal{PT}$ to be weakly broken so that both Berry curvature dipole and quantum metric dipole are allowed. We show how our data shows that the Berry curvature dipole contribution is insignificant.

\begin{itemize}
\item \textbf{AFM order}: The two AFM states (related by time-reversal $\mathcal{T}$) are expected to have the same Berry curvature dipole (see Fig. 3, E and F), and therefore the same Berry curvature dipole induced nonlinear Hall signal. By contrast, our observed nonlinear Hall signal flips sign upon reversing the AFM order (see Fig. 3, G and H).

\item \textbf{Vertical $E_z$ field}: The Berry curvature dipole is expected to be antisymmetric around $E_z=0$, i.e., $D_{\textrm{Berry}}(+E_z)=-D_{\textrm{Berry}}(-E_z)$. By contrast, our observed nonlinear Hall signal is symmetric about $E_z=0$, as shown in Fig.~\ref{E_Berry_Metric}.

We can also experimentally determine the direction of the Berry curvature dipole. Intuitively, the direction of the Berry curvature dipole $D_{\textrm{Berry}}$ is determined by the sign of Berry curvature. In MnBi$_2$Te$_4$, we can measure the sign of Berry curvature by measuring the linear Hall signals $\sigma_{xy}$ \cite{gao2021layer}. We simultaneously measured $\sigma_{xy}$ and $V_{yxx}^{2\omega}$ as a function of $E_z$ field (Fig.~\ref{Additional_E}). The $\sigma_{xy}$ is antisymmetric as a function of $E_z$ which indicates that $D_{\textrm{Berry}}$ is also antisymmetric as a function of $E_z$. However, the $V_{yxx}^{2\omega}$ is symmetric as a function of $E_z$. Therefore, the nonlinear Hall signals observed here are not induced by Berry curvature dipole.

\item \textbf{Mirror symmetry}: In Device-BMB1, by aligning the $a$ axis of the top BP, MnBi$_2$Te$_4$ and bottom BP layers, we also preserve the mirror plane $\mathcal{M}_y$. In the AFM state, $\mathcal{M}_y$ is broken but $\mathcal{M}_y\mathcal{T}$ is a good symmetry. For Berry curvature dipole, one expects $\sigma_{yxx}\equiv 0$ and $\sigma_{xyy}\neq0$. For quantum metric dipole, one expects $\sigma_{yxx}\neq 0$ and $\sigma_{xyy}\equiv 0$. Our data (Fig.~\ref{Mirror_Exclude_Berry}) is consistent with the latter.

\item \textbf{Scaling property}: For Berry curvature dipole, the nonlinear Hall conductivity is expected to be proportional to the scattering time, $\sigma^{2\omega}_{yxx}\propto \tau$. For quantum metric dipole, we have, $\sigma^{2\omega}_{yxx}\propto \tau^0$. Our data (Fig. 3A and Fig.~\ref{Additional_Temp}) is consistent with the latter.

\end{itemize}

\subsection*{\underline{II.2. Second-order Drude conductivity}}

\hspace{4mm}Another important mechanism is the second-order Drude effect \cite{Wang2021Intrinsic}. From the symmetry point of view, the second-order Drude effect and the quantum metric dipole induced nonlinear signals are both allowed in $\mathcal{PT}$-symmetric AFMs. Therefore, our goal is to investigate which effect is more dominant in our sample.

\subsubsection*{\underline{II.2.1 Anti-symmetric vs symmetric}}

\hspace{4mm}For the quantum metric Hall effect, its nonlinear conductivity is expected to be antisymmetric, $\sigma^{2\omega}_{yxx}=-\sigma^{2\omega}_{xyx}$. For the second-order Drude effect, its nonlinear conductivity is expected to be symmetric, $\sigma^{2\omega}_{yxx}=\sigma^{2\omega}_{xyx}$. $\sigma^{2\omega}_{ijk}$ is defined as $V^{2\omega}_{i}\propto\sigma^{2\omega}_{ijk}I^{\omega}_{j}I^{\omega}_{k}$. $\sigma^{2\omega}_{xyx}$ means that we flow currents along both $x$ and $y$ directions, which is difficult to implement experimentally. 

Therefore, here we present our electrical sum frequency generation (SFG) measurements, where we inject two currents with frequencies $\omega_1$ and $\omega_2$, and we detect the SFG voltage $V^{\omega_1+\omega_2}$. By separating the two currents in frequency domain, it is easier to also control their directions separately.

\vspace{3mm}

\textbf{Data:} Figure~\ref{Drude_Exclude}A shows $\sigma^{2\omega}_{yxx}$ measurement, which is achieved by passing both currents at $\omega_1$ and $\omega_2$ along $x$ direction and measure $V^{\omega_1+\omega_2}$ along $y$ direction. Fig.~\ref{Drude_Exclude}B shows $\sigma^{2\omega}_{xyx}$ measurement, which is achieved by passing a current at $\omega_1$ along $y$ direction while passing another current at $\omega_2$ along $x$ direction and measure $V^{\omega_1+\omega_2}$ along $x$ direction. Indeed, our data show $\sigma^{2\omega}_{yxx}=-\sigma^{2\omega}_{xyx}$ (Fig.~\ref{Drude_Exclude}C), which demonstrates that the quantum metric Hall effect is dominant in our BP/MnBi$_2$Te$_4$ samples.

\vspace{3mm}
\textbf{Methods:} The SFG voltage $V^{\omega_1+\omega_2}$ are detected by the following method: The SFG voltage can be expressed as $V^{\omega_1+\omega_2}\propto \sin(\omega_1+\omega_2)t$. The standard lock-in detectors (SR830) cannot directly lock to the $\omega_1+\omega_2$ frequency. Instead, we measured $\sin\omega_1t\sin\omega_2t$, which directly relates to $\sin(\omega_1+\omega_2)t$ by the angle addition theorem. We chose $\omega_1=547$ Hz and $\omega_2=1.37$ Hz, so that $\omega_1\gg\omega_2$. To probe $\sin\omega_1t\sin\omega_2t$, the SFG voltage signal was fed through the first lock-in (lock-in A), which was locked to $\omega_1$ and its integral time was set to $\frac{2\pi}{w_1}<t_1\ll\frac{2\pi}{w_2}$. The output was then fed through the second lock-in (lock-in B), which was locked to $\omega_2$ and its integral time was set to $t_2>\frac{2\pi}{w_2}$. The output of the second lock-in was $\sin\omega_1t\sin\omega_2t$. The measurement setups for $\sigma^{2\omega}_{yxx}$ and $\sigma^{2\omega}_{xyx}$ are shown in Fig.~\ref{Drude_Method}.

\subsubsection*{\underline{II.2.2 Scaling property}}

\hspace{4mm} For the second-order Drude effect, its nonlinear conductivity is expected to be quadratic with respect to the scattering time, $\sigma^{2\omega}\propto \tau^2$ ~\cite{sodemann2015quantum,Watanabe2020Nonlinear,Wang2021Intrinsic}. For quantum metric dipole, we have $\sigma^{2\omega}\propto \tau^0$. Our data (Fig. 3A and Fig.~\ref{Scaling_Temperature}) is consistent with the latter.

\subsection*{\underline{II.3. Joule heating induced Anomalous Nernst effect}}

\hspace{4mm} We now consider a Joule heating induced anomalous Nernst effect: (1) Joule heating leads to a temperature gradient $\Delta T\propto I^2R$; (2) Let us assume that our even-layered MnBi$_2$Te$_4$ is not fully-compensated, so there is a small $M$. The combination of the $\Delta T$, and the magnetization $M$ can lead to an anomalous Nernst current, $J=\sigma^{\textrm{Nernst}}\Delta T\propto\sigma^{\textrm{Nernst}} (I^2R)$.

\begin{itemize}
\item \textbf{Mirror symmetry}: The Joule heating induced anomalous Nernst effect is insensitive to crystalline symmetry. By contrast, as described above, our data in Device-BMB1 shows clear dependence with respect to the mirror plane $\mathcal{M}_y$. Specifically, we observed $\sigma_{yxx}\neq 0$ and $\sigma_{xyy}\equiv 0$. 

\item \textbf{Fermi level dependence}: For the anomalous Nernst effect, in the presence of a fixed $M$, we expect the electrons and holes to deflect toward opposite directions. By contrast, in our data, the signals from electrons and holes have the same sign.

\item \textbf{Scaling property}: The Joule heating induced anomalous Nernst effect is $J=\sigma^{\textrm{Nernst}}\Delta T\propto\sigma^{\textrm{Nernst}} (I^2R)$. Because $\sigma^{\textrm{Nernst}} \propto \tau^0$ and $I^2R \propto \tau^1$, this effect is expected to be proportional to $\tau$. By contrast, our data is independent of $\tau$.

\item \textbf{Full compensated AFM}: The magnetization is necessary to generate anomalous Nernst effect. However, the BP/MnBi$_2$Te$_4$ heterostructure is a fully compensate AFM system.  The linear magnetotransport data of the BP/MnBi$_2$Te$_4$ (Fig.~\ref{Magnetotransport_MBT_vs_BP_MBT}) shows a zero Hall resistance at $B=0$ which confirms that there is no global magnetization without external magnetic field. 
\end{itemize}

\subsection*{\underline{II.4. Other extrinsic effects}}

\hspace{4mm} Finally, we consider various extrinsic effects such as accidental contact junctions, flake shape, etc. 

\begin{itemize}
\item \textbf{Contact junction}: An accidental contact junction can lead to a nonlinear effect. (1) The nonlinear signals induced by contact junction should not relate to AFM states. (2) The nonlinear signals induced by contact junction should not be sensitive to Neel temperature. (3) The nonlinear signals induced by contact junction should not show Hall dominance. (4) The nonlinear signals induced by contact junction should not relate to the mirror symmetry of BP/MnBi$_2$Te$_4$/BP heterostructure.

The nonlinear signals induced by contact junction can be excluded by the following observations: (1) The nonlinear signals have opposite sign for the different AFM states (Fig. 2, G and H). (2) The nonlinear signals decrease to zero when the temperature is higher than Neel temperature (Fig. 2, I and J). (3) The nonlinear signals show clear Hall dominance (Fig. 2B). (4) The nonlinear signals are significantly enhanced after covering BP on the same MnBi$_2$Te$_4$ sample (Fig. 1I). (5) For a careful aligned BP/MnBi$_2$Te$_4$/BP device, the nonlinear Hall signals can only be observed when the current applied in mirror plane $\mathcal{M}_a$ (Fig.~\ref{Mirror_Exclude_Berry}). 

\item \textbf{Flake shape}: The asymmetric global shape of the sample can lead directional movement of the carriers by colliding against the asymmetric sample boundaries, which can also induce nonlinear signals. The nonlinear signals induced by flake shape should highly depend on flake shape and should not depend on AFM states, temperature and carrier density.

The flake shape induced nonlinear signals can be excluded by following observations: (1) Most MnBi$_2$Te$_4$ in the Hall bar devices are shaped into rectangular-like shape (Fig.~\ref{Mirror_Exclude_Berry}A). As shown in Fig.~\ref{Mirror_Exclude_Berry}, the asymmetry of the MnBi$_2$Te$_4$ is similar in $x$ and $y$ directions, but the nonlinear Hall signal can only be observed when current is in the mirror plane $\mathcal{M}_a$. (2) The nonlinear signals are highly dependent on AFM states, temperature and carrier density.
\end{itemize}

Therefore, our systematic data as a function of temperature, AFM states, crystalline direction, doping, etc. allow us to show that these extrinsic effects are not important.

\subsection*{\underline{II.5. Addressing the nonlinear Hall signals induced by skew scattering}}

\hspace{4mm} By now, it has been firmly established that the linear AHE consists of the intrinsic AHE due to Berry curvature and the extrinsic AHE from defect scattering \cite{nagaosa2010anomalous}. The defect scattering induced linear anomalous Hall effect has further been categorized into two kinds, skew scattering and side jump. Prior studies have made an important conclusion that scaling law can be used to differentiate the skew scattering induced AHE from intrinsic AHE. Skew scattering induced AHE is proportional to longitudinal conductivity $\sigma _{xx} ^2$ \cite{tian2009proper}, while the Berry curvature induced AHE only depends on Berry curvature. Because the side jump induced AHE has the same scaling properties as Berry curvature induced AHE, there is currently no effective way to differentiate the side jump from Berry curvature contribution. The working principle adopted by the community so far is to combine the Berry curvature and side jump contributions and consider them as the intrinsic Berry phase contribution \cite{nagaosa2010anomalous}. Therefore, we only talk about the skew scattering induced AHE here. \\

Recent theory has extended the studies from linear AHE to the nonlinear AHE. Specifically, for the intrinsic nonlinear Hall effect studied here, the scaling is predicted to be the same as the linear AHE \cite{Wang2021Intrinsic}: The intrinsic component induced by quantum metric scales as $\tau^0$, while the extrinsic component induced by skew scattering scales as $\tau^2$. Therefore, we use scaling laws to exclude the skew scattering induced nonlinear Hall current.

\begin{itemize}
\item \textbf{Temperature dependence}: As shown in Fig. 3A, the $\sigma_{yxx} ^{2\omega}$ does not depend on $\sigma_{xx}$ in the low temperature range ($<$15 K). In addition, Fig.~\ref{Scaling_Temperature} shows more details about temperature dependence of $\sigma_{xx} ^2$ and $\sigma_{yxx} ^{2\omega}$. It is notable that $\sigma_{xx} ^2$ and $\sigma_{yxx} ^{2\omega}$ show different temperature dependence. $\sigma_{xx} ^2$ decreases with temperature increasing when temperature is lower than 21 K. However, $\sigma_{yxx} ^{2\omega}$ does not depend on temperature when temperature is lower than 15 K but quickly decrease to zero when temperature is higher than 15 K. To be more clear, we plot $\sigma_{yxx} ^{2\omega}$ versus $\sigma_{xx} ^2$ (Fig.~\ref{Scaling_Temperature}C) which clear show that $\sigma_{yxx} ^{2\omega}$ does not depend on $\sigma_{xx} ^2$ when temperature is lower than 15 K.  

\item \textbf{Carrier density $n$ dependence}: In BP/MnBi$_2$Te$_4$/BP device, both $\sigma_{yxx} ^{2\omega}$ and $\sigma_{xx}$ depend on $n$. Therefore, it is good way to check the scaling properties of $\sigma_{yxx} ^{2\omega}$ with $\sigma_{xx}$ by tuning $n$. As shown in Fig.~\ref{Scaling_n_E}, A and B, $\sigma_{xx} ^2$ keeps increasing with $n$ increasing. By contrast, the $\sigma_{yxx} ^{2\omega}$ first increase with $n$ but eventually decrease to negative values. The contrasting behaviors between $\sigma_{yxx} ^{2\omega}$ and $\sigma_{xx} ^2$ suggest that skew scattering is not the main contribution to nonlinear Hall signals. Moreover, the sign reversal of $\sigma_{yxx} ^{2\omega}$ in higher $n$ doped regime can be explained by quantum metric dipole contribution.
\end{itemize}

All scaling laws related to temperature and carrier density show that the nonlinear Hall signals observed here are independent of $\tau$ ($\sigma_{xx}$). The result of the scaling law is consistent with the prediction of the quantum metric dipole induced intrinsic nonlinear Hall effect. Therefore, we can conclude that the nonlinear Hall signals observed here are not induced by skew scattering.

\section*{\underline{III. Additional data}}

\hspace{4mm} In this section, we present additional experimental data that did not appear in the main text. These data help us to further confirm the major conclusions.

\begin{itemize}

\item Figure~\ref{Additional_Dev_Sum} shows the summary of the measured 26 MnBi$_2$Te$_4$ heterostructure and 7 MnBi$_2$Te$_4$ devices. The nonlinear Hall signals in BP/MnBi$_2$Te$_4$ heterostructures are highly reproducible. The nonlinear Hall signals in MnBi$_2$Te$_4$ are induced by nonlinear Drude conductivity which is two orders smaller than the nonlinear Hall signals induced by quantum metric dipole. This further confirms that nonlinear Drude conductivity is not the main contribution to the nonlinear Hall signals. This result is consistent with the data in Fig. 1I. 

\item Figure~\ref{Additional_hysB} shows the nonlinear Hall signals as a function of out of plane magnetic field. The vertical electric field $E_z$ is -0.17 V/nm. With a finite $E_z$, we can choose AFM states by sweeping magnetic field \cite{gao2021layer}. The nonlinear Hall signals $V_{yxx}^{2\omega}$ have opposite signals for the different AFM states. 

\item Figure~\ref{Additional_AFM} shows the transport data for different AFM states in another device (device-BMB2). The linear longitudinal signals, $R_{xx}$ and $V_{xx}^{\omega}$, are identical for the two AFM states. However, the nonlinear Hall signals are opposite for two AFM states. The experiment data on the new device reproduces the main data features, and further confirms our major conclusion.

\item Figure~\ref{Additional_Temp_Fig3} shows the additional data related to Fig. 3A. Figures ~\ref{Additional_Temp_Fig3}, A and B show the nonlinear Hall voltage $V_{yxx}^{2\omega}$ and conductivity $\sigma_{yxx}^{2\omega}$ as a function of temperature. When temperature is lower than 15 K, $\sigma_{yxx}^{2\omega}$ is constant and independent of temperature. 

\item Figure~\ref{Additional_Temp} shows the temperature dependence of linear $\sigma_{xx}$, nonlinear Hall signals $V_{yxx}^{2\omega}$ and $\sigma_{yxx}^{2\omega}$ on another device (Device-BM21). The $\sigma_{xx}$ and $\sigma_{yxx}^{2\omega}$ show notable different temperature dependencies. The $\sigma_{yxx}^{2\omega}$ is constant when temperature is lower than 15 K, which means that $\sigma_{yxx}^{2\omega}$ is independent of $\tau$ ($\sigma_{xx}$). The data reproduced in this device further confirms that the nondissipative nonlinear Hall signals are induced by the intrinsic mechanism, i.e., quantum metric dipole. 

\item Figure~\ref{Additional_4SL} shows the nonlinear Hall signals in a BP/4SL MnBi$_2$Te$_4$ heterostructure (Device-BM3). The nonlinear Hall signals as a function of carrier density shows the similar behavior as the main text. 

\item Figure~\ref{Additional_RF} shows the microwave rectification in another BP/MnBi$_2$Te$_4$ heterostructure (Device-BM23). The DC signals as a function of carrier density $n_e$ and RF power $P$ are similar as the nonlinear signals in the main text, which confirms that the DC rectification effect is also induce by quantum metric dipole. The DC rectification effect can work at very broad frequency range. The frequency range is limited by the antenna and source generator rather than the BP/MnBi$_2$Te$_4$ device.

\item Figure~\ref{DFT} shows the DFT computed band structure of the BP/MnBi$_2$Te$_4$/BP heterostructure. The heterostructure considered here is created by a rectangular supercell of 2L-MnBi$_2$Te$_4$ and a 1x2 supercell of BP. The armchair direction of the MnBi$_2$Te$_4$ is aligned along the zigzag direction of the BP. The MnBi$_2$Te$_4$ bands are marked in red, and the BP bands are shown in blue. As evident from the plot, MnBi$_2$Te$_4$ and BP share similar work functions, and as a result, the low-energy BP bands hybridize with MnBi$_2$Te$_4$ bands around the Fermi level of MnBi$_2$Te$_4$, forming a straddling gap (type-I) arrangement.

\end{itemize}

\setcounter{figure}{0}
\renewcommand{\thefigure}{S\arabic{figure}}
\renewcommand{\theequation}{S\arabic{equation}}
\renewcommand{\thetable}{S\arabic{table}}
\renewcommand{\figurename}{\textbf{Fig.}}

\setcounter{section}{1}

\section*{\underline{IV. Theoretical studies}}

\subsection{\underline{IV. 1. An intuitive picture of quantum metric}}\label{ssec:QM}

\hspace{4mm} Here, we give an intuitive picture to under stand the quantum metric. The Bloch wavefunction $\psi_\mathbf{k}(\mathbf{r})$ is the product of two terms,  $\psi_\mathbf{k}(\mathbf{r})=u_\mathbf{k}e^{i\mathbf{k}\mathbf{r}}$, where $e^{i\mathbf{k}\mathbf{r}}$ describes how the Bloch wave propagates from one unit cell to the next, and $u_\mathbf{k}$ captures the physics within a unit cell. The quantum metric ($g$) describes the the difference between neighboring $u_\mathbf{k}$, $|u_\mathbf{k+\delta k}-u_\mathbf{k}|^2$, which becomes the distance between two points on a Bloch sphere when we map $u_\mathbf{k}$ to a specific point on a Bloch sphere.

Specifically, let us consider a two-band system, which arises from two atomic positions $\alpha$ and $\beta$ in the unit cell. Within a unit cell, an electron has probability to occupy both positions. So the two atomic positions serve as two orthogonal quantum bases $\arrowvert\alpha\rangle$ and $\arrowvert\beta\rangle$, whose linear combination constructs the $\arrowvert u_\mathbf{k}\rangle$, $\arrowvert u_\mathbf{k}\rangle=C_{\alpha}(\mathbf{k})\arrowvert\alpha\rangle+C_{\beta}(\mathbf{k})\arrowvert\beta\rangle$. $\arrowvert u_\mathbf{k}\rangle$ can be represented by a specific point on the Bloch sphere (Fig.~\ref{Theory_QM}A). The simplest scenario is that the electrons of the conduction band (CB) only occupy $\alpha$ ($C_{\alpha}^{\textrm{CB}}(\mathbf{k})\equiv1$ and $C_{\beta}^{\textrm{CB}}(\mathbf{k})\equiv0$, so, $\arrowvert u_\mathbf{k}^{\textrm{CB}}\rangle\equiv\arrowvert\alpha\rangle$) and the electrons of the valence band (VB) only occupy $\beta$ ($C_{\alpha}^{\textrm{VB}}(\mathbf{k})\equiv0$ and $C_{\beta}^{\textrm{VB}}(\mathbf{k})\equiv1$, so, $\arrowvert u_\mathbf{k}^{\textrm{VB}}\rangle\equiv\arrowvert\beta\rangle$). Taking the CB as an example, according to the definition above, its quantum metric is the difference of $\arrowvert u_\mathbf{k}^{\textrm{CB}}\rangle$ at $\mathbf{k}$ and $\mathbf{k}+\delta\mathbf{k}$. Clearly, because $\arrowvert u_\mathbf{k}^{\textrm{CB}}\rangle\equiv\arrowvert\alpha\rangle$ at every $\mathbf{k}$ point, the quantum metric is zero. Equivalently, $\arrowvert u_\mathbf{k}^{\textrm{CB}}\rangle$ at every $\mathbf{k}$ point is mapped onto the north pole of the Bloch sphere. So the distance on the Bloch sphere between $u_\mathbf{k}$ and $u_\mathbf{k+\delta k}$ is zero (Fig.~\ref{Theory_QM}B).

A nontrivial case is realized when CB and VB go through a band inversion. In this case, each band becomes a $\mathbf{k}$-dependent superposition of $\alpha$ and $\beta$, i.e., both $C_{\alpha}(\mathbf{k})$ and $C_{\beta}(\mathbf{k})$ depend strongly on $\mathbf{k}$. The change of $\arrowvert u\rangle$ is dramatic near $\mathbf{k}=0$ where band inversion occurs but weak at large $\mathbf{k}$. Equivalently, $u_\mathbf{k}$ and $u_\mathbf{k+\delta k}$ are mapped onto different points on the Bloch sphere. So the distance on the Bloch sphere between $u_\mathbf{k}$ and $u_\mathbf{k+\delta k}$ (i.e., the quantum metric) is nonzero (Fig.~\ref{Theory_QM}C). Interestingly, if we correspond $\alpha$ and $\beta$ to Bi and Te, the nontrivial case above can help us to understand the large quantum metric in the low-energy electronic states of MnBi$_2$Te$_4$.

\subsection{\underline{IV.2. Symmetry analysis}}\label{ssec:symm}

\subsubsection{\underline{IV.2.1. Symmetric nonlinear Hall conductivity as a function of $E_z$}}\label{NLH_E}

\hspace{4mm} In the main text and SM Sec. II.1 our data showed that nonlinear Hall signals $\sigma_{yxx}^{2\omega}$ are symmetric as a function of vertical electric field $E_z$. Here we explain this behavior by symmetry analysis. We start from the generic expression of our nonlinear Hall effect:

\begin{equation}
J^{2\omega}_{yxx}=\sigma_{yxx}^{2\omega}E_x^{\omega}E_x^{\omega} 
\label{eq:NLH}
\end{equation} 

Now, in order to understand the $E_z$ dependence of the nonlinear Hall effect, we can expand $\sigma_{yxx}^{2\omega}$ as a function of $E_z^n$ ($n$ is an integer.) 

\begin{equation}
 \sigma_{yxx}^{2\omega}=A_0 E_z^0 + A_1 E_z^1 + A_2 E_z^2 + A_3 E_z^3 + \cdot\cdot\cdot
 \label{eq:Expand}
\end{equation}

Substituting Eq.(\ref{eq:Expand}) into Eq.(\ref{eq:NLH}), $J^{2\omega}_{yxx}$ can be wrote:

\begin{equation}
 J_{yxx}^{2\omega}=A_0 E_z^0(E_x^{\omega})^2 + A_1 E_z^1(E_x^{\omega})^2 + A_2 E_z^2(E_x^{\omega})^2 + A_3 E_z^3(E_x^{\omega})^2 + \cdot\cdot\cdot
 \label{eq:NLH_Ex}
\end{equation}

The $\mathcal{PT}$ symmetry dictates that all odd-power terms of $E_z$ to vanish and only even-power terms of the $E_z$ are allowed, i.e.,

\begin{equation}
 J_{yxx}^{2\omega}=A_0 E_z^0(E_x^{\omega})^2 + A_2 E_z^2(E_x^{\omega})^2 + A_4 E_z^4(E_x^{\omega})^2 + A_6 E_z^6(E_x^{\omega})^2 + \cdot\cdot\cdot
 \label{eq:NLH_Even}
\end{equation}

\begin{equation}
 \sigma_{yxx}^{2\omega}=A_0 E_z^0 + A_2 E_z^2 + A_4 E_z^4 + A_6 E_z^6 + \cdot\cdot\cdot
 \label{eq:sigma_Even}
\end{equation}

Equation (\ref{eq:sigma_Even}) clearly shows that the nonlinear Hall conductivity $\sigma_{yxx}^{2\omega}$ is symmetric as a function of $E_z$ which is consistent with our observation in Fig. 3D.

\subsubsection{\underline{IV.2.2. Symmetry comparison of second-order Drude and quantum metric Hall nonlinear conductivities }}\label{Symm_Drude}
\hspace{4mm} The second-order Drude conductivity can be express as \cite{Wang2021Intrinsic}:

\begin{equation}
 \sigma^{\textrm{Drude}}_{\alpha\beta\gamma}= -\frac{e^3 \tau^2}{h^3} \sum_n \int_\mathbf{k} \bigl( \partial_{k_{\alpha}} \partial_{k_{\beta}} \partial_{k_{\gamma}} \varepsilon_n \bigr) f(\varepsilon_n)
 \label{eq:NL_Drude}
\end{equation}

One can see that the second-order Drude conductivity requires the breaking both $\mathcal{P}$ and $\mathcal{T}$, because the momentum derivative $\partial_{k_{i}}$ is odd under $\mathcal{P}$ and $\mathcal{T}$ and there are three of them in the expression. In addition, exchanging $\alpha$ and $\beta$ changes the sequence of derivatives. Therefore, the second-order Drude conductivity $\sigma_{\alpha\beta\gamma}^{\textrm{Drude}}$ is symmetric, i.e., $\sigma_{\alpha\beta\gamma}^{\textrm{Drude}}=\sigma_{\beta\alpha\gamma}^{\textrm{Drude}}$. 


The quantum metric nonlinear Hall conductivity can be express as \cite{Wang2021Intrinsic}:
\begin{equation}
\sigma_{\alpha\beta\gamma}^{2\omega}= -2 e^3 \sum^{\varepsilon_n \neq \varepsilon_m}_{n,m} \text{Re} \int_\mathbf{k} \bigl( \frac{v_{\alpha}^n \langle u_n\arrowvert i\partial_{k_{\beta}}u_m\rangle \langle u_m\arrowvert i\partial_{k_{\gamma}}u_n\rangle}{\varepsilon_n - \varepsilon_m} -  \frac{v_{\beta}^n \langle u_n\arrowvert i\partial_{k_{\alpha}}u_m\rangle \langle u_m\arrowvert i\partial_{k_{\gamma}}u_n\rangle}{\varepsilon_n - \varepsilon_m} \bigr)\delta(\varepsilon_n-\varepsilon_{\mathrm{F}})
\label{eq:QMH} 
\end{equation}

Here, exchanging $\alpha$ and $\beta$ flips the signs of the quantum metric nonlinear Hall conductivity. Therefore, the quantum metric nonlinear Hall conductivity is antisymmetric, i.e., $\sigma_{\alpha\beta\gamma}^{\textrm{Metric}}=-\sigma_{\beta\alpha\gamma}^{\textrm{Metric}}$. Therefore, we can use electrical sum frequency generation measurements to differentiate the quantum metric nonlinear Hall conductivity and second-order Drude conductivity.

\subsection{\underline{IV.3. Quantum metric dipole contribution dominated nonlinear Hall signal}}\label{ssec:QMH}

\hspace{4mm} In the main text we already showed that the nonlinear Hall signals in BP/MnBi$_2$Te$_4$/BP heterostructure could be exactly calculated using the general expression. This general expression contains the quantum metric dipole $D_{\mathrm{Metric}}$ contribution plus additional inter-band contributions (AIC). 

\begin{equation}
\sigma_{yxx}^{2\omega}= -2 e^3 \sum_{n} \int_\mathbf{k}\frac{v_y^n g^n_{xx}-v_x^n g^n_{yx}}{\varepsilon_n - \varepsilon_{\bar{n}}}\delta(\varepsilon_n-\varepsilon_{\mathrm{F}})+ \textrm{AIC}
\label{eq:Multiband}
\end{equation}
 \begin{equation}
 \textrm{AIC}= -2 e^3 \sum_{n,m}^{\varepsilon_m \neq \varepsilon_n, \varepsilon_{\bar{n}}} \text{Re} \int_\mathbf{k} \bigl( \frac{v_y^n \langle u_n\arrowvert i\partial_{k_x}u_m\rangle \langle u_m\arrowvert i\partial_{k_x}u_n\rangle}{\varepsilon_n - \varepsilon_m} -  \frac{v_x^n \langle u_n\arrowvert i\partial_{k_y}u_m\rangle \langle u_m\arrowvert i\partial_{k_x}u_n\rangle}{\varepsilon_n - \varepsilon_m} \bigr)\frac{\varepsilon_m-\varepsilon_{\bar{n}}}{\varepsilon_n - \varepsilon_{\bar{n}}}\delta(\varepsilon_n-\varepsilon_{\mathrm{F}})
 \end{equation}

Where the first term in Eq.~(\ref{eq:Multiband}) is the quantum metric dipole contribution, the second term is the additional inter-band contributions. $n$ and $m$ are band induces. $\bar{n}$ is the partner band of band $n$. Because the energy difference term $\varepsilon_n - \varepsilon_{\bar{n}}$ is in the denominator, to maximize the $D_{\mathrm{Metric}}$ contribution, we always choose the band whose energy is the closest to $n$ as the partner band. 

Figure~\ref{Theory_Twobands} shows the nonlinear Hall signals calculated with general expression (blue) and $D_{\mathrm{Metric}}$ contribution (red). Interestingly, the nonlinear Hall signals calculated with $D_{\mathrm{Metric}}$ contribution are approximate to the result calculated with general expression. Therefore, the nonlinear Hall signals observed in the BP/MnBi$_2$Te$_4$ heterostructure are dominated by the quantum metric dipole contribution.

\subsection{\underline{IV.4. Low-energy model for BP/MnBi$_2$Te$_4$/BP heterostructure}}\label{ssec:intro}

\hspace{4mm} In this section, we give an overview of the structure of our minimal model for the heterostructure composed by 6SL MnBi$_2$Te$_4$ (MBT) encapsulated by monolayers of black phosphorus (BP). The following sections contain the details of the different parts (or blocks) of the model. The MBT-BP Hamiltonian we use to model the heterostructure in Fig.~\ref{fig:SM_theory_1}(a) consists of a $40$-band model, 

\begin{equation}
    \hat{H}(\mathbf{k})=\begin{pmatrix}
        \hat{h}_{\text{MBT}}(\mathbf{k}) & \hat{U}_t(\mathbf{k}) & \hat{U}_b(\mathbf{k})\\ 
        \hat{U}_t(\mathbf{k})^{\dag} & \hat{h}_{\text{BP},t}(\mathbf{k}) & 0\\
        \hat{U}_b^{\dag}(\mathbf{k}) & 0 & \hat{h}_{\text{BP},b}(\mathbf{k})
    \end{pmatrix}\text{ ,}
    \label{eq:H_MBT_BP}
\end{equation}

\noindent where $\mathbf{k}=(k_x,k_y)$. We now discuss the structure of each of the building blocks of Eq.(\ref{eq:H_MBT_BP}), and we refer to the corresponding later section about the details of each block.

\subsubsection{\underline{IV.4.1 MnBi$_2$Te$_4$ septuple Hamiltonian}}
\hspace{4mm} We start with the $24 \times 24$ block Hamiltonian $\hat{h}_{\text{MBT}}(\mathbf{k})$, which describes the set of six MBT SLs:

\begin{equation}
    \hat{h}_{\text{MBT}}(\mathbf{k})=\begin{pmatrix}
    \hat{h}_{\text{SL},1}(\mathbf{k}) & \hat{T}_0(\mathbf{k}) & 0 & 0 & 0 & 0\\
    \hat{T}_0^{\dag}(\mathbf{k}) & \hat{h}_{\text{SL},2}(\mathbf{k}) & \hat{T}_0(\mathbf{k}) & 0 & 0 & 0 \\
    0 & \hat{T}_0^{\dag}(\mathbf{k}) & \hat{h}_{\text{SL},3}(\mathbf{k}) & \hat{T}_0(\mathbf{k}) & 0 & 0 \\
    0 & 0 & \hat{T}_0^{\dag}(\mathbf{k}) & \hat{h}_{\text{SL},4}(\mathbf{k}) & \hat{T}_0(\mathbf{k}) & 0 \\
    0 & 0 & 0 & \hat{T}_0^{\dag}(\mathbf{k}) & \hat{h}_{\text{SL},5}(\mathbf{k}) & \hat{T}_0(\mathbf{k}) \\
    0 & 0 & 0 & 0 & \hat{T}_0^{\dag}(\mathbf{k}) & 
    \hat{h}_{\text{SL},6}(\mathbf{k})
    \end{pmatrix} \text{ .}
    \label{h_MBT}
\end{equation}

\noindent Here, $\hat{h}_{\text{SL},i}$ is the Hamiltonian of the SL $i$ and $\hat{T}_0$ denotes the hoping between nearest neighbor SLs. Both $\hat{h}_{\text{SL},i}$ and $\hat{T}_0$ are obtained from the bulk MBT Hamiltonian by taking explicitly into account the system's periodicity along the stacking direction $\hat{z}$~\cite{lian2020flat}. This procedure is highlighted in Sec. IV.5 and yields a SL Hamiltonian that is independent of $k_z$ \cite{lian2020flat},

\begin{equation}
    \hat{h}_{\text{SL},i}(\mathbf{k})=\hat{h}_{\text{N}}(\mathbf{k})-\gamma_{af}\, \hat{h}_{\text{AFM},i}(\mathbf{k}) \text{ .} 
    \label{eq:hSL}
\end{equation}

\noindent The term $\hat{h}_{\text{N}}$ describes the normal state dispersion close to the $\Gamma$ point and $\hat{h}_{\text{AFM},i}$ accounts for terms that arise from AFM order. Their explicit $\mathbf{k}$-dependence is shown in Sec. IV.5. Both are 4-band $k\cdot p$ Hamiltonians in the basis

\begin{equation}
\left\{\left|p_{z,\text{Bi}}^{+},\uparrow\right\rangle,\left|p_{z,\text{Te}}^{-},\downarrow\right\rangle,\left|p_{z,\text{Te}}^{-},\uparrow\right\rangle,\left|p_{z,\text{Bi}}^{+},\downarrow\right\rangle\right\} \,.
    \label{eq:MBT_basis}
\end{equation}

This basis is formed by the symmetric ($+$) superposition of the $p_z$ orbitals of the two Bi layers within a given SL and by the anti-symmetric superposition of the $p_z$ orbitals of the Te atoms in the top and bottom layers of a given SL (see Fig.~\ref{fig:SM_theory_1}). These orbitals carry the largest weight in the low-energy bands of MBT close to the $\Gamma$ point~\cite{Zhang2019a}. The constant $\gamma_{af}$ controls the strength of the magnetic order. 

The inter-SL hopping matrix $\hat{T}_0(\mathbf{k})$, on the other hand, only contains contributions from the normal state since it is difficult to obtain from first principles the momentum-dependence of the exchange fields, which are thus set to constants~\cite{lian2020flat} (see also Sec.IV. 5). Importantly, in this work, we go beyond the $k \cdot p$ expansion considered in Ref.~\cite{lian2020flat} and include cubic terms in $\mathbf{k}$. This is needed in order to capture the 3-fold symmetry of the Fermi surface of the six MBT layers as it appears fully isotropic to order $\mathcal{O}(k^2)$. The explicit $\mathbf{k}$-dependences of $\hat{h}_{\text{SL},i}$ and $\hat{T}_0$ are shown in Sec.IV.5.

\subsubsection{\underline{IV.4.2. BP Hamiltonian}}
\hspace{4mm} The $8 \times 8$ blocks $\hat{h}_{\text{BP},t}$ and $\hat{h}_{\text{BP},b}$ in the heterostructure Hamiltonian in Eq.(\ref{eq:H_MBT_BP}) contain the tight-binding Hamiltonian for BP monolayer located on the top and bottom of the heterostructure, respectively (see Fig.~\ref{fig:SM_theory_1}). Note that top and bottom layer Hamiltonians are related by inversion and the two blocks are thus identical: $\hat{h}_{\text{BP},t}=\hat{h}_{\text{BP},b}$. We discuss the explicit momentum dependence of $\hat{h}_{\text{BP},t}$ in Sec.IV.7. 

\subsubsection{\underline{IV.4.3. MnBi$_2$Te$_4$ and BP coupling Hamiltonian}}
\hspace{4mm} The off-diagonal blocks in the heterostructure Hamiltonian in Eq.\eqref{eq:H_MBT_BP} account for the electronic coupling between MBT and BP. These couplings are obtained from the Fourier transform of the hopping integral between the $p_z$ orbitals of the P atoms in the BP layers, and the $p_z$ orbitals of Bi and Te atoms in the MBT SLs. Since the hopping amplitude decays exponentially as a function of the distance between the orbitals, the coupling is truncated beyond the nearest neighbors layers (see Fig.~\ref{fig:SM_theory_1}A). We have checked explicitly that this is an extremely accurate approximation. If desired it is straightforward to include couplings to further neighbor layers within our formalism. Within this approximation, $\hat{U}_t(\mathbf{k})$ denotes the coupling between the top BP monolayer and the top-most MBT SL, while $\hat{U}_b(\mathbf{k})$ accounts for the coupling between the bottom BP layer and the bottom-most MBT SL. The detailed derivation of the coupling terms and their explicit momentum dependence is discussed in Sec. IV.8.  

\subsection{\underline{IV.5. Modeling MnBi$_2$Te$_4$ septuple layers}}\label{ssec:warping}

\hspace{4mm} In this section, we derive the MBT SL Hamiltonian $\hat{h}_{\text{SL},i}(\mathbf{k})$ in Eq.(\ref{eq:hSL}) and the inter-SL hopping $\hat{T}_0(\mathbf{k})$. As mentioned previously, they are obtained from the bulk MBT Hamiltonian. This procedure is explained in Ref.~\cite{lian2020flat}, but for completeness we briefly describe it here as well. To be consistent with the notation, we denote $\mathbf{k}\equiv (k_x,k_y)$ and keep the $k_z$ dependence explicit in the bulk Hamiltonian.

The first step in obtaining $\hat{h}_{\text{SL},i}$ and $\hat{T}_0$ is to write a $k \cdot p$ Hamiltonian for bulk MBT centered at the $\Gamma$ point, which is where the low-energy bands are located~\cite{Zhang2019a}. The form of the bulk $k \cdot p$ Hamiltonian is determined by symmetry. In its paramagnetic phase, MBT belongs to the space group $R\bar{3}m$ (No. 166). Therefore, its Hamiltonian is invariant under all operations of the point group $\bar{3}m1'$, which is generated by time-reversal ($\mathcal{T}$), spatial inversion ($\mathcal{I}$), three-fold rotation around $\hat{z}$ ($C_{3z}$), and two-fold rotation around $\hat{x}$ ($C_{2x}$), where $\hat{x}$ is aligned with one of the in-plane primitive vectors of the underlying triangular lattice (see Fig.~\ref{fig:SM_theory_1}E). In the basis $\left\{\left|p_{z,\text{Bi}}^{+},\uparrow\right\rangle,\left|p_{z,\text{Te}}^{-},\downarrow\right\rangle,\left|p_{z,\text{Te}}^{-},\uparrow\right\rangle,\left|p_{z,\text{Bi}}^{+},\downarrow\right\rangle\right\}$ these symmetry operations are represented by

\begin{equation}
    \hat{U}_{\mathcal{T}}=\begin{pmatrix}
    0 & 0 & 0 & 1\\
    0 & 0 & -1 & 0\\
    0 & 1 & 0 & 0\\
    -1 & 0 & 0 & 0
    \end{pmatrix} ,
    \hat{U}_{\mathcal{I}}=\begin{pmatrix}
    1 & 0 & 0 & 0 \\
    0 & -1 & 0 & 0 \\
    0 & 0 & -1 & 0\\
    0 & 0 & 0 & 1
    \end{pmatrix} ,
    \hat{U}_{3z}=\begin{pmatrix}
    e^{-i\frac{\pi}{3}} & 0 & 0 & 0 \\
    0 & e^{i\frac{\pi}{3}} & 0 & 0 \\
    0 & 0 & e^{-i\frac{\pi}{3}} & 0\\
    0 & 0 & 0 & e^{i\frac{\pi}{3}}
    \end{pmatrix}, \hat{U}_{2x}=\begin{pmatrix}
    0 & 0 & 0 & i\\
    0 & 0 & -i & 0\\
    0 & -i & 0 & 0\\
    i & 0 & 0 & 0
    \end{pmatrix}\,.
\end{equation}

\noindent We derive the generic form of the Hamiltonian up to order $k^3$ that is consistent with these symmetries using the Python toolkit QSymm~\cite{Varjas_2018}: 

\begin{multline}
    \hat{H}_0(\mathbf{k},k_z)=E_0(\mathbf{k},k_z)+
    \begin{pmatrix}
    M(\mathbf{k},k_z) & iA_2k_{-} & i A_1 k_z & 0 \\
    -i A_2 k_{+} & -M(\mathbf{k},k_z) & 0 & -iA_1 k_z\\
    -iA_1k_z &  0 & -M(\mathbf{k},k_z) & iA_2k_{-}\\
    0 & iA_1k_z& -iA_2k_+ & M(\mathbf{k},k_z)
    \end{pmatrix} \\ +\begin{pmatrix}
    0 & A_3k_zk_+^2 & -P(\mathbf{k}) & 0 \\
    A_3k_zk_-^2 & 0 & 0 & P(\mathbf{k})\\
    -P(\mathbf{k}) & 0 & 0 & A_3k_zk_+^2\\
    0 & P(\mathbf{k}) & A_3k_zk_-^2 & 0
    \end{pmatrix} \text{ ,}
    \label{eq:H0}
\end{multline}

\noindent where we define $k_{\pm}\equiv k_x\pm ik_y$ and 
\begin{align}
    & E_0(\mathbf{k},k_z)=C+D_1k_z^2+D_2(k_x^2+k_y^2)\text{ ,}\\
    & M(\mathbf{k},k_z)=M_0+B_1k_z^2+B_2(k_x^2+k_y^2) \text{ .}\\
    & P(\mathbf{k})=\beta_1 k_x\left(k_x^2-3k_y^2\right) \text{ .}\label{eq:P}
\end{align}

\noindent The constants $A_1$, $A_2$, $A_3$, $B_1$, $B_2$, $C$, $D_1$, $D_2$, $M_0$, and $\beta_1$ are obtained by fitting the $k \cdot p$ model to a DFT band structure of MBT. We discuss the numerical values of these parameters in Sec. IV.9. The first two terms in the right-hand-side of Eq.(\ref{eq:H0}) are equivalent to the bulk MBT Hamiltonian shown in Refs.~\cite{Zhang2019a,lian2020flat} after a unitary transformation using the matrix $\tilde{U}=\text{diag}(i,1,1,-i)$. The third term on the right-hand-side of Eq.(\ref{eq:H0}) is of order $k^3$ and is called the \textit{warping term}. 

In bulk MBT the SLs are stacked along the $\hat{z}$ direction with lattice constant $c_0=40.91$~\AA \cite{lian2020flat}. Therefore, by replacing $k_z\rightarrow\frac{1}{c_0}\sin(c_0k_z)$ and $k_z^2\rightarrow \frac{2}{c_0^2}\left[1-\cos(c_0k_z)\right]$ in Eq.(\ref{eq:H0}), we can reinterpret this compactified model as a one-dimensional (1D) hopping model along $\hat{z}$, 
\begin{equation}
    \hat{H}_0(\mathbf{k},k_z)=\hat{h}_{\text{N}}(\mathbf{k})+\cos(c_0k_z)\left[\hat{T}_0(\mathbf{k})+\hat{T}_0^{\dag}(\mathbf{k})\right]+i\sin(c_0k_z)\left[\hat{T}_0^{\dag}(\mathbf{k})-\hat{T}_0(\mathbf{k})\right] \text{ .}
    \label{eq:H0k}
\end{equation}

\noindent In this scenario, terms that are independent of $k_z$ give the normal state contribution to a given SL
\begin{equation}
    \hat{h}_{N}(\mathbf{k})=\varepsilon_0(\mathbf{k})+\begin{pmatrix}
    m(\mathbf{k}) & i\alpha k_- & -P(\mathbf{k}) & 0 \\
    -i\alpha k_+& -m(\mathbf{k}) & 0 & P(\mathbf{k}) \\
    -P(\mathbf{k}) & 0 & -m(\mathbf{k}) & i\alpha k_-\\
    0 & P(\mathbf{k}) & -i\alpha k_+ & m(\mathbf{k})
    \end{pmatrix} \text{ .}
    \label{eq:hN}
\end{equation}

\noindent Here, $P(\mathbf{k})$ is defined in Eq.(\ref{eq:P}) and
\begin{align}
&\varepsilon_0(\mathbf{k})=\gamma_0+\gamma\left(k_x^2+k_y^2\right)\text{ ,}\\
&m(\mathbf{k})=m_0+\beta_0\left(k_x^2+k_y^2\right) \text{ ,}
\end{align}

\noindent with $\gamma_0=C+\frac{2D_1}{c_0^2}$, $\gamma=D_2$, $m_0=M_0+\frac{2B_1}{c_0^2}$, $\beta_0=B_2$, and $\alpha=A_2$. The remaining terms that are proportional to $\sin(ck_z)$ and $\cos(ck_z)$ define the hopping between two nearest-neighbour SLs, 
\begin{equation}
\hat{T}_0=\begin{pmatrix}
t_1 & i\frac{\alpha_3}{2}k_+^2 & -\lambda & 0 \\
i\frac{\alpha_3}{2}k_-^2 & t_2 & 0 & \lambda\\
\lambda & 0 & t_2 & i\frac{\alpha_3}{2}k_+^2 \\
0 & -\lambda & i\frac{\alpha_3}{2}k_-^2 & t_1
\end{pmatrix} 
\label{eq:T0}
\end{equation}
\noindent with $t_1=-\frac{D_1+B_1}{c_0^2}$, $t_2=-\frac{D_1-B_1}{c_0^2}$, $\lambda=\frac{A_1}{2c_0}$, and $\alpha_3=\frac{A_3}{c_0}$. We note that other cubic terms are allowed by symmetry in Eq.(\ref{eq:H0}), but we verified explicitly that the leading terms promoting the formation of a threefold symmetric Fermi surface are those proportional to $\beta_1$ and $\alpha_3$ that were kept in Eqs.(\ref{eq:hN}) and~(\ref{eq:T0}).

At low temperatures MBT develops A-type AFM order, where the Mn local moments order ferromagnetically along $\pm\hat{z}$ within each SL, while moments in nearest-neighbor SLs point antiparallel (see Fig.~\ref{fig:SM_theory_1}A). To find the magnetic contribution to the SL Hamiltonian, we repeat the previous steps starting from a FM bulk MBT, which has point group $\bar{3}m'$. This point group is generated by the elements $C_{3z}$, inversion $\mathcal{I}$ and the combination of time-reversal and two-fold rotation $\mathcal{T}C_{2x}$. A generic Hamiltonian up to order $\mathcal{O}(k^0)$ that obeys these symmetry constraints reads 

\begin{equation}
    \hat{H}_{\text{mag}}=\begin{pmatrix}
        m_1 & 0 & 0 & 0 \\
        0 & -m_2 & 0 & 0\\
        0 & 0 & m_2 & 0 \\
        0 & 0 & 0 & -m_1
    \end{pmatrix} \text{ ,}
    \label{eq:Hmag}
\end{equation}

\noindent Here we kept only the constant terms in the $k \cdot p$ expansion because the $\mathbf{k}$ dependence of the exchange fields is difficult to reliably calculate from first principles~\cite{lian2020flat}. We note that Eq.(\ref{eq:Hmag}) acquires a global minus sign if the FM order points towards $-\hat{z}$. Since consecutive nearest-neighbors SLs are ordered AFM, the magnetic contribution for the Hamiltonian of the SL $i$ defined in Eq.(\ref{eq:hSL}) is simply
  
\begin{equation}
    \hat{h}_{\text{AFM},i}=(-1)^i\hat{H}_{\text{mag}} \text{\,.}
    \label{eq:hAFM}
\end{equation}

\noindent As we approximate $\hat{H}_{\text{mag}}$ to be independent of $\mathbf{k}$, there is no magetic contribution to the interlayer hopping $\hat{T}_0$.

\subsection{\underline{IV.6.Effects of strain on low-energy model of MnBi$_2$Te$_4$}}\label{ssec:strain}
\hspace{4mm} Encapsulating the MBT SLs by BP monolayer introduces strain due to the large mismatch between their lattice constants. The effects of strain can be straightforwardly incorporated in our model using symmetry arguments similar to those in Sec. IV.5. 

In this section, we focus on the effects of strain on the MBT part of the heterostructure. We show here that strain modifies the SL Hamiltonian and the inter-SL hopping according to 

\begin{align}
    &\hat{h}_{\text{SL},i}(\mathbf{k})=\hat{h}_{\text{N}}(\mathbf{k})+\gamma_{s}\hat{h}_{\text{N},s}(\mathbf{k})-\gamma_{af}\left[\hat{h}_{\text{AFM},i}(\mathbf{k}) +\gamma_s\hat{h}_{\text{AFM},s}(\mathbf{k})\right]\text{ .} 
    \label{eq:newh_SL}\\
    &\hat{T}(\mathbf{k})=\hat{T}_0(\mathbf{k})+\gamma_s\hat{T}_{s}(\mathbf{k}) \text{ .}\label{eq:newT}
\end{align}

\noindent Here, $\hat{h}_{\text{N}}$, $\hat{h}_{\text{AFM},i}$ and $\hat{T}_0$ are the same as defined in Eq.(\ref{eq:hN}), Eq.(\ref{eq:hAFM}) and Eq.(\ref{eq:T0}), respectively. The constant $\gamma_s>0$ tunes the strength of strain. The new terms $\hat{h}_{\text{N},s}$, $\hat{h}_{\text{AFM},s}$ and $\hat{T}_s$ are identical for all SLs and are obtained from compactifying the $k \cdot p$ MBT bulk Hamiltonian along $\hat{z}$ in the presence of strain following the same steps shown in Sec. IV.5.

We focus on the case of uniaxial strain along $\hat{x}=[100]$, which breaks the threefold rotation around $\hat{z}=[001]$ ($C_{3z}$) as well as the twofold rotations around $[010]$ and $[110]$ originally present in the unstrained $R\bar{3}m$ space group of the material. Note that this strain preserves the two-fold rotation around $[100]$.
We obtain a generic term consistent with these symmetries using Qsymm \cite{Varjas_2018}:
\begin{equation}
    \hat{H}_{s}(\mathbf{k})=\begin{pmatrix}
    s_1k_x^2+s_2k_yk_z & is_5k_x+s_6k_z & s_7k_x+is_8k_y & 0 \\
    -is_5k_x+s_6k_z & s_3k_x^2+s_4k_yk_z & 0 & -s_7k_x-is_8k_y \\
    s_7k_x-is_8k_y & 0 & s_3k_x^2+s_4k_yk_z & is_5k_x+s_6k_z \\
    0 & -s_7k_x+is_8k_y & -is_5k_x+s_6k_z & s_1k_x^2+s_2k_yk_z
    \end{pmatrix} \text{ \,.}
    \label{eq:Hs}
\end{equation}

\noindent This term is added to the bulk MBT Hamiltonian in Eq.(\ref{eq:H0}). Similarly, the magnetic part of the Hamiltonian (Eq.(\ref{eq:Hmag})) gains a term of the form
\begin{equation}
    \hat{H}_{\text{mag},s}=s_9\begin{pmatrix}
    0& 0& 0 & i\\
    0 & 0 & i & 0\\
    0 & -i & 0 & 0 \\
    -i & 0 & 0 & 0
    \end{pmatrix} \text{ .}
    \label{eq:Hmag_s}
\end{equation}

\noindent For simplicity, we carried the $k \cdot p$ expansion up to $k^2$ order in Eqs.(\ref{eq:Hs}) and (\ref{eq:Hmag_s}). The quantities $a_j$ are real parameters. Replacing $k_z\rightarrow \frac{1}{c_0}\sin{c_0k_z}$ in Eqs.(\ref{eq:Hs}) and (\ref{eq:Hmag_s}) and separating the terms proportional to $\sin(c_0k_z)$ from those indenpendent of $k_z$ as in Eq.(\ref{eq:H0k}), we obtain the strain correction to the normal state Hamiltonian of the SL,
\begin{equation}
    \hat{h}_{\text{N},s}(\mathbf{k})=\begin{pmatrix}
    s_{1}k_x^2 & is_{5} k_x & s_{7} k_x+i s_8 k_y & 0\\
    -i s_5 k_x& s_3 k_x^2 & 0 & -s_7 k_x - i s_8 k_y\\
    s_7 k_x - i s_8 k_y & 0 & s_3 k_x^2 & i s_5 k_x\\
    0 & -s_7 k_x + i s_8 k_y & -i s_5 k_x & s_1 k_x^2
    \end{pmatrix}\text{ .}
    \label{eq:hNs}
\end{equation}

\noindent  The strain contribution to the magnetic part of the SL Hamiltonian $\hat{h}_{\text{AFM},s}(\mathbf{k})$ is identical to $\hat{H}_{\text{mag},s}$ defined in Eq.(\ref{eq:Hmag_s}), and the modification to the inter-SL hopping reads
\begin{equation}
    \hat{T}_{s}(\mathbf{k})=\frac{i}{2c_0} \begin{pmatrix}
    s_2 k_y & s_6 & 0 & 0 \\
    s_6 & s_4 k_y & 0 & 0 \\
    0 & 0 & s_4 k_y & s_6\\
    0 & 0 & s_6 & s_2 k_y
    \end{pmatrix} \text{ .}
    \label{eq:Ts}
\end{equation}

The low-energy  bands for a single MBT SL modeled by Eq.(\ref{eq:newh_SL}) and for a set of six SL described by Eq.(\ref{h_MBT}) are shown in Fig.\ref{fig:SM_theory_2}. In the paramagnetic state, the SL is a topological insulator with protected Dirac cones in the top and bottom surfaces normal to $\hat{z}$. As a result of the small thickness of the SL, these surface states hybridize and a energy gap develops at the $\Gamma$ point, as shown in Fig.~\ref{fig:SM_theory_2}A. When multiple SLs are stacked along $\hat{z}$, the hybridization of the surface states suppressed and the stacking of six SLs is already enough to recover the surface Dirac cone at $\Gamma$. At the onset of the AFM order, however, time-reversal symmetry is lost, which culminates in the emergence of a gap in the surface states. 

The magnetic order preserves $C_{3z}$, which is reflected in a threefold symmetric Fermi surface of six SL system shown in Fig.~\ref{fig:SM_theory_2}D, as long as warping terms discussed in Sec. IV.5 are included in the low energy model. This 3-fold symmetry of the Fermi surface is further lowered in the presence of uniaxial strain as highlighted in Fig.~\ref{fig:SM_theory_2}E and Fig.~\ref{fig:SM_theory_2}F.  

\subsection{\underline{IV.7. Modeling the BP monolayers}}\label{ssec:modelBP}
\hspace{4mm} In this section we derive the minimal model for the BP monolayer, $\hat{h}_{\text{BP},t}$ and $\hat{h}_{\text{BP},b}$ in Eq.(\ref{eq:H_MBT_BP}). Since the BP layers on the top and bottom of the heterostructure are identical, $\hat{h}_{\text{BP},t}=\hat{h}_{\text{BP},b}$, we hereafter drop the layer subindex. 

Following Refs.~\cite{rudenko2015toward, ezawa2014topological}, the low-energy features of the material is captured by a tight-biding model where the hopping processes involve only the $p_z$ orbitals of the P atoms. The BP layer is corrugated rectangular lattice with primitive vectors $\mathbf{a}_{1}^{(\text{BP})}=a\hat{\mathbf{x}}$ and $\mathbf{a}_{2}^{(\text{BP})}=b\hat{\mathbf{y}}$, with lattice constants $a=4.43$~\AA, $b=3.27$~\AA. There are four nonequivalent P atoms per unit cell, and its low-energy tight-biding model is thus a 4-band Hamiltonian, 
\begin{equation}
\hat{h}_{\text{BP},\sigma}(\mathbf{k})=\begin{pmatrix}
0 & f_{1}(k_x,k_y) & 0 & f_{2}(k_x,k_y)\\
f_{1}^*(k_x,k_y) & 0 & f_{2}(k_x,k_y) & 0\\
0 & f_{2}^*(k_x,k_y) & 0 & f_{1}(k_x,k_y)\\
f_{2}^*(k_x,k_y) & 0 & f_{1}^*(k_x,k_y) & 0
\end{pmatrix} \text{ \,.}
\label{eq:hBP_sigma}
\end{equation}
This matrix is written in the basis $\{\left|p_{z,\text{P}_1}\right\rangle,\left|p_{z,\text{P}_2}\right\rangle,\left|p_{z,\text{P}_3}\right\rangle,\left|p_{z,\text{P}_4}\right\rangle\}$, which is ordered according to the fractional positions $\boldsymbol{\rho}_i$ of the phosphorus atoms in the unit cell (see Table \ref{tab:Ppositon}). Since the hopping processes are spin independent, the Hamiltonian in Eq.(\ref{eq:hBP_sigma}) is identical for spin up ($\sigma=\uparrow$) and spin down ($\sigma=\downarrow$). The full BP Hamiltonian that enters in Eq.(\ref{eq:H_MBT_BP}) is block-diagonal, 
\begin{equation}
    \hat{h}_{\text{BP},t}(\mathbf{k})=\hat{h}_{\text{BP},b}(\mathbf{k})=\begin{pmatrix}
    \hat{h}_{\text{BP},\uparrow}(\mathbf{k}) & 0 \\
    0 & \hat{h}_{\text{BP},\downarrow}(\mathbf{k})
    \end{pmatrix} \text{ .}
    \label{eq:hBP_full}
\end{equation}

In Eq.(\ref{eq:hBP_sigma}) we kept only two distinct hopping parameters $\tilde{t}_1$ and $\tilde{t}_2$ defined in the functions
\begin{align}
   & f_{1}(k_x,k_y)=2\tilde{t}_1e^{i\frac{ak_x}{2\sqrt{3}}}\cos\left(\frac{bk_y}{2}\right)\text{ ,}\\[0.2cm]
   & f_{2}(k_x,k_y)=\tilde{t}_2e^{-i\frac{ak_x}{\sqrt{3}}} \text{ ,}
\end{align}

\noindent and thus our BP model is a simplified version of the tight-binding Hamiltonian considered in Ref.~\cite{rudenko2015toward}. This simplification based on the fact that our minimal model for MBT, and by construction also the model for the restructure, is defined around the $\Gamma$ point ($\mathbf{k}=0)$. In this region of the  Brillouin zone, $\tilde{t}_1$ and $\tilde{t}_2$ are the dominant contributions the band gap and energy dispersion of BP~\cite{rudenko2015toward}. Note that the $\Gamma$ point in BP Brillouin zone coincides with the $\Gamma$ point in MBT Brillouin zone since we assume that the origin of these two lattices are aligned. 

To account for the experimentally observed electronic gap of the BP monolayers in our system, we choose values for the hopping amplitudes that are slightly different from Ref.~\cite{rudenko2015toward}. The BP band gap at $\Gamma$ is given by
\begin{equation}
\Delta=4\left|\tilde{t}_1\right|\left|1- \frac12 \left|\frac{\tilde{t}_2}{\tilde{t}_1}\right|\right| \text{ ,}    
\end{equation}

\noindent and we choose $\tilde{t}_1=-1.7575$~eV and $\tilde{t}_2=3.665$~eV to obtain $\Delta=0.3$~eV in agreement with experimental observation. The energy bands of our minimal model for BP [Eq.(\ref{eq:hBP_sigma})] is shown in Fig.~\ref{fig:SM_theory_3}.

\subsection{\underline{IV.8. Modeling the coupling between MnBi$_2$Te$_4$ and BP}}\label{ssec:coupling}

\hspace{4mm} In this section we derive the expressions for the couplings between the MBT SLs and the BP monolayer, which are denoted as as $\hat{U}_t(\mathbf{k})$ and $\hat{U}_b(\mathbf{k})$ in Eq.~(\ref{eq:H_MBT_BP}). These off-diagonal blocks results from the electronic hopping between the $p_z$ orbitals of the BP and MBT layers. Due to the exponential decrease of the electronic coupling with distance between orbitals, we restrict the coupling to nearest-neighbor layers. The BP monolayer at the bottom is coupled only to the lowest MBT layer (SL index $l=6$ in Fig.~\ref{fig:SM_theory_1}A) and the BP monolayer at the top is only coupled to the top SL (SL index $l=1$). 

\subsubsection{\underline{IV.8.1. Construction of the MnBi$_2$Te$_4$ and BP coupling Hamiltonian}}

\hspace{4mm} Let us first consider the hopping involving the bottom BP layer. In the site basis it reads

\begin{align}
    \hat{H}_{\text{int}}^{(b)}&=\sum\limits_{i\in \text{SL}}\sum\limits_{j\in\text{BP}}\sum\limits_{\sigma}\left(\hat{\psi}_{i,\sigma}^{\dag}\,\hat{t}_{ij}^{(b)}\,\hat{\phi}_{j,\sigma}+\text{h.c.}\right) \text{ .}
    \label{eq:Hint_site}
\end{align}

\noindent Here, the sum over $i$ runs over the unit cells of the bottommost MBT SL ($l=6$) with position $\mathbf{R}_i\equiv n_1\mathbf{a}_1^{(\text{MBT})}+n_2\mathbf{a}_2^{(\text{MBT})}$ ($n_1$, $n_2$ integers). The sum over $j$ runs over the unit cells of the bottom BP layer, with $\mathbf{R}_j=m_1\mathbf{a}_1^{(\text{BP})}+m_2\mathbf{a}_2^{(\text{BP})}$ ($m_1$, $m_2$ integrers) and $\mathbf{a}_{i}^{(\text{MBT})}\neq \mathbf{a}_{j}^{(\text{BP})}$. Besides, $\hat{\phi}_{j,\sigma}^{\dag}$ is a 4-dimensional spinor whose $\nu$th component corresponds to creating an electron with spin $\sigma$ in the $p_z$ orbital of the P atom located at position $\boldsymbol{\rho}_{\nu}$ in the $j$-th unit cell of the BP lattice (see Table \ref{tab:Ppositon}). Similarly, $\hat{\psi}_{i,\sigma}^{\dag}$ ($l=1,2$) is a 7-dimensional spinor whose $\mu$th component creates an electron with spin $\sigma$ on Mn (if $\mu=7$), Bi (if $\mu=5,6$) or Te (if $\mu=1,2,3,4$) atoms that are located at position $\boldsymbol{\tau}_{\mu}$ in the $i$-th unit cell of the bottom most MBT SL. 

The quantity $\hat{t}_{ij}^{(b)}$ is a $7\times 4$ dimensional hopping matrix, whose matrix elements 

\begin{equation}
    \left(\hat{t}_{ij}^{(b)}\right)_{\mu,\nu}\equiv t_b\left(\mathbf{R}_i-\mathbf{R}_j+\boldsymbol{\tau}_\mu-\boldsymbol{\rho}_{\nu}\right)
    \label{eq:tb}
\end{equation}

\noindent depend only on the relative distance between the sites involved in the hopping process. It is important to note that while $\mathbf{R}_i$ are two-dimensional vectors denoting the unit cell position in the layers, the basis vectors $\boldsymbol{\tau}_\mu$ and $\boldsymbol{\rho}_\mu$ are three-dimensional. We can decompose the basis vectors as $\boldsymbol{\tau}_\mu = \boldsymbol{\tau}_\mu^\parallel + \boldsymbol{\tau}_\mu^\perp$ and $\boldsymbol{\rho}_\nu = \boldsymbol{\rho}_\nu^\parallel + \boldsymbol{\rho}_\nu^\perp$, where the subscript $\parallel$ denotes the component in the layer and $\perp$ denotes the component perpendicular to the layer, i.e., along the stacking direction $\hat{\mathbf{z}}$. In this work we calculate the hopping elements $t_b$ using the Slater-Koster approach. The Fourier transform of Eq.(\ref{eq:Hint_site}) gives $\hat{U}_{b}(\mathbf{k})$. This procedure has subtleties because of the lattice mismatch between BP and MBT and is thoroughly presented in the remainder of this section. 

To go from site to momentum space, we use convention to include the basis atom positions in the Fourier transformation

\begin{align}
    & \psi_{i\mu,\sigma}^{\dag}=\frac{1}{\sqrt{N_{\text{SL}}}}\sum\limits_{\mathbf{k}\in \text{BZ}_{\text{SL}}}e^{-i\mathbf{k}\cdot\left(\mathbf{R}_i+\boldsymbol{\tau}_\mu^{\parallel}\right)} \psi_{\mathbf{k}\mu,\sigma}^{\dag} \text{ ,}\label{eq:cI}\\[0.2cm]
    & \phi_{j\nu,\sigma}^{\dag}=\frac{1}{\sqrt{N_{\text{BP}}}}\sum\limits_{\mathbf{q}\in \text{BZ}_{\text{BP}}}e^{-i\mathbf{q}\cdot\left(\mathbf{R}_j+\boldsymbol{\rho}_\nu^{\parallel}\right)} \phi_{\mathbf{q}\nu,\sigma}^{\dag} \text{ .}\label{eq:dI}
\end{align}

\noindent 
Here, $N_{\text{SL}}$ ($N_{\text{BP}}$) corresponds to the number of unit cells in the MBT SL (BP layer), and  $\boldsymbol{\tau}_{\mu}^{\parallel}$ and $\boldsymbol{\rho}_{\nu}^{\parallel}$ are the components of $\boldsymbol{\tau}_{\mu}$ and $\boldsymbol{\rho}_{\nu}$ normal to the stacking direction $\hat{\mathbf{z}}$. 

Substituting Eqs.(\ref{eq:cI}) and (\ref{eq:dI}) into Eq.(\ref{eq:Hint_site}), we find

\begin{equation}
    \hat{H}_{\text{int}}^{(b)}=\sum\limits_{\mathbf{k}\in \text{BZ}_{\text{SL}} \atop \mathbf{q}\in \text{BZ}_{\text{BP}}} \sum\limits_{\sigma}\biggl[\frac{e^{-i\left(\mathbf{k}\cdot\boldsymbol{\tau}_{\mu}^{\parallel}-\mathbf{q}\cdot\boldsymbol{\rho}_{\nu}^{\parallel}\right)}}{\sqrt{N_{\text{SL}}N_{\text{BP}}}}\psi_{\mathbf{k}\mu,\sigma}^{\dag}\biggl\{\sum\limits_{i\in \text{SL}_6\atop j\in\text{BP}}t_b\bigl(\mathbf{R}_i-\mathbf{R}_j+\boldsymbol{\tau}_{\mu}-\boldsymbol{\rho}_{\nu}\bigr)e^{-i\mathbf{k}\cdot\mathbf{R}_i}e^{i\mathbf{q}\cdot\mathbf{R}_j}\biggr\}\phi_{\mathbf{q}\nu,\sigma} +\text{h.c}\biggr] \text{ .}
    \label{eq:Hsite_FT}
\end{equation}

\noindent The $\mathbf{k}$ and $\mathbf{q}$ sums run over the Brillouin zones of the MBT SL and BP layer, respectively. To deal with the remaining sums over $i$ and $j$, we note that 

\begin{align}
    &\sum\limits_{i\in \text{SL}_6}t_b\left(\mathbf{R}_i-\mathbf{R}_j+\boldsymbol{\tau}_{\mu}-\boldsymbol{\rho}_\nu\right)e^{-i\mathbf{k}\cdot \mathbf{R}_i}
    =\sum\limits_{(n_1,n_2)\in \mathbb{Z}^2}t_b\left(n_1\mathbf{a}_1^{(\text{MBT})}+n_2\mathbf{a}_2^{(\text{MBT})}-\mathbf{R}_j+\boldsymbol{\tau}_{\mu}-\boldsymbol{\rho}_\nu\right) \nonumber \\ & \qquad \qquad \qquad \qquad \qquad \qquad \qquad \qquad \qquad \qquad \qquad \times  e^{-i\mathbf{k}\cdot \left(n_1\mathbf{a}_1^{(\text{MBT})}+n_2\mathbf{a}_2^{(\text{MBT})}\right)}
 \nonumber\\
   &=\sum\limits_{(n_1,n_2)\in \mathbb{Z}^2}\int\limits_{-\infty}^{\infty} dx_1\int\limits_{-\infty}^{\infty} dx_2\,\delta(x_1-n_1)\delta(x_2-n_2)t_b\left(x_1\mathbf{a}_1^{(\text{MBT})}+x_2\mathbf{a}_2^{(\text{MBT})}-\mathbf{R}_j+\boldsymbol{\tau}_{\mu}-\boldsymbol{\rho}_\nu\right)\nonumber \\ & \qquad \qquad \qquad \qquad \qquad \qquad \qquad \qquad \qquad \qquad \qquad \times e^{-i\mathbf{k}\cdot \left(x_1\mathbf{a}_1^{(\text{MBT})}+x_2\mathbf{a}_2^{(\text{MBT})}\right)} \text{ ,}\label{eq:sum_n}
\end{align}

\noindent where $x_1$ and $x_2$ are real dimensionless variables. Using the Poisson summation formula, we can rewrite

\begin{equation}
    \sum\limits_{n=-\infty}^{\infty}\delta(x-n)=\sum\limits_{m=-\infty}^{\infty}e^{-i2\pi mx} \text{ ,}
\end{equation}

\noindent and then 

\begin{align}
    \sum\limits_{(n_1,n_2)\in \mathbb{Z}^2}\delta(x_1-n_1)\delta(x_2-n_2) &=\sum\limits_{(m_1,m_2)\in \mathbb{Z}^2}e^{-i2\pi\left(m_1x_1+m_2x_2\right)} \nonumber \\ 
    & =\sum\limits_{(m_1,m_2)\in \mathbb{Z}^2}e^{-i\left(x_1\mathbf{a}_1^{(\text{MBT})}+x_2\mathbf{a}_2^{(\text{MBT})}\right)\cdot \left(m_1\mathbf{b}_1^{(\text{MBT})}+m_2\mathbf{b}_2^{(\text{MBT})}\right)}\nonumber\\[0.3cm]
    & = \sum\limits_{i}e^{-i\left(x_1\mathbf{a}_1^{(\text{MBT})}+x_2\mathbf{a}_2^{(\text{MBT})}\right)\cdot\mathbf{G}_i^{(\text{MBT})}} \text{ \,.}\label{eq:sumGi}
\end{align}

\noindent Here, $\mathbf{G}_i^{(\text{MBT})}=i_1\mathbf{b}_1^{(\text{MBT})}+i_2\mathbf{b}_2^{(\text{MBT})}$ denotes the unit cell position of the reciprocal MBT SL lattice, and the sum over $m$ runs over the reciprocal MBT SL lattice unit cells. 

Substituting Eq.(\ref{eq:sumGi}) into Eq.(\ref{eq:sum_n}), we obtain

\begin{multline}
    \sum\limits_{i\in \text{SL}_6}t_b\left(\mathbf{R}_i-\mathbf{R}_j+\boldsymbol{\tau}_{\mu}-\boldsymbol{\rho}_\nu\right)e^{-i\mathbf{k}\cdot \mathbf{R}_i} =\sum\limits_{i}\int\limits_{-\infty}^{\infty} dx_1\int\limits_{-\infty}^{\infty} dx_2\,t_b\left(x_1\mathbf{a}_1^{(\text{MBT})}+x_2\mathbf{a}_2^{(\text{MBT})}-\mathbf{R}_j+\boldsymbol{\tau}_{\mu}-\boldsymbol{\rho}_\nu\right) \\ \times  e^{-i\left(\mathbf{k}+\mathbf{G}_i^{(\text{MBT})}\right)\cdot \left(x_1\mathbf{a}_1^{(\text{MBT})}+x_2\mathbf{a}_2^{(\text{MBT})}\right)}
    \label{eq:res_i}
\end{multline}

\noindent Similarly

\begin{multline}
    \sum\limits_{j\in \text{BP}}t_b\left(\mathbf{R}_i-\mathbf{R}_j+\boldsymbol{\tau}_{\mu}-\boldsymbol{\rho}_\nu\right)e^{i\mathbf{q}\cdot \mathbf{R}_j}=\sum\limits_{j}\int\limits_{-\infty}^{\infty} dx_3\int\limits_{-\infty}^{\infty} dx_4\,t_b\left(\mathbf{R}_i-x_3\mathbf{a}_1^{(\text{BP})}-x_4\mathbf{a}_2^{(\text{BP})}+\boldsymbol{\tau}_{\mu}-\boldsymbol{\rho}_\nu\right) \\ \times e^{i\left(\mathbf{q}-\mathbf{G}_j^{(\text{BP})}\right)\cdot \left(x_3\mathbf{a}_1^{(\text{BP})}+x_4\mathbf{a}_2^{(\text{BP})}\right)}
  \label{eq:res_j}
\end{multline}

\noindent where again $x_3$ and $x_4$ are real dimensionless variables and the $j$ sum runs over the unit cells of the reciprocal BP lattice $\mathbf{G}_j^{(\text{BP})}=j_1\mathbf{b}_1^{(\text{BP})}+j_2\mathbf{b}_2^{(\text{BP})}$. 

Combining Eqs.~(\ref{eq:res_i}) and (\ref{eq:res_j}) and substituting them into Eq.~(\ref{eq:Hsite_FT}) yields 

\begin{multline}
    \hat{H}_{\text{int}}^{(b)}=\frac{1}{\sqrt{N_{\text{SL}}N_{\text{BP}}}}\frac{1}{\Omega_{\text{SL}}\Omega_{\text{BP}}} 
   \sum\limits_{\mu,\nu,\sigma}\sum\limits_{\mathbf{k},i}\sum\limits_{\mathbf{q},j}\biggl[\int d^2r\int d^2r'\, t_b\left(\mathbf{r}-\mathbf{r}'+\boldsymbol{\tau}_{\mu}-\boldsymbol{\rho}_{\nu}\right) \\ \times e^{i\left(-\mathbf{k}\cdot\boldsymbol{\tau}_{\mu}^{\parallel}+\mathbf{q}\cdot\boldsymbol{\rho}_{\nu}^{\parallel}\right)}e^{-i\left(\mathbf{k}+\mathbf{G}_i^{(\text{MBT})}\right)\cdot\mathbf{r}}e^{i\left(\mathbf{q}-\mathbf{G}_j^{(\text{BP})}\right)\cdot\mathbf{r}'} \psi_{\mathbf{k}\mu,\sigma}^{\dag}\phi_{\mathbf{q}\nu,\sigma}+\text{h.c.}\biggr] \text{ ,}
    \label{eq:Hint_momentum}
\end{multline}

\noindent where we applied the change of variables $\mathbf{r}=x_1\mathbf{a}_1^{\text{(MBT)}}+x_2\mathbf{a}_2^{\text{(MBT)}}$ and $\mathbf{r}'=x_3\mathbf{a}_1^{\text{(BP)}}+x_4\mathbf{a}_2^{\text{(BP)}}$. Recall that here the sum over $i$ ($j$) runs over the reciprocal unit cells of MBT SL (BP), while the sum over $\mathbf{k}$ ($\mathbf{q}$) are restricted to the first BZ of MBT SL (BP). Besides, $\Omega_{\text{SL}}$ ($\Omega_{\text{BP}}$) denotes the area of the unit cell of the SL (BP) Bravais lattices.

We focus now on the arguments of the hopping amplitude $t_b$. While $\mathbf{r}$ and $\mathbf{r}'$ are 2D vectors, $\boldsymbol{\tau}_{\mu}$ and $\boldsymbol{\rho}_{\nu}$ are in general 3D, as mentioned before. Defining 

\begin{equation}
    s_{\mu\nu}\equiv \left(\boldsymbol{\tau}_{\mu}-\boldsymbol{\rho}_{\nu}\right)\cdot \hat{\mathbf{z}} = \tau^\perp_\mu - \rho^\perp_\nu \,,
\end{equation}

we can rewrite $ \boldsymbol{\tau}_{\mu}-\boldsymbol{\rho}_{\nu}= \boldsymbol{\tau}_{\mu}^{\parallel}-\boldsymbol{\rho}_{\nu}^{\parallel}+s_{\mu\nu}\hat{\mathbf{z}}$. Performing the change of variables $\mathbf{y}=\mathbf{r}-\mathbf{r}'+\boldsymbol{\tau}_{\mu}^{\parallel}-\boldsymbol{\rho}_{\nu}^{\parallel}$, we obtain 

\begin{multline}
    \hat{H}_{\text{int}}^{(b)}=\sqrt{\frac{N_{\text{SL}}}{N_{\text{BP}}}}\frac{1}{\Omega_{\text{BP}}}\sum\limits_{\mu,\nu,\sigma}^{(b)}\sum\limits_{i,\mathbf{k}}\sum\limits_{j,\mathbf{q}}\biggl[ t_{\mu\nu}(\mathbf{k}+\mathbf{G}_i^{(\text{MBT})})e^{i\left(\mathbf{G}_i^{(\text{MBT})}\cdot\boldsymbol{\tau}_{\mu}^{\parallel}+\mathbf{G}_j^{(\text{BP})}\cdot\boldsymbol{\rho}_{\nu}^{\parallel}\right)} \\ \times \psi_{\mathbf{k}\mu,\sigma}^{\dag}\phi_{\mathbf{q}\nu,\sigma}\delta_{\mathbf{q},\mathbf{k}+\mathbf{G}_i^{(\text{MBT})}+\mathbf{G}_j^{(\text{BP})}}+\text{h.c.}\biggr] \,.
    \label{eq:Hint2_final}
\end{multline}

\noindent Here,

\begin{equation}
    t_{\mu\nu}^{(b)}(\mathbf{q}) = \int d^2y\, t_b\left(\mathbf{y}+s_{\mu\nu}\hat{\mathbf{z}}\right)e^{-i\mathbf{q}\cdot \mathbf{y}}\text{ ,}
    \label{eq:hopping_Fourier}
\end{equation}

\noindent is the Fourier transform of the hopping amplitude, which we explicitly calculate, and we used
\begin{equation}
\int d^2r e^{i\mathbf{r}\cdot\left(-\mathbf{k}-\mathbf{G}_i^{(\text{MBT})}-\mathbf{G}_j^{(\text{BP})}+\mathbf{q}\right)}=N_{\text{SL}}\Omega_{\text{SL}}\,\delta_{\mathbf{q}-\mathbf{k},\mathbf{G}_i^{(\text{MBT})}+\mathbf{G}_j^{(\text{BP})}} \text{ .}
\end{equation}

Since we here focus on a minimal model for the MBT-BP heterostructure that is valid around the $\Gamma$ point and $t_{\mu\nu}^{(b)}(\mathbf{k}+\mathbf{G}_i)$ decays rapidly with $|\mathbf{k}+\mathbf{G}_i|$ (on a scale determined by the size of the orbitals), only $\mathbf{G}_i^{(\text{MBT})}=0$ contributes significantly in the summation over $i$ in Eq.~\eqref{eq:Hint2_final}. Momentum conservation in  Eq.(\ref{eq:Hint2_final}) then enforces $\mathbf{q}=\mathbf{G}_j^{(\text{BP})}+\mathbf{k}$, which can only be satisfied for $\mathbf{G}_j^{\text{(BP)}}=0$ as $\mathbf{q}$ and $\mathbf{k}$ lie in the first Brillouin zone. 
Therefore, the contribution to the continuum model of the heterostructure around $\Gamma$ due to MBT-BP coupling takes the final form 

\begin{equation}
   \hat{H}_{\text{int}}^{(b)}=\sqrt{\frac{N_{\text{SL}}}{N_{\text{BP}}}}\frac{1}{\Omega_{\text{BP}}}\sum\limits_{\mu,\nu,\sigma}\sum\limits_{\mathbf{k}}\left[t_{\mu\nu}^{(b)}(\mathbf{k})\psi_{\mathbf{k}\mu,\sigma}^{\dag}\phi_{\mathbf{k}\nu,\sigma}+\text{h.c.}\right] \text{ .}
   \label{eq:Hint_b}
\end{equation}

In the following, we set $N_{\text{SL}} = N_{\text{BP}}$ and we recall that $\Omega_{\text{BP}}$ denotes the are of the BP unit cell. 
This interaction Hamiltonian involves the full $7 \times 4$ dimensional hopping matrix $t^{(b)}_{\mu \nu}$. However, since the minimum low-energy model for MBT SL contains only four specific even and odd combinations of selected Bi and Te orbitals [see low-energy MBT basis in Eq.~\eqref{eq:MBT_basis}], we only need to consider a part of the matrix $t_{\mu\nu}^{(b)}(\mathbf{k})$ to construct $\hat{U}_{b}(\mathbf{k})$. Projecting $\hat{H}_{\text{int}}^{(b)}$ into the low-energy basis in Eq.~\eqref{eq:MBT_basis} yields

\begin{equation}
    \mathcal{P}\hat{H}_{\text{int}}^{(b)}\mathcal{P}^{-1}=\sum\limits_{\mathbf{k}}\left[\hat{\varphi}_{\mathbf{k}}^{\dag}\hat{V}_{b,\uparrow}(\mathbf{k})\hat{\phi}_{\mathbf{k},\uparrow}+\hat{\varphi}_{\mathbf{k}}^{\dag}\hat{V}_{b,\downarrow}(\mathbf{k})\hat{\phi}_{\mathbf{k},\downarrow}+\text{h.c.}\right]\text{\,.}
    \label{eq:Hint_projected}
\end{equation}

\noindent Here, $\hat{\varphi}_{\mathbf{k}}$ is a 4-dimensional spinor that creates electrons in the MBT basis states in Eq.~\eqref{eq:MBT_basis} and 

\begin{align}
   \hat{V}_{b,\uparrow}(\mathbf{k})&=\frac{1}{\sqrt{2} \Omega_{\text{BP}}}\begin{pmatrix}
   t_{51}^{(b)}(\mathbf{k})+t_{61}^{(b)}(\mathbf{k}), & t_{52}^{(b)}(\mathbf{k})+t_{62}^{(b)}(\mathbf{k}), &  t_{53}^{(b)}(\mathbf{k})+t_{63}^{(b)}(\mathbf{k}), &  t_{54}^{(b)}(\mathbf{k})+t_{64}^{(b)}(\mathbf{k})\\
   0 & 0 & 0 & 0 \\
    t_{11}^{(b)}(\mathbf{k})-t_{41}^{(b)}(\mathbf{k}), & t_{12}^{(b)}(\mathbf{k})-t_{42}^{(b)}(\mathbf{k}), &  t_{13}^{(b)}(\mathbf{k})-t_{43}^{(b)}(\mathbf{k}), &  t_{14}^{(b)}(\mathbf{k})-t_{44}^{(b)}(\mathbf{k})\\
    0 & 0 & 0 & 0 
   \end{pmatrix} \\
   \hat{V}_{b,\downarrow}(\mathbf{k})&=\frac{1}{\sqrt{2} \Omega_{\text{BP}}}\begin{pmatrix}
    0 & 0 & 0 & 0 \\
    t_{11}^{(b)}(\mathbf{k})-t_{41}^{(b)}(\mathbf{k}), & t_{12}^{(b)}(\mathbf{k})-t_{42}^{(b)}(\mathbf{k}), &  t_{13}^{(b)}(\mathbf{k})-t_{43}^{(b)}(\mathbf{k}), &  t_{14}^{(b)}(\mathbf{k})-t_{44}^{(b)}(\mathbf{k})\\
    0 & 0 & 0 & 0 \\ 
   t_{51}^{(b)}(\mathbf{k})+t_{61}^{(b)}(\mathbf{k}), & t_{52}^{(b)}(\mathbf{k})+t_{62}^{(b)}(\mathbf{k}), &  t_{53}^{(b)}(\mathbf{k})+t_{63}^{(b)}(\mathbf{k}), &  t_{54}^{(b)}(\mathbf{k})+t_{64}^{(b)}(\mathbf{k})
   \end{pmatrix} \,.
\end{align}

\noindent Here, we have set $N_{\text{SL}} = N_{\text{BP}}$. The relation between the matrices $\hat{V}_{b,\sigma}$ and $\hat{U}_b$ becomes evident when connecting Eq.(\ref{eq:Hint_projected}) with Eq.(\ref{eq:H_MBT_BP}): $\hat{U}_b$ is a $24\times 8$ matrix with zero matrix elements everywhere other than its last four lines, which are given by the $4 \times 8$ matrix $\left(\hat{V}_{b,\uparrow}, \hat{V}_{b,\downarrow}\right)$. Note that $t_{\mu,1}^{(b)}=t_{\mu,2}^{(b)}$ because the $z$-component of the fractional positions of phosphorus $1$ and $2$ are the same, $\rho_{1,z}=\rho_{2,z}$ (see Table \ref{tab:Ppositon} and Fig.~\ref{fig:SM_theory_1}). Similarly, $t_{\mu,3}^{(b)}=t_{\mu,4}^{(b)}$. Morepver, the dominant matrix element of $\hat{V}_{b,\sigma}$ is $t_{13}^{(b)}-t_{43}^{(b)}$, as shown in Fig.~\ref{fig:SM_theory_4}.

Similar steps apply for calculating the coupling between the top BP layer and the top-most MBT SL (SL number one in, which is described by a Hamiltonian as Eq.(\ref{eq:Hint_projected}) with $b\rightarrow t$. Due to inversion symmetry, the hopping amplitudes $t_{\mu\nu}^{(t)}$ are related to $t_{\mu\nu}^{(b)}$. More specifically, $t_{\mu\nu}^{(t)}=t_{\mu'\nu'}^{(b)}$, where $\mu'=4,1,6,5$ if $\mu=1,2,5,6$, respectively and $\nu'=3,4,1,2$ if $\nu=1,2,3,4$. We thus obtain
 
\begin{align}
    \hat{V}_{t,\uparrow}(\mathbf{k})&=\frac{1}{\sqrt{2} \Omega_{\text{BP}}}\begin{pmatrix}
   t_{53}^{(b)}(\mathbf{k})+t_{63}^{(b)}(\mathbf{k}) & t_{54}^{(b)}(\mathbf{k})+t_{64}^{(b)}(\mathbf{k}) &  t_{51}^{(b)}(\mathbf{k})+t_{61}^{(b)}(\mathbf{k}) &  t_{52}^{(b)}(\mathbf{k})+t_{62}^{(b)}(\mathbf{k})\\
   0 & 0 & 0 & 0 \\
    -t_{13}^{(b)}(\mathbf{k})+t_{43}^{(b)}(\mathbf{k}) & -t_{14}^{(b)}(\mathbf{k})+t_{44}^{(b)}(\mathbf{k}) &  -t_{11}^{(b)}(\mathbf{k})+t_{41}^{(b)}(\mathbf{k}) &  -t_{12}^{(b)}(\mathbf{k})+t_{42}^{(b)}(\mathbf{k})\\
    0 & 0 & 0 & 0 
   \end{pmatrix} \\
\hat{V}_{t,\downarrow}(\mathbf{k})&=\frac{1}{\sqrt{2} \Omega_{\text{BP}}}\begin{pmatrix}
    0 & 0 & 0 & 0 \\
    -t_{13}^{(b)}(\mathbf{k})+t_{43}^{(b)}(\mathbf{k}) & -t_{14}^{(b)}(\mathbf{k})+t_{44}^{(b)}(\mathbf{k}) &  -t_{11}^{(b)}(\mathbf{k})+t_{41}^{(b)}(\mathbf{k}) &  -t_{12}^{(b)}(\mathbf{k})+t_{42}^{(b)}(\mathbf{k})\\
    0 & 0 & 0 & 0 \\ 
  t_{53}^{(b)}(\mathbf{k})+t_{63}^{(b)}(\mathbf{k}) & t_{54}^{(b)}(\mathbf{k})+t_{64}^{(b)}(\mathbf{k}) &  t_{51}^{(b)}(\mathbf{k})+t_{61}^{(b)}(\mathbf{k}) &  t_{52}^{(b)}(\mathbf{k})+t_{62}^{(b)}(\mathbf{k})
   \end{pmatrix} \text{ .}
\end{align}

\noindent Here, we have set $N_{\text{SL}} = N_{\text{BP}}$. The matrices $\hat{V}_{t,\uparrow}$ and $\hat{V}_{t,\downarrow}$ define the blocks $\hat{U}_t$ in Eq.~(\ref{eq:H_MBT_BP}). Like $\hat{U}_b$, the matrix $\hat{U}_t$ is also a $24\times 8$ matrix, but here the only non-zero elements are in the first four lines of this matrix, which are formed by the $4 \times 8$ matrix $\left(\hat{V}_{t,\uparrow},\hat{V}_{t,\downarrow}\right)$. 

\subsubsection{\underline{IV.8.2. MnBi$_2$Te$_4$-BP hopping integrals using Slater-Koster approach}}\label{sssec:hopping_int}

\hspace{4mm} In this section we determine the hopping amplitude $t_{\mu \nu}^{(b)}(\mathbf{q})$ that enters the final expression for the coupling Hamiltonian in Eq.~\eqref{eq:Hint_b}. We obtain the real-space hopping amplitudes between the $p_z$ orbitals of BP and its nearest-neighbor MBT SL using the Slater-Koster approach, and then Fourier transform to obtain the desired expression for $t_b(\mathbf{q})$ in Eq.(\ref{eq:hopping_Fourier}).

The Slater-Koster approach~\cite{slater1954simplified} consist of parametrizing the real-space hopping amplitude between two $p_z$ orbitals separated by a distance vector $\brc=\rc\left(\sin\theta\cos\phi,\sin\theta\sin\phi,\cos\theta\right)$ as 

\begin{equation}
    t_b(\brc)=V_{\sigma}(\rc)\cos^2\theta+V_{\pi}(\rc)\sin^2\theta \,.
    \label{eq:tSK}
\end{equation}

\noindent Here, the subscripts $\sigma$ and $\pi$ denote the two orthogonal projections of the $p_z$ orbitals along the direction of $\brc$ ($\sigma$ and $\pi$ bonding). The radial functions $V_{\sigma}(\rc)=A_{\sigma}e^{-\rc/a_{\sigma}}$ and $V_{\pi}(\rc)=A_{\pi}e^{-\rc/a_{\pi}}$  decay exponentially with the distance $\rc$ between the orbitals. The parameters $A_{\sigma}$, $A_{\pi}$, $a_{\sigma}$, and $a_{\pi}$ are adjustable parameters in the Slater-Koster approach.  While $a_{\sigma}$ and $a_{\pi}$ play the role of a effective Bohr radii that set the characteristic radial decay range of the MBT-BP coupling: larger $a_{\sigma}$ and $a_{\pi}$ results in a slower $t(\rc)$ decays and coupling functions $\hat{U}_{b}(\mathbf{k})$ and $\hat{U}_{t}(\mathbf{k})$ that are more localized in momentum space. In this work, we used $a_{\sigma}=a_{\pi}=2$~\AA.

The parameters $A_{\sigma}$ and $A_{\pi}$ have dimension of energy and determine the hopping strength, which are physically given by the overlaps of the orbital wavefunctions. Since the $\sigma$ projection of two orbitals is generally larger than the $\pi$ projection, we choose $A_{\sigma} \gtrsim A_{\pi}$. In this work, we used $A_\sigma = 0.34$~eV and $A_\pi = 0.0$. 

In Eq.~(\ref{eq:hopping_Fourier}), we take a partial Fourier transform of Eq.(\ref{eq:tSK}) with respect to the in-plane components, leaving the $z$ coordinate unchanged,

\begin{equation}
    t(\mathbf{q},z)=\int d^2 y \left[A_{\sigma}\frac{z^2}{y^2+z^2}e^{-a_{\sigma}^{-1}\sqrt{y^2+z^2}}+A_{\pi}\frac{y^2}{y^2+z^2}e^{-a_{\pi}^{-1}\sqrt{y^2+z^2}}\right]e^{-i\mathbf{q}\cdot \mathbf{y}} \text{ .} 
\end{equation}

\noindent We numerically calculate this integral and fit the result to a sum of two decaying exponentials of the form 

\begin{equation}
t(\mathbf{q},z)=z^2\left[A_{\sigma}\gamma_1e^{-\sqrt{\gamma_2+\gamma_3q^2z^2}}+A_{\pi}\gamma_4\left(1-\gamma_5q^2z^2\right)e^{-\sqrt{\gamma_6+\gamma_7q^2z^2}}\right] \text{ .}
    \label{eq:fit}
\end{equation}

\noindent Here, $\gamma_i$ ($i=1,\ldots,6$) are real fitting constants that implicitly depend on the Bohr radii $a_{\sigma}$ (for $i=1,2,3$) and $a_{\pi}$ (for $i=4,5,6$). Substituting $z$ by the $z$ component distance of the orbitals $s_{\mu\nu} = \tau_\mu^\perp - \rho_\mu^\perp$ in Eq.(\ref{eq:fit}) finally yields $t_{\mu\nu}^{(b)}(\mathbf{q})$. 

\subsection{\underline{IV.9. Parameter set used in the main text}}\label{ssec:parameter_set_main}

\hspace{4mm} The model for the heterostructure contains a total of 33 parameters. In Table~\ref{tab:kp_parameter_values} we provide the numerical parameter values that we have used in the main text to reproduce the experimental observations.

\normalsize
\clearpage
\begin{figure*}[h]
\centering
\includegraphics[width=16cm]{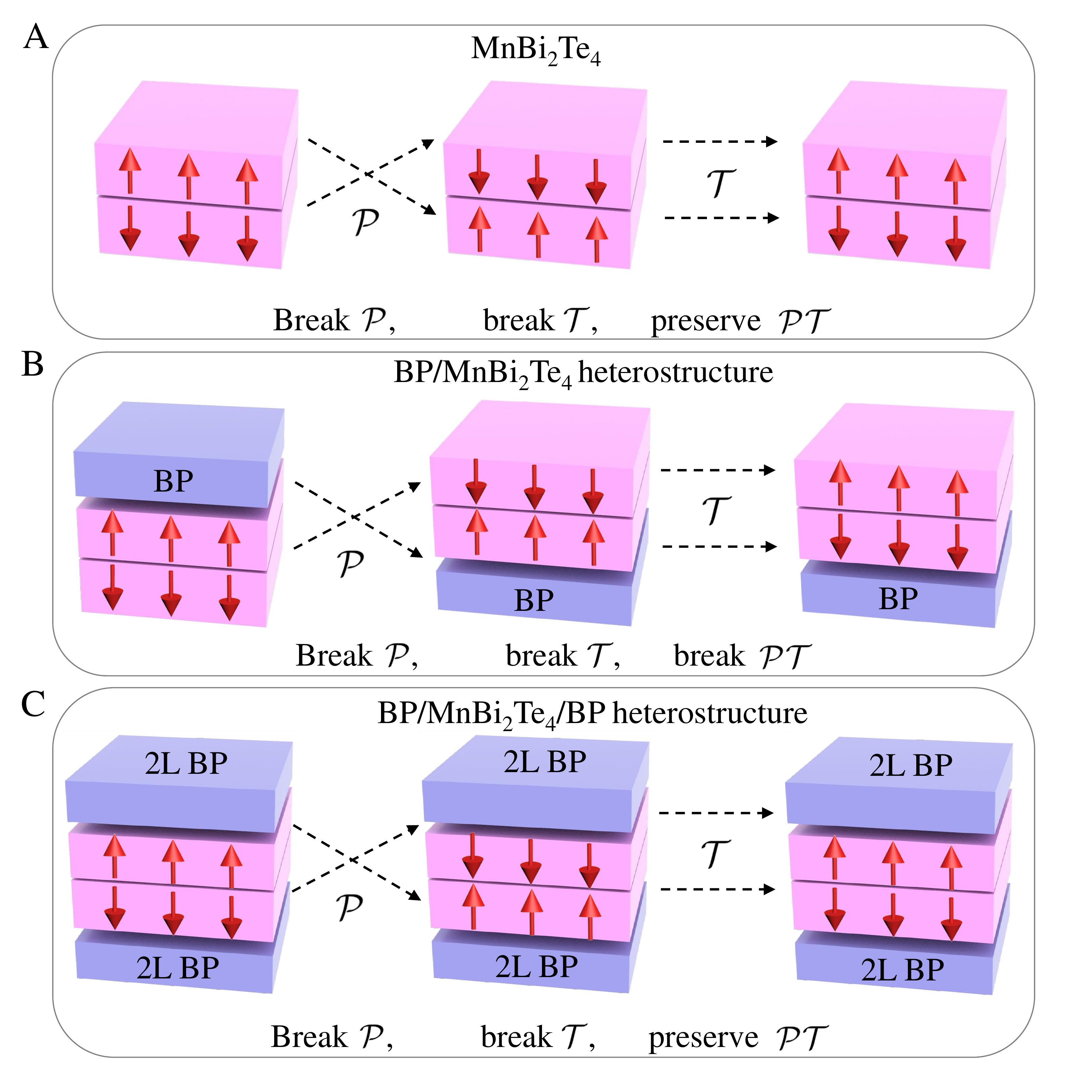}
\caption{\textbf{The symmetry properties of even-layered MnBi$_2$Te$_4$, BP/MnBi$_2$Te$_4$ and BP/MnBi$_2$Te$_4$/BP heterostructures at low temperature.} (A) The symmetry of even-layered MnBi$_2$Te$_4$. (B) The symmetry of BP/MnBi$_2$Te$_4$ heterostructure. (C) The symmetry of BP/MnBi$_2$Te$_4$/BP heterostructure. The mirror planes of top BP and bottom BP are aligned.}
\label{MBT_Symmetry}
\end{figure*}

\normalsize
\clearpage
\begin{figure*}[h]
\centering
\includegraphics[width=15cm]{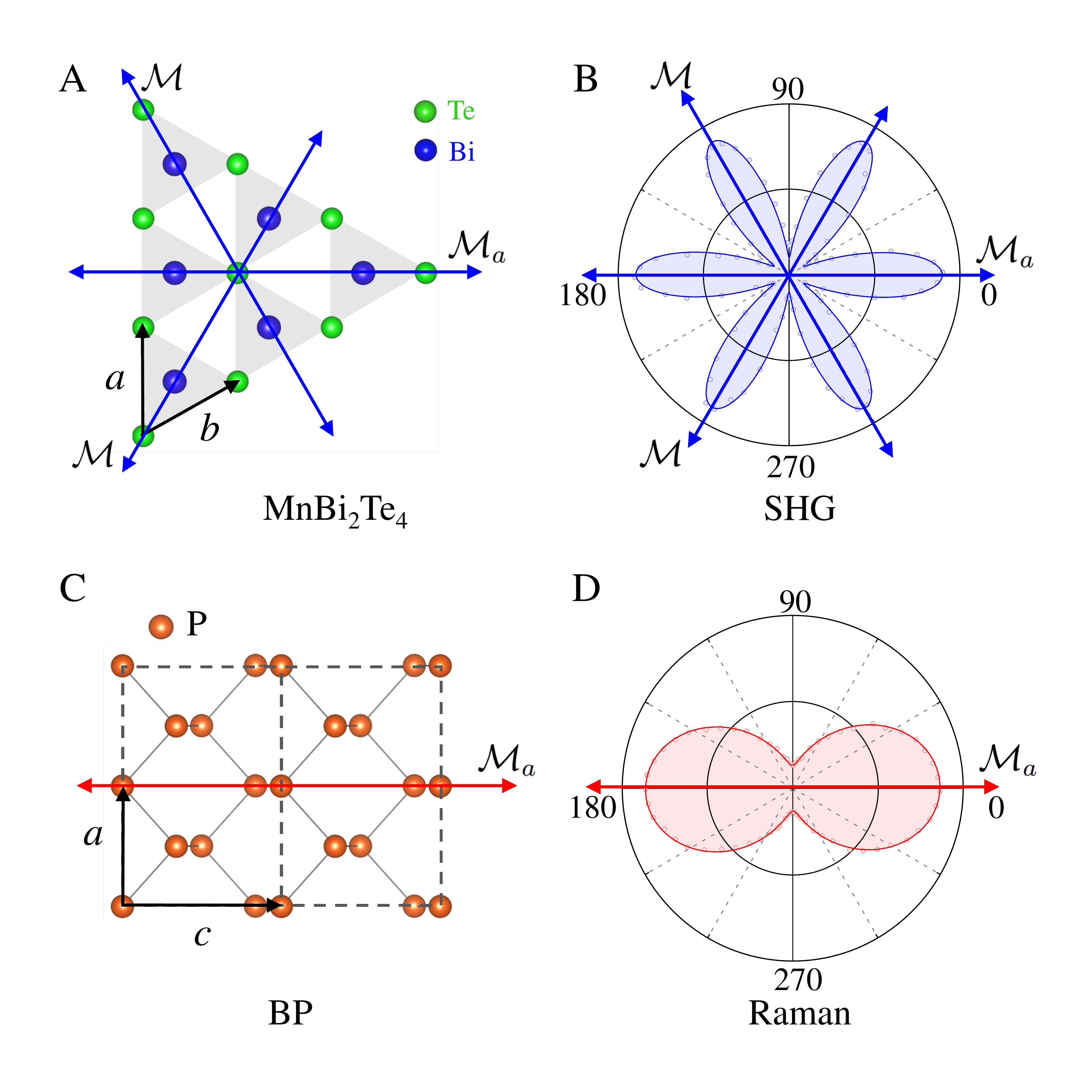}
\caption{\textbf{The determination of crystalline directions of the MnBi$_2$Te$_4$ and BP.} (A) Top view of the MnBi$_2$Te$_4$ lattice. It shows three-fold rotational symmetry $C_{3z}$. The three mirror planes are noted by the blue double arrow lines. (B) The SHG signals of the MnBi$_2$Te$_4$ measured at room temperature. The largest signals corresponding to the mirror planes of the MnBi$_2$Te$_4$. (C) Top view of the BP lattice. It has a mirror plane $\mathcal{M}_a$ which is noted by the red double arrow line. (D) The angular-resolved Raman signals of the BP measured with a 532-nm laser. The largest signals corresponding to the mirror plane (armchair direction) of the BP.}
\label{SHG_And_Raman}
\end{figure*}

\normalsize
\clearpage
\begin{figure*}[h]
\centering
\includegraphics[width=18cm]{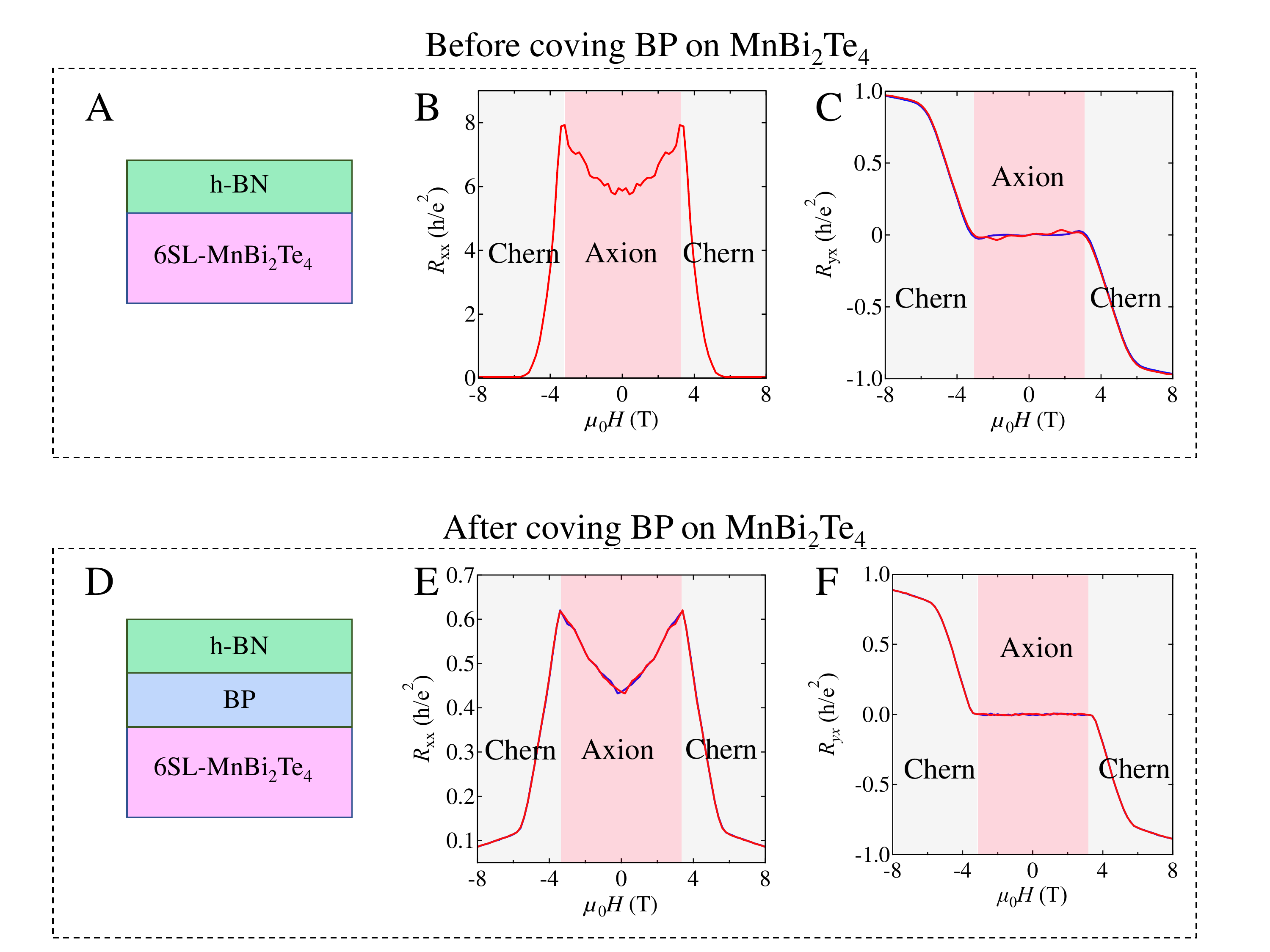}
\caption{\textbf{The magneto-transport data of MnBi$_2$Te$_4$ with and without BP.} (A to C) The magneto-transport data of the MnBi$_2$Te$_4$ at charge neutral point. (D to F) The magneto-transport data of the BP/MnBi$_2$Te$_4$ heterostructure at charge neutral point. The BP/MnBi$_2$Te$_4$ heterostructure exhibits the same topological phases as the MnBi$_2$Te$_4$, i.e. Chern insulator in FM phase and Axion insulator in AFM phase.}
\label{Magnetotransport_MBT_vs_BP_MBT}
\end{figure*}

\normalsize
\clearpage
\begin{figure*}[h]
\centering
\includegraphics[width=14cm]{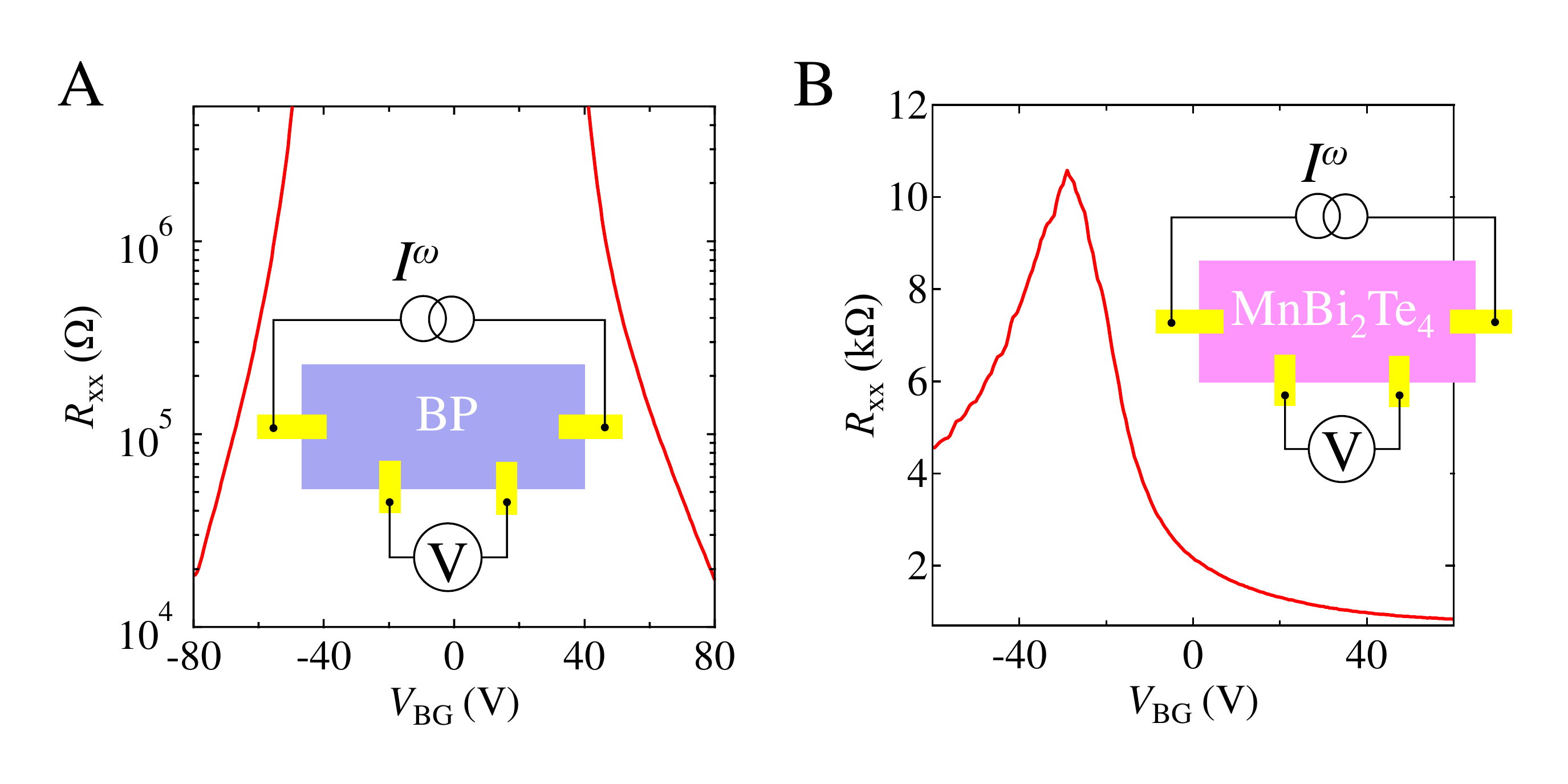}
\caption{\textbf{The typical four-probe resistance of BP and MnBi$_2$Te$_4$ as a function of back gate voltages at 1.8 K.} The thickness of BP is around 10 nm and MnBi$_2$Te$_4$ is 6SL. The contact geometry of BP and MnBi$_2$Te$_4$ is identical. The four-probe resistance of BP is two orders of magnitude larger than that of MnBi$_2$Te$_4$.}
\label{Gate_BP_MBT}
\end{figure*}

\normalsize
\clearpage
\begin{figure*}[h]
\centering
\includegraphics[width=16cm]{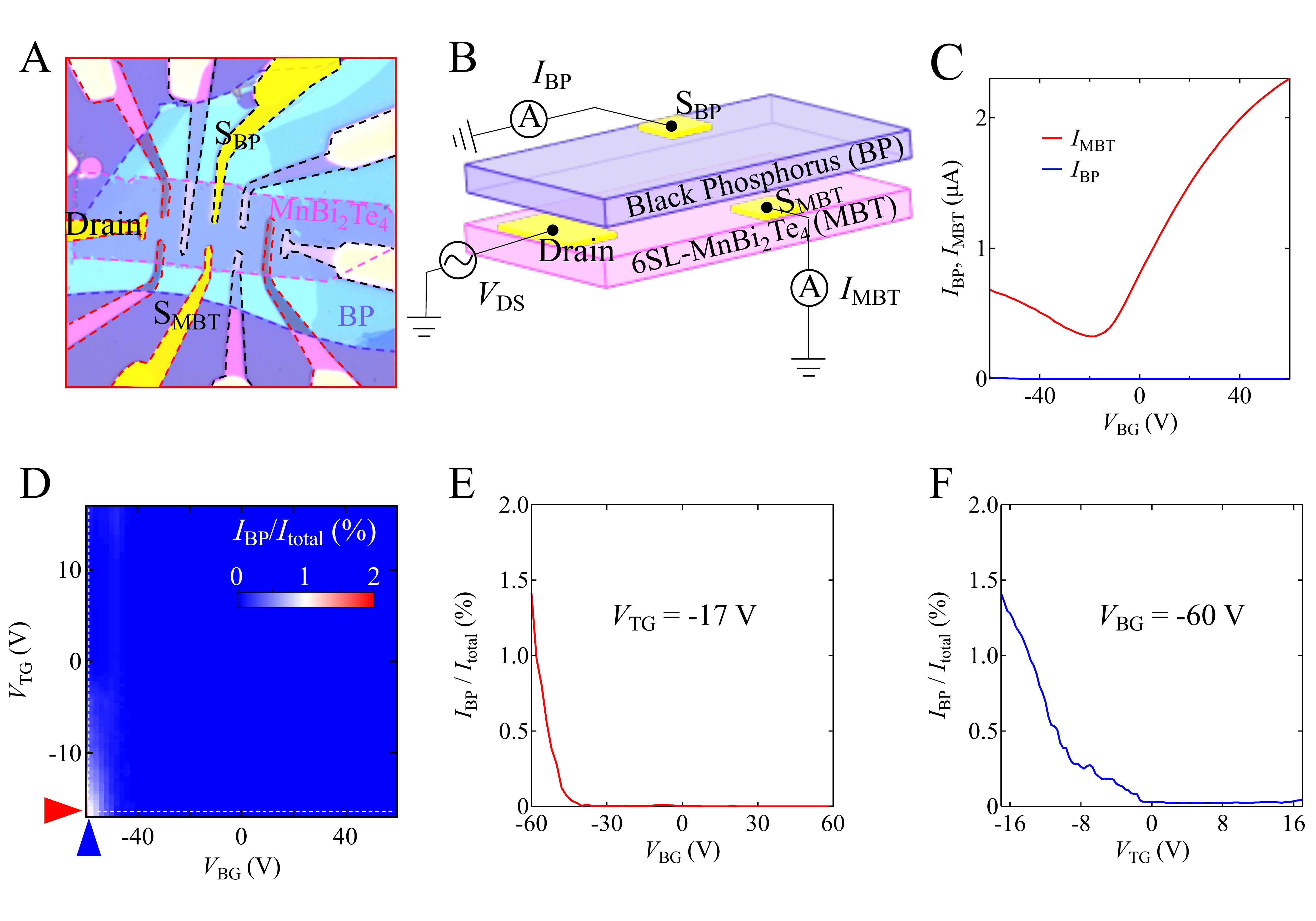}
\caption{\textbf{Identifying the current flowing in the BP and MnBi$_2$Te$_4$ layers.} (A and B) Optical microscopic image and schematic of the special BP/MnBi$_2$Te$_4$ device. The contacts outlined by red dashed lines are deposited on the MnBi$_2$Te$_4$ layer (which also contact with BP layer) while the contacts outlined by black dashed lines are deposited on the BP layer. The measurement electrodes are highlighted in yellow. The measurement circuit is shown in panel (B). A constant AC current was applied on the drain contact and was collected from two source electrodes S$_\textrm{{MBT}}$ and S$_\textrm{{BP}}$, respectively.  (C) $I_{\textrm{BP}}$ and $I_{\textrm{MBT}}$ as a function of $V_\textrm{BG}$. (D) The percentage of current flowing in BP layer ($I_{\textrm{BP}}/I_{\textrm{total}}$) as a  function of $V_\textrm{BG}$ and $V_\textrm{TG}$. (E and F) The line-cut data in panel (D). The current flowing in BP layer is always less than 2$\%$ of the total current.} 
\label{Two_pair_contacts}
\end{figure*}

\normalsize
\clearpage
\begin{figure*}[h]
\centering
\includegraphics[width=15cm]{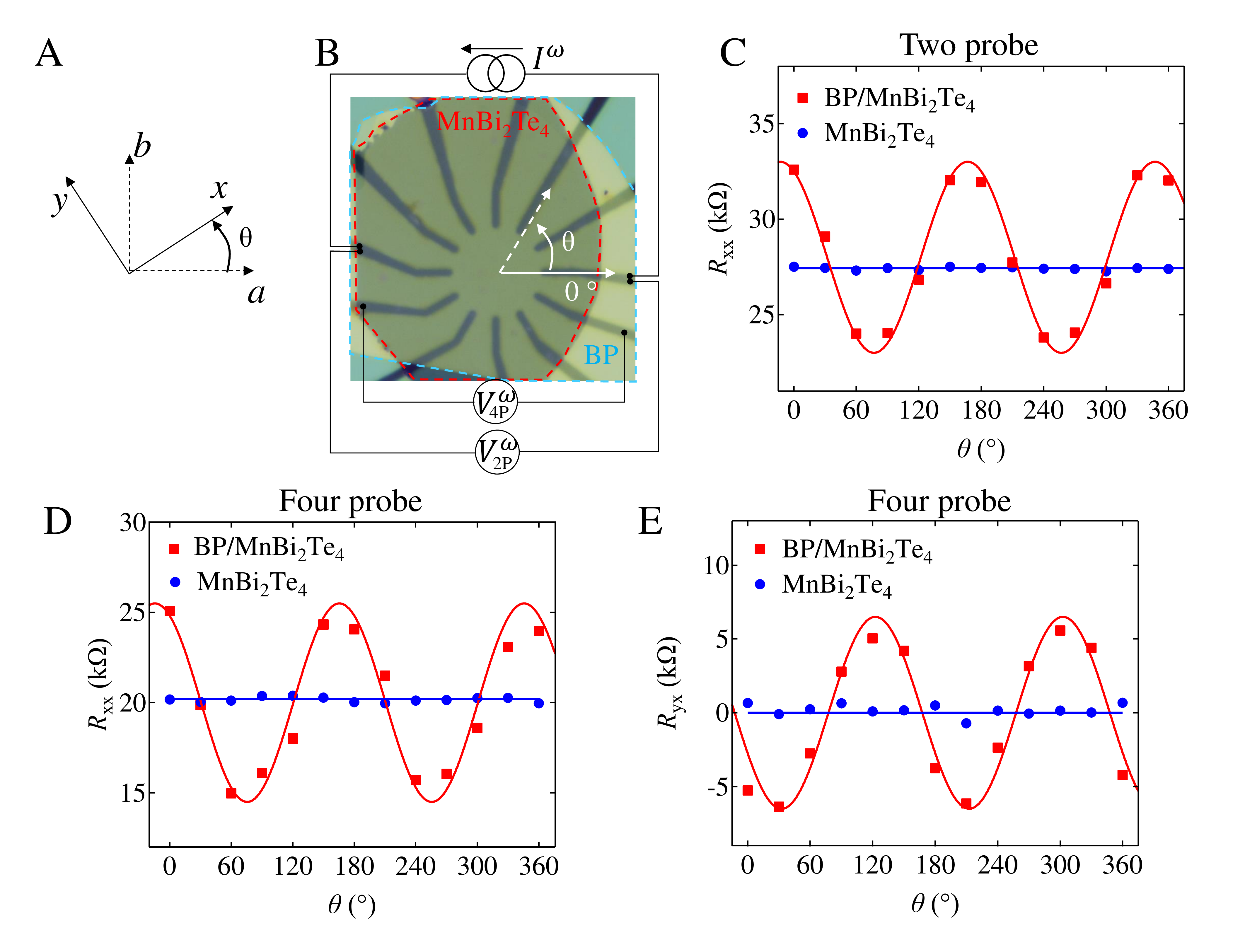}
\caption{\textbf{The demonstration of $\mathcal{C}_{3z}$ breaking by electrical transport measurements.} (A) The crystalgraphic bases are noted by $a$ and $b$. The Cartesian bases are noted by $x$ and $y$. The angle difference between crystalgraphic and Cartesian basis is noted by $\theta$. (B) A BP/MnBi$_2$Te$_4$ heterostructure with the circular-disc contacts. MnBi$_2$Te$_4$ was roughly scratched into a circular shape with a tip in glovebox. Two-probe and four-probe measurements configuration are shown in the panel. (C and D) The longitudinal resistance ($R_{xx}$) of two-probe and four-probe measurements. (E) The transverse resistance ($R_{yx}$) of four-probe measurements. The blue and red curves correspond to the resistance of the same MnBi$_2$Te$_4$ sample before and after interfacing with BP.}
\label{Angular_Resolved_Transport}
\end{figure*}

\normalsize
\clearpage
\begin{figure*}[h]
\centering
\includegraphics[width=15cm]{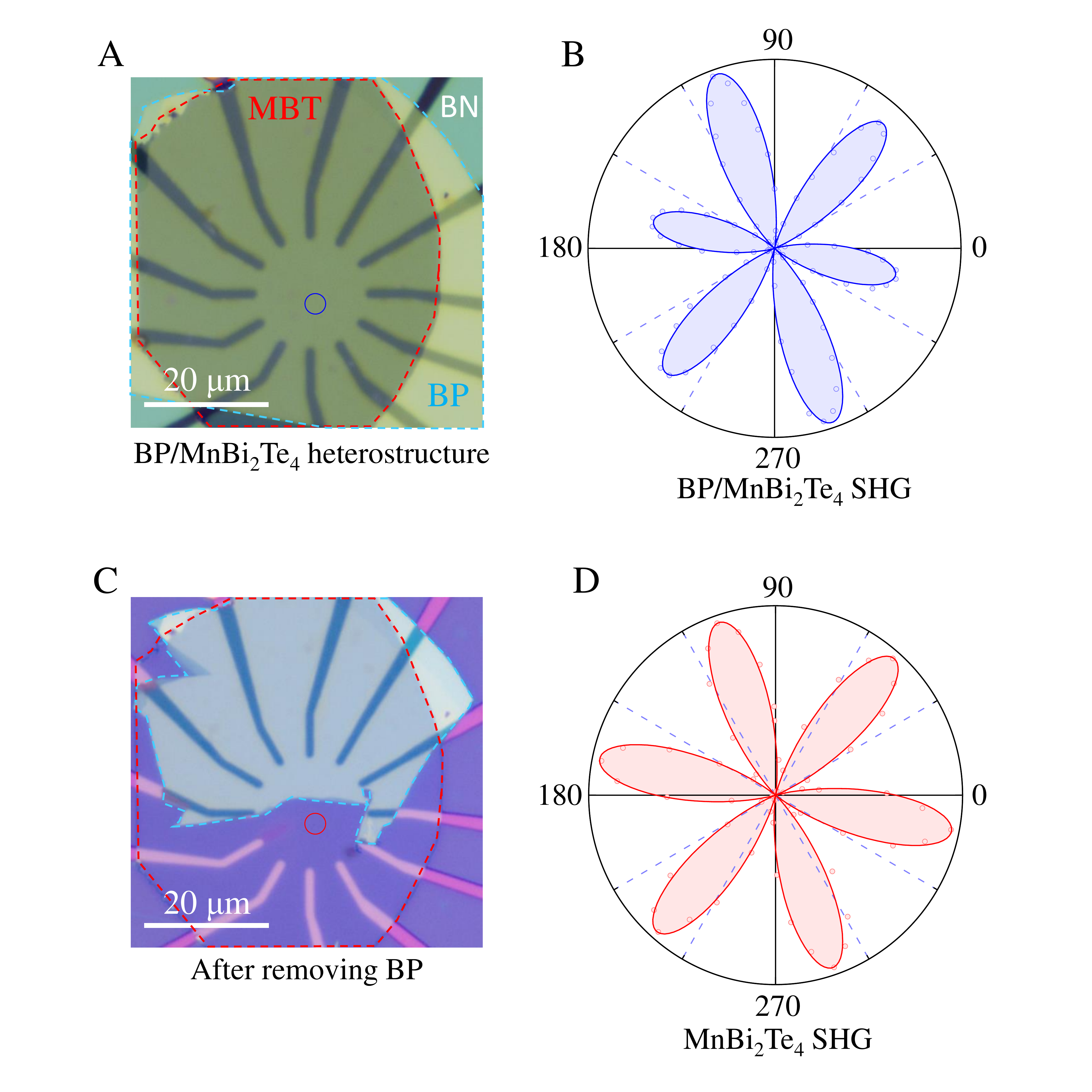}
\caption{\textbf{The demonstration of $\mathcal{C}_{3z}$ breaking by optical SHG measurements at room temperature.} (A) One of the BP/MnBi$_2$Te$_4$ heterostructures for optical SHG measurements. The MnBi$_2$Te$_4$ and BP are outline by red and cyan dashed line. The measured area is marked by a blue circle. (B) The optical SHG signals of BP/MnBi$_2$Te$_4$ heterostructure. (C) The same device as in (A) but removed the BN and part of BP with Scotch tape. The laser shone on the same area as in (A), noted by the red circle. (D) The optical SHG signals of the MnBi$_2$Te$_4$.}
\label{Angular_Resolved_SHG}
\end{figure*}

\normalsize
\clearpage
\begin{figure*}[h]
\centering
\includegraphics[width=18cm]{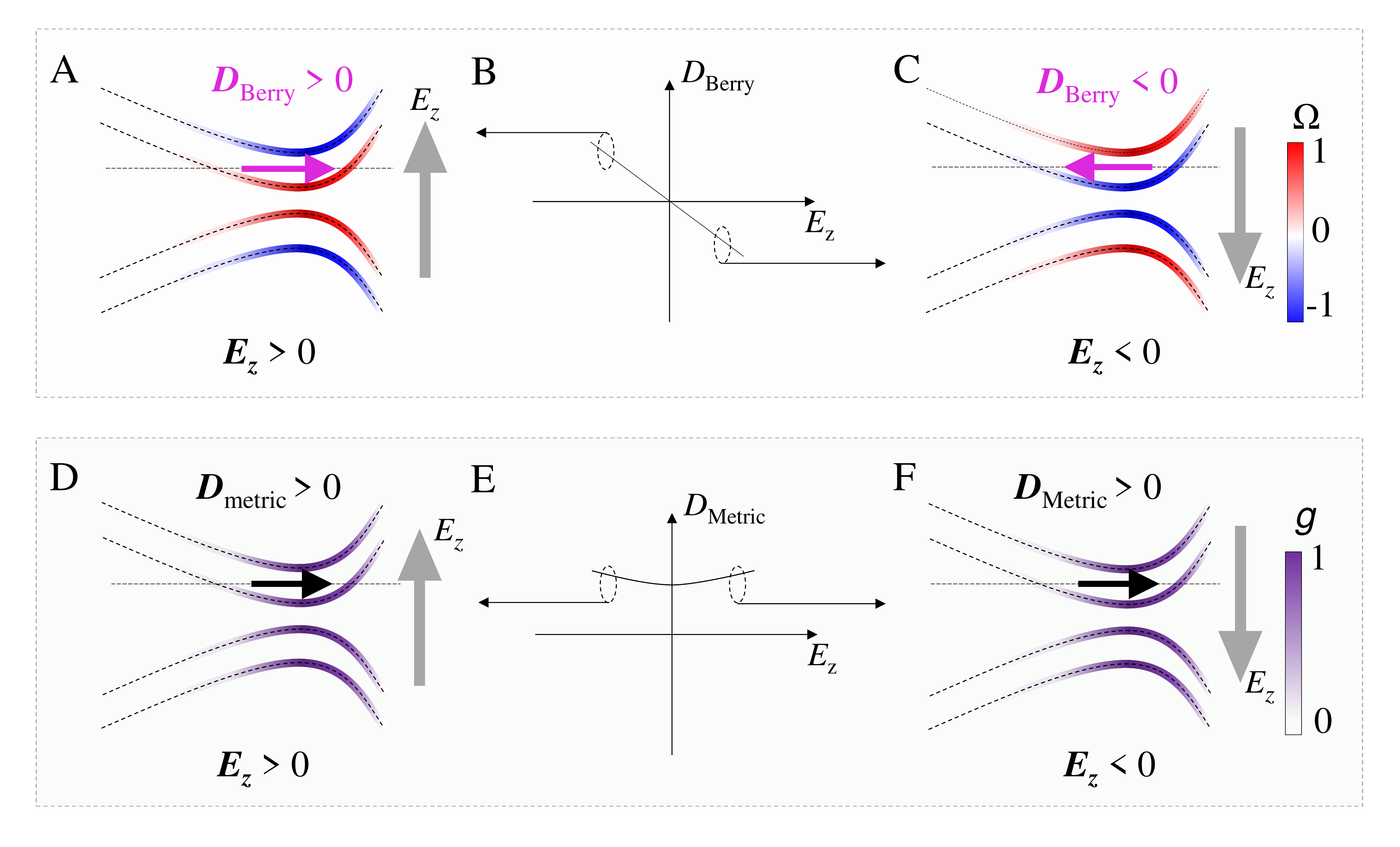}
\caption{\textbf{The electric field dependence of Berry curvature dipole and quantum metric dipole.} (A to C) The Berry curvature dipoles under the out of plane electrical field $E_z > 0$ and $E_z < 0$. The bands for top and bottom layers of MnBi$_2$Te$_4$ have the opposite Berry curvature. The electrical field $E_z > 0$ and $E_z < 0$ break the band degeneration in opposite ways \cite{gao2021layer}. Therefore, Berry curvature dipoles have different signs for opposite electrical fields. (D to F) The quantum metric dipoles under the out of plane electrical field $E_z > 0$ and $E_z < 0$. The bands for top and bottom layers of MnBi$_2$Te$_4$ have the same quantum metric. Although the opposite electrical field shift the bands of top and bottom layers of MnBi$_2$Te$_4$ in opposite ways, the quantum metric distribution is identical. Therefore, the quantum metric dipoles are symmetric as a function of $E_z$. }
\label{E_Berry_Metric}
\end{figure*}

\normalsize
\clearpage
\begin{figure*}[h]
\centering
\includegraphics[width=15cm]{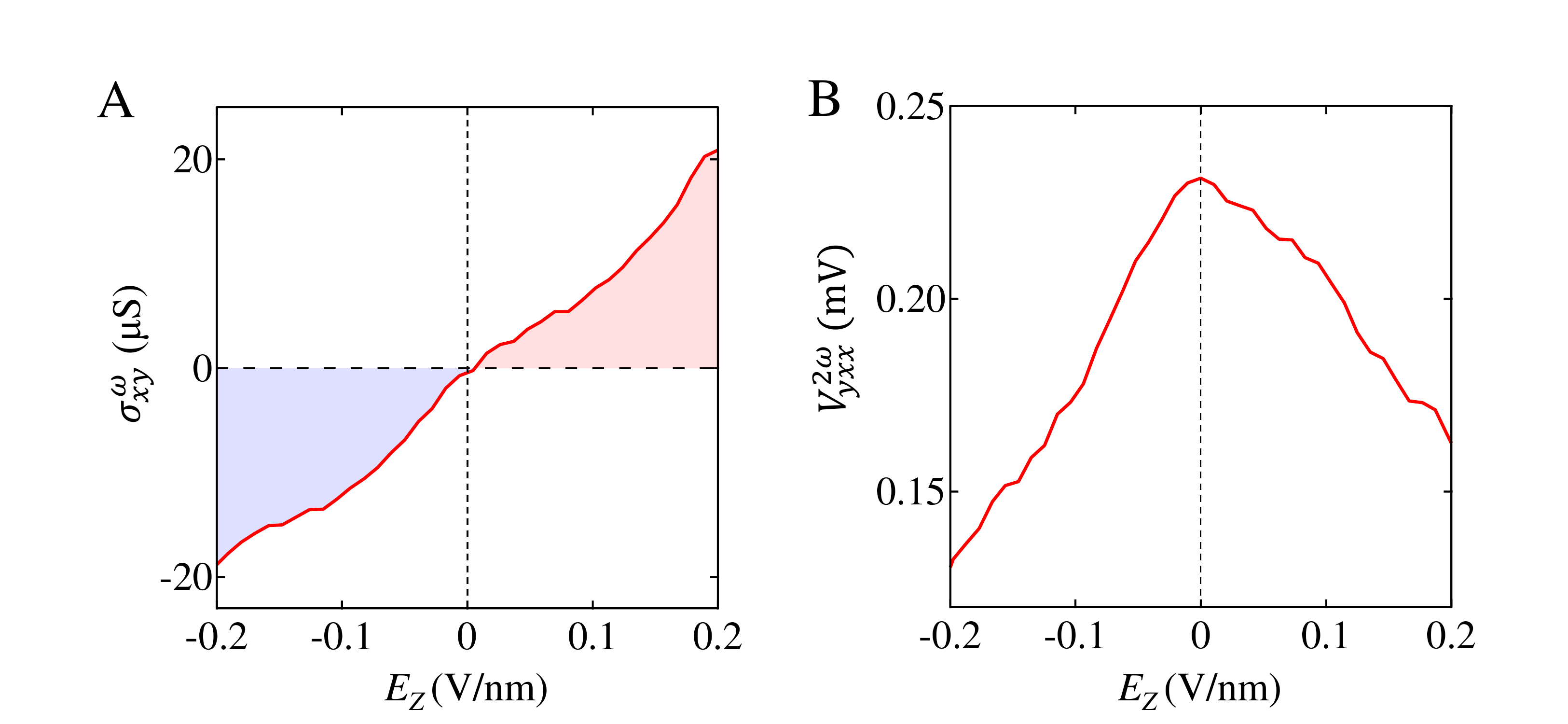}
\caption{\textbf{The electric filed dependence of linear and nonlinear Hall signals.} (A) The linear Hall conductivity $\sigma_{xy}^{\omega}$ as a function of vertical electric field $E_z$. (B) The nonlinear Hall conductivity $\sigma _{yxx}^{2\omega}$ as a function of $E_z$. $\sigma_{xy}^{\omega}$ is antisymmetric as a function of $E_z$, while $\sigma _{yxx}^{2\omega}$ is symmetric as a function of $E_z$. At $E_z = 0$, the $\mathcal{PT}$ confirms the $\sigma_{xy}^{\omega}=0$ while $\sigma _{yxx}^{2\omega} \neq 0$. 
}
\label{Additional_E}
\end{figure*}

\normalsize
\clearpage
\begin{figure*}[h]
\centering
\includegraphics[width=18cm]{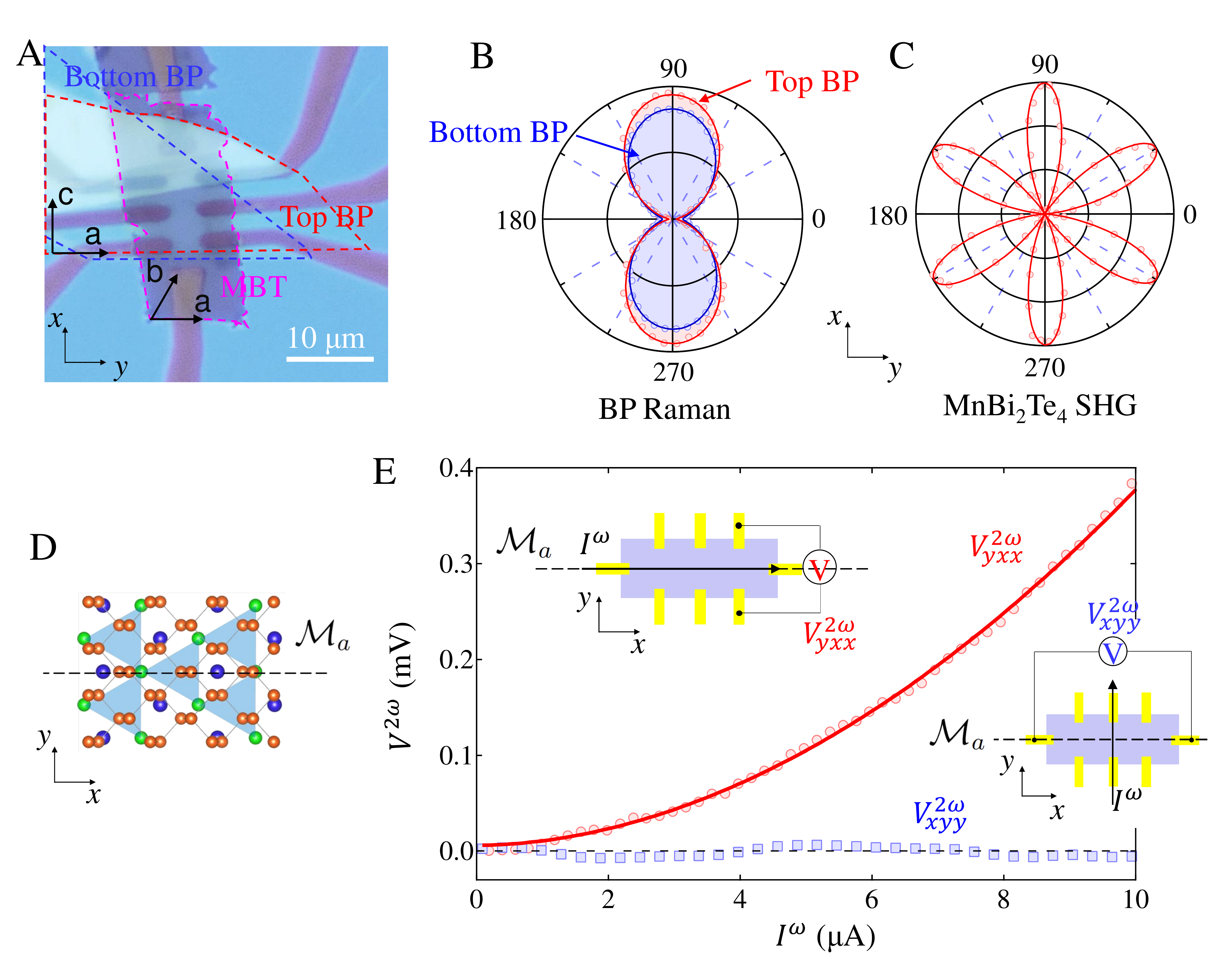}
\caption{\textbf{A mirror plane aligned 2L BP/6SL MnBi$_2$Te$_4$/2L BP heterostructure.} (A) The microscopic image of the BP/MnBi$_2$Te$_4$/BP heterostructure. The mirror planes of top BP, MnBi$_2$Te$_4$ and bottom BP are all aligned along $x$ direction in Cartesian axis. MnBi$_2$Te$_4$, top and bottom BP are outlined by pink, red and blue dashed lines, respectively. The top and bottom BP are both bilayer. MnBi$_2$Te$_4$ is 6SL. (B) The polarized Raman signals of top (red curve) and bottom (blue curve) BP at room temperature. (C) Room-temperature SHG signals of the 6SL MnBi$_2$Te$_4$. (D) The schematic of aligned BP/MnBi$_2$Te$_4$/BP lattice structure. The mirror plane is along $a$ ($x$). (E) The nonlinear Hall signals of the heterostructure for driving current along ($V_{yxx}^{2 \omega}$) and vertical to ($V_{xyy}^{2 \omega}$) the mirror direction.}
\label{Mirror_Exclude_Berry}
\end{figure*}

\normalsize
\clearpage
\begin{figure*}[h]
\centering
\includegraphics[width=18cm]{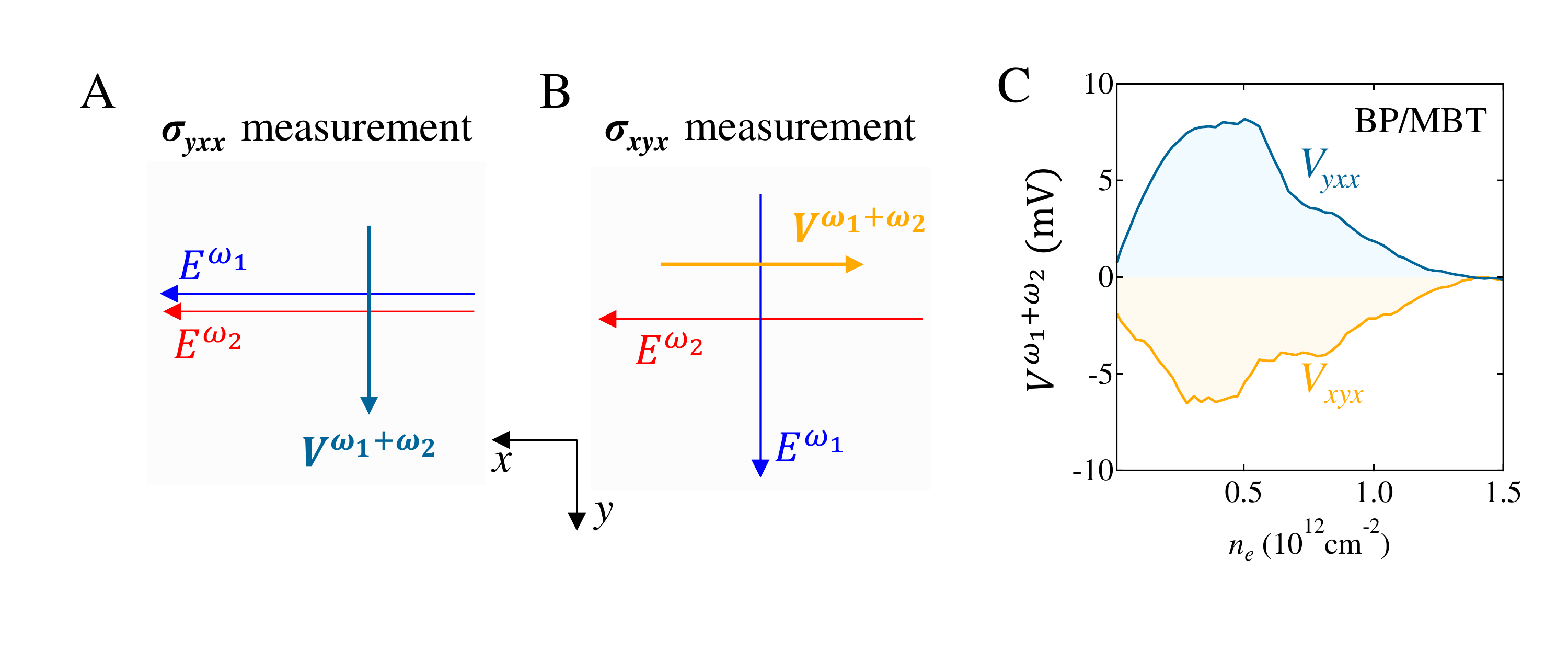}
\caption{\textbf{The sum frequency generation measurements.} (A and B) The schematic of $\sigma _{yxx}$ and $\sigma _{xyx}$ measurements. $\omega_1 \gg \omega_2$. (C) The experiment results of the $\sigma _{yxx}$ and $\sigma _{xyx}$ as a function of carrier density $n_e$ on a BP/MnBi$_2$Te$_4$ device. It is notable that $\sigma _{yxx}$ and $\sigma _{xyx}$ are antisymmetric, i.e., $\sigma_{yxx} = - \sigma_{xyx}$.
}
\label{Drude_Exclude}
\end{figure*}

\normalsize
\clearpage
\begin{figure*}[h]
\centering
\includegraphics[width=18cm]{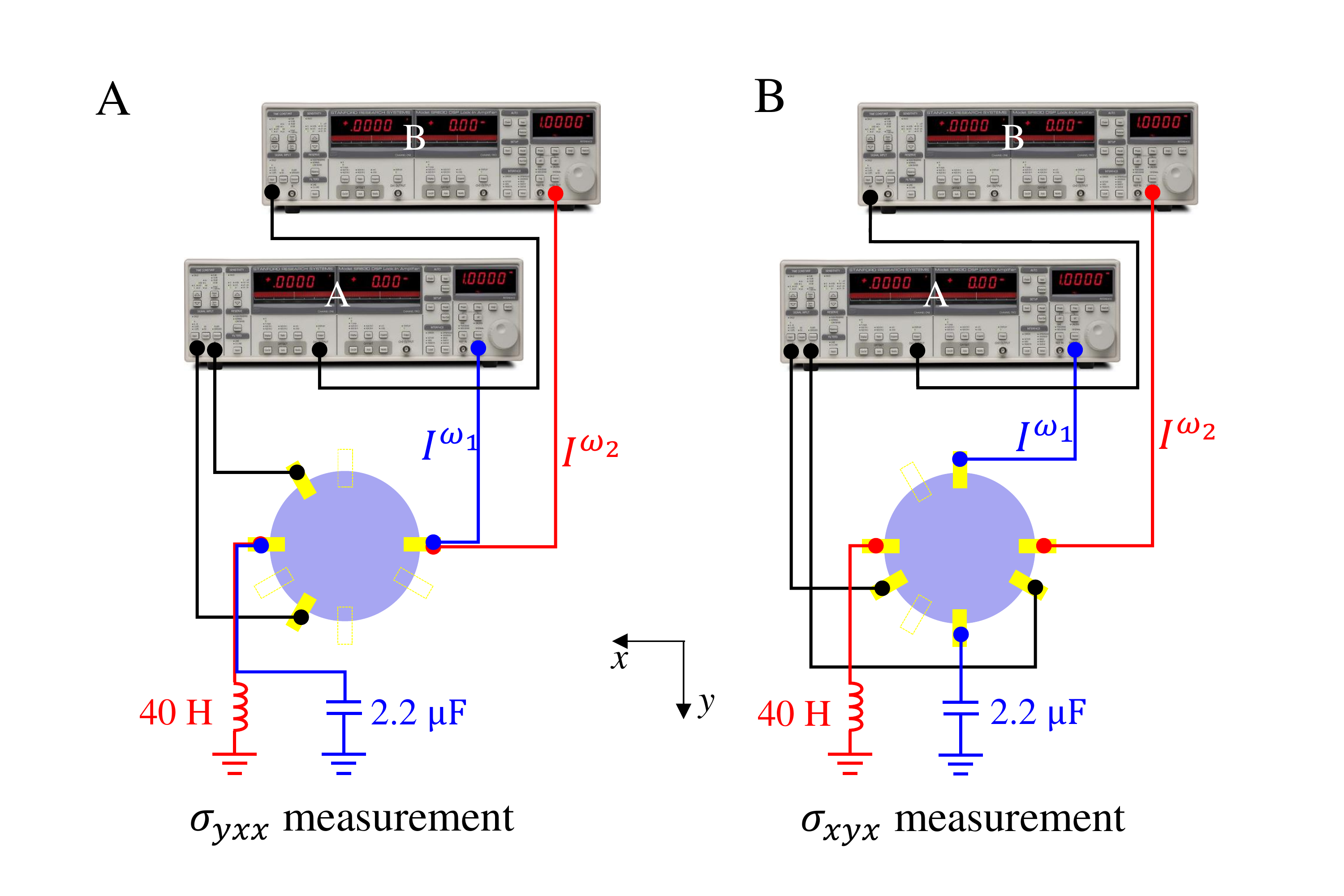}
\caption{\textbf{The measurement setup of sum frequency generation.} (A and B) The circuit of $\sigma _{yxx}$ and $\sigma _{xyx}$ measurements. The time constant $t$ for lock-in A and B are 30 ms and 3 s, respectively. $\omega _1 = 547$ Hz, $\omega _2 = 1.37$ Hz. A 2.2 $\mu$F capacitor and 40 H inductor were chosen to confirm the current flow in target direction.
}
\label{Drude_Method}
\end{figure*}

\newpage
\begin{figure*}[h]
\centering
\includegraphics[width=18cm]{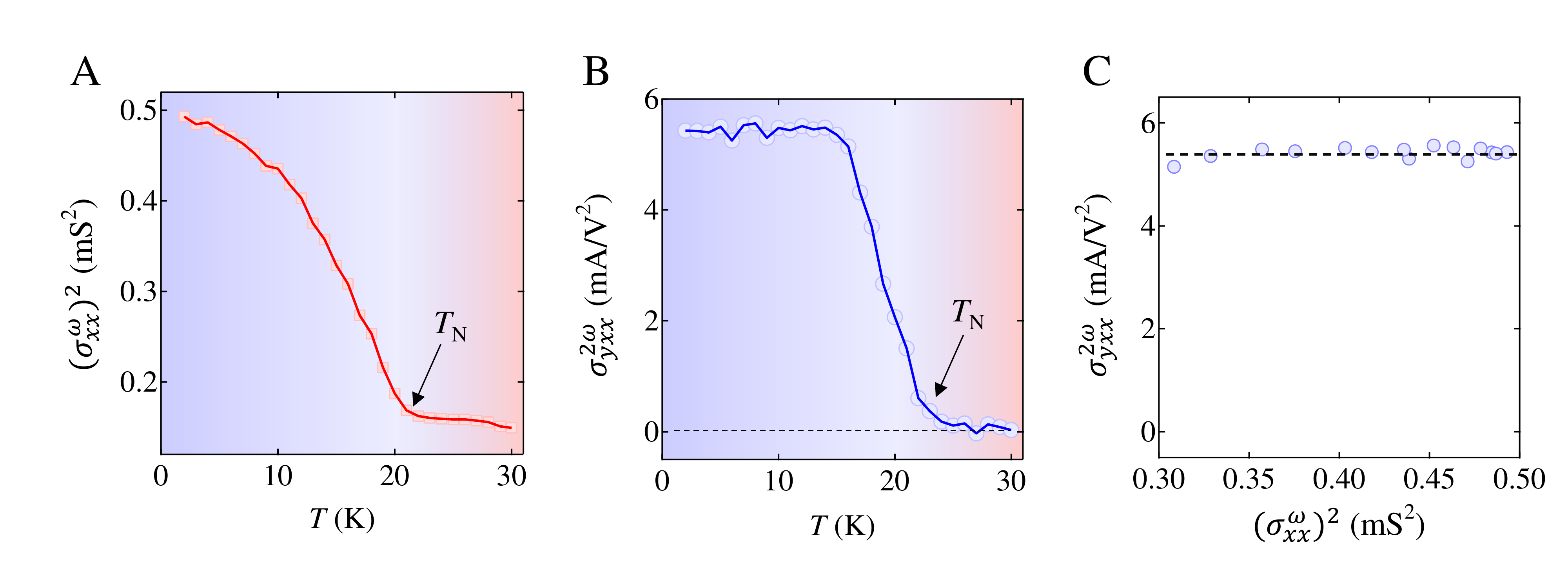}
\caption{\textbf{The temperature dependence of nonlinear Hall signals.} (A and B) The square of linear conductivity ($\sigma_{xx}^{\omega}$)$^2$ and nonlinear Hall conductivity $\sigma _{yxx}^{2\omega}$ as a function of temperature. The scaling properties of $\sigma _{yxx}^{2\omega}$ with ($\sigma_{xx}^{\omega}$)$^2$ for temperature lower the 15 K. 
}
\label{Scaling_Temperature}
\end{figure*}

\newpage
\begin{figure*}[h]
\centering
\includegraphics[width=14cm]{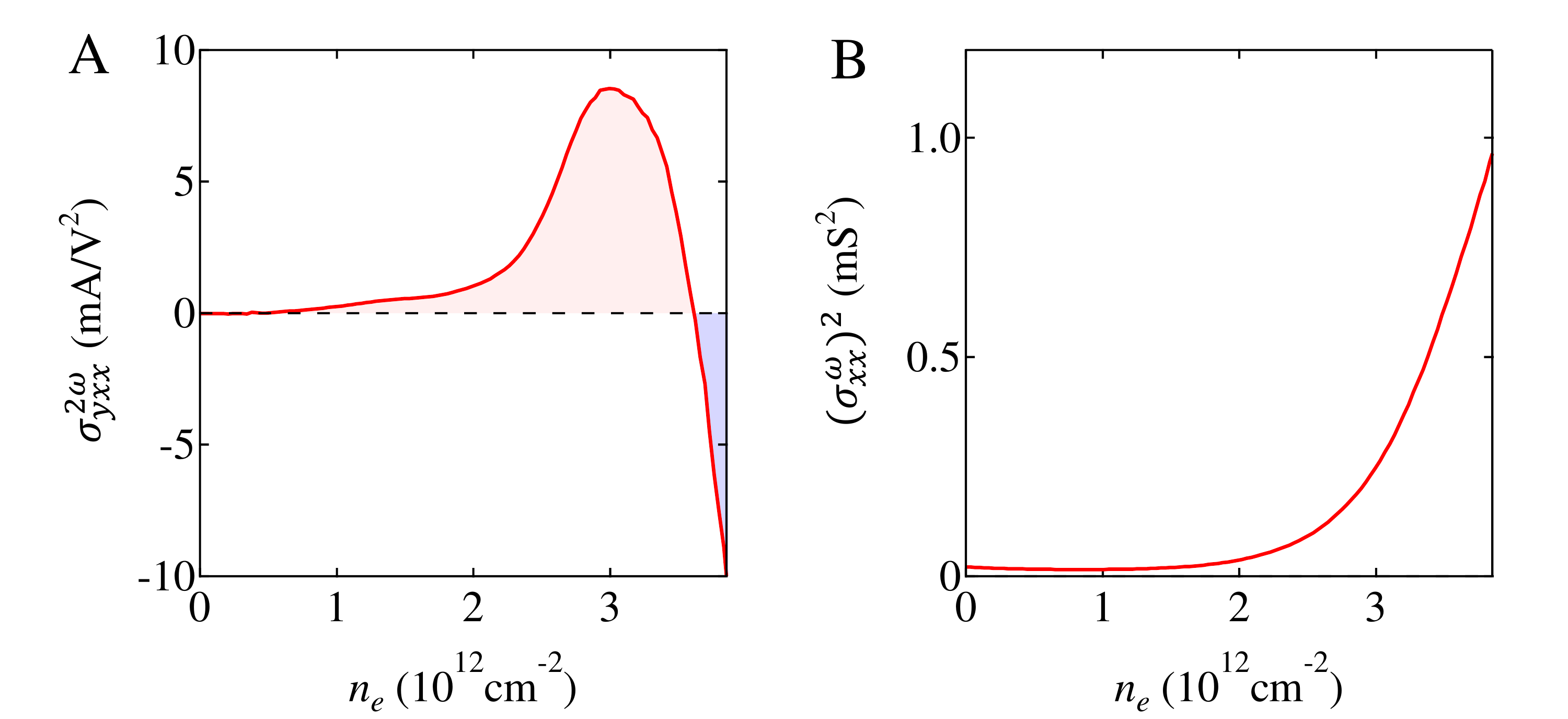}
\caption{\textbf{The carrier density dependence of the nonlinear Hall signals.} (A and B) The experimental results of $\sigma _{yxx}^{2\omega}$ and ($\sigma_{xx}^{\omega}$)$^2$ as a function of carrier density $n_e$. 
}
\label{Scaling_n_E}
\end{figure*}

\newpage
\begin{figure*}[h]
\centering
\includegraphics[width=16cm]{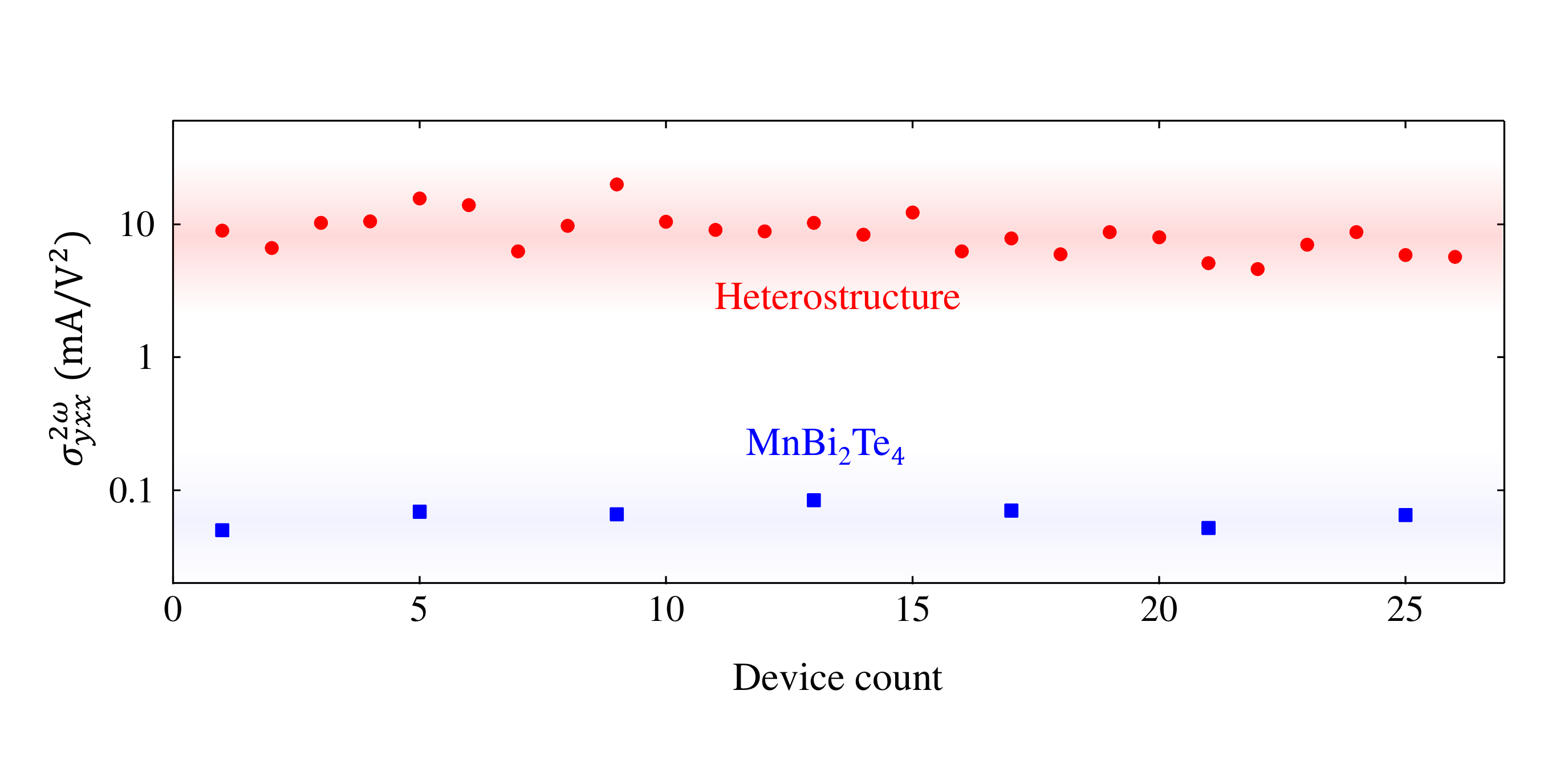}
\caption{\textbf{The summary of $\sigma _{yxx}^{2\omega}$ of 26 MnBi$_2$Te$_4$ heterostructures and 7 MnBi$_2$Te$_4$ devices.} \\
The $\sigma _{yxx}^{2\omega}$ in MnBi$_2$Te$_4$ device is induced by nonlinear Drude conductivity which is two orders of amplitude smaller then that in MnBi$_2$Te$_4$ heterostructures.
}
\label{Additional_Dev_Sum}
\end{figure*}

\clearpage
\begin{figure*}[h]
\centering
\includegraphics[width=12cm]{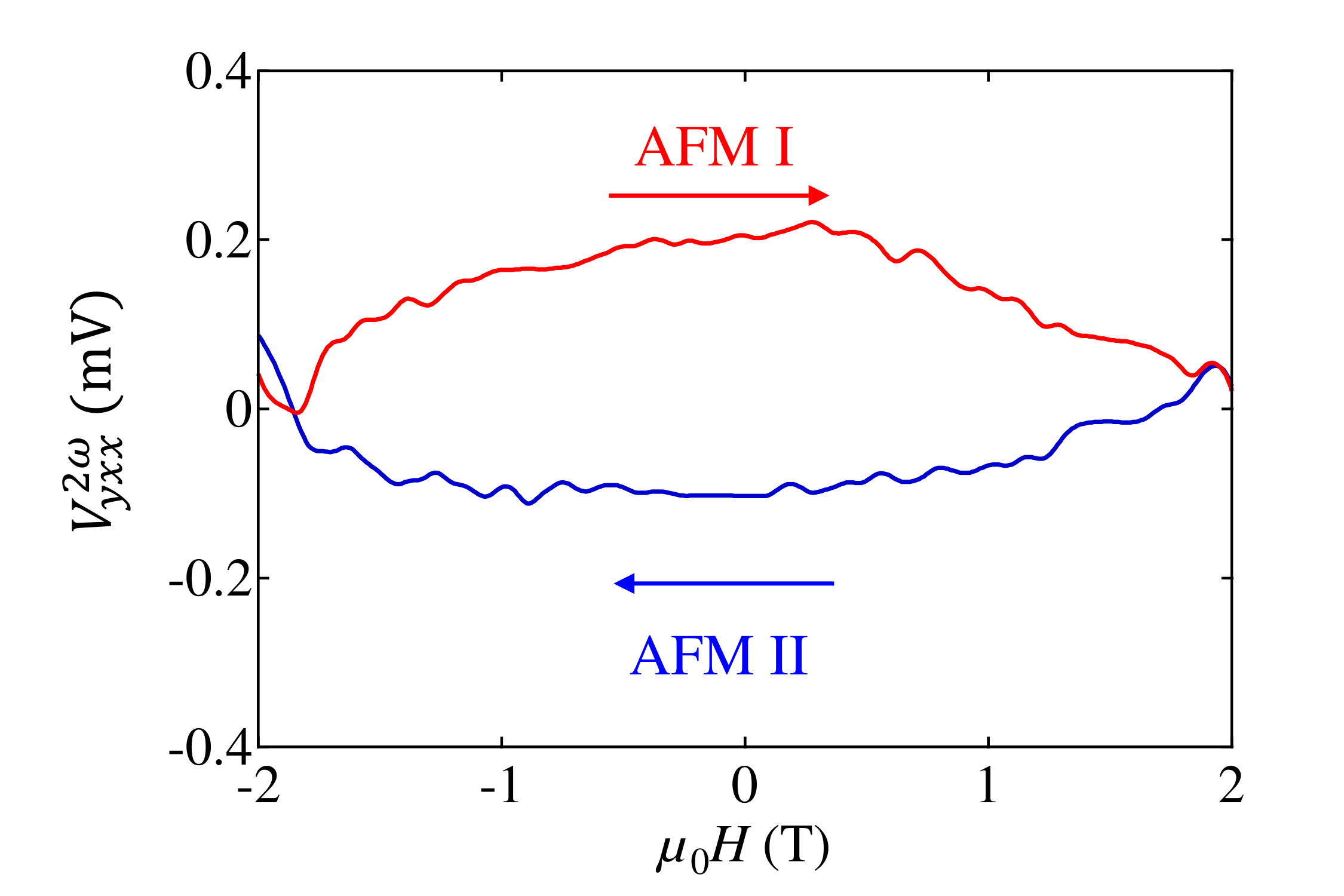}
\caption{\textbf{The magnetic field dependence of nonlinear Hall signals in the BP/MnBi$_2$Te$_4$/BP heterostructure.} The nonlinear transverse voltages $V_{yxx}^{2\omega}$ as a function of out of plane magnetic field $\mu_0H$. At $\mu_0H = 0$, the $V_{yxx}^{2\omega}$ have opposite signals for different AFM states. The forward and backward scans and corresponding AFM states are noted by the red and blue curves, respectively.
}
\label{Additional_hysB}
\end{figure*}

\clearpage
\begin{figure*}[h]
\centering
\includegraphics[width=18cm]{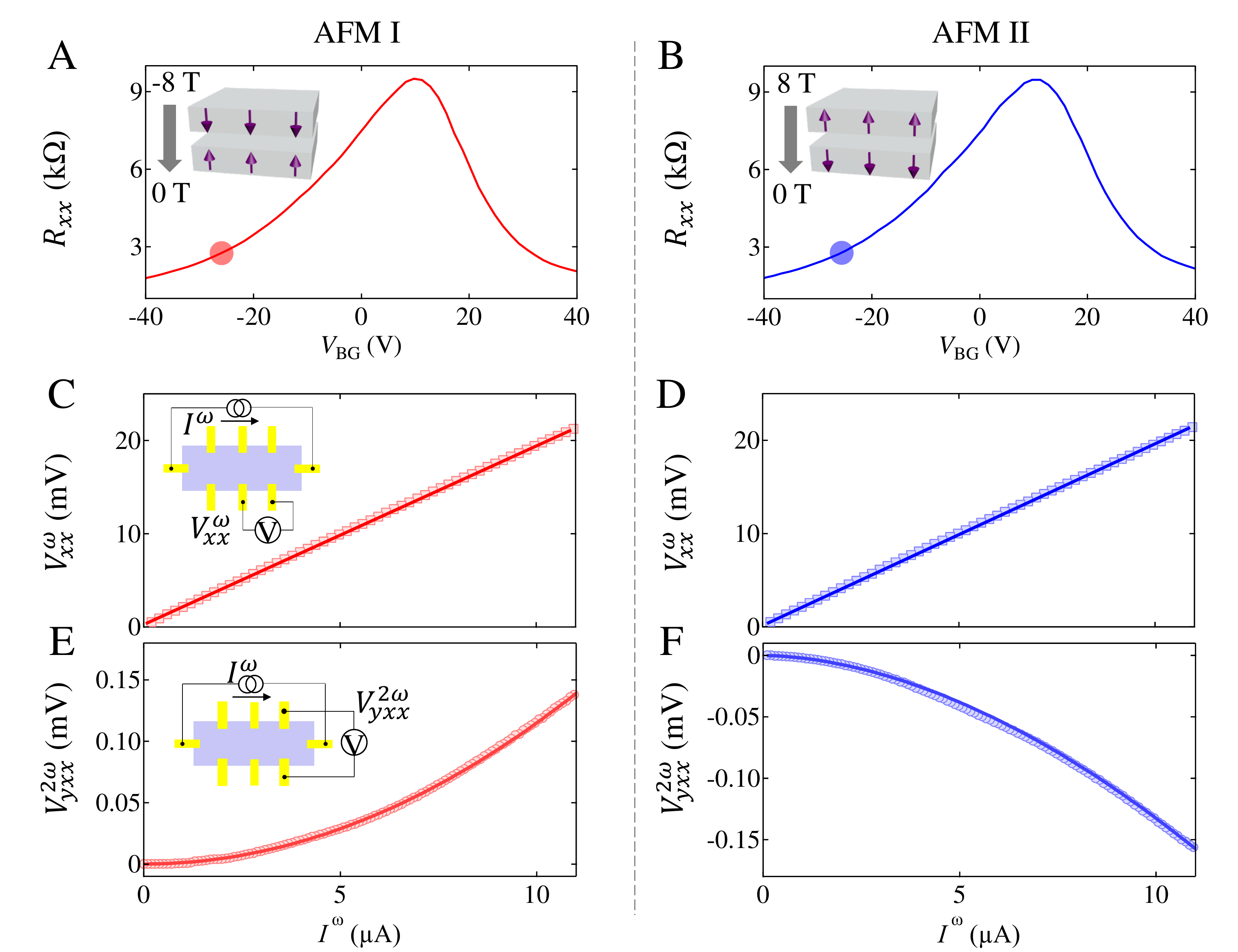}
\caption{\textbf{The observation of antiferromagnetic nonlinear Hall in another BP/MnBi$_2$Te$_4$/BP device.} (A and B) The linear longitudinal resistance $R_{xx}$ as a function of back gate voltages for different AFM states. (C and E) The linear longitudinal voltages $\sigma _{xx}^{\omega}$ and nonlinear Hall voltages $\sigma _{yxx}^{2\omega}$ as a function of supply current for an AFM state. Insets show the corresponding measurement circuits. (D and F) The same as the panels (C) and (E) but for another AFM state.
}
\label{Additional_AFM}
\end{figure*}

\clearpage
\begin{figure*}[h]
\centering
\includegraphics[width=16cm]{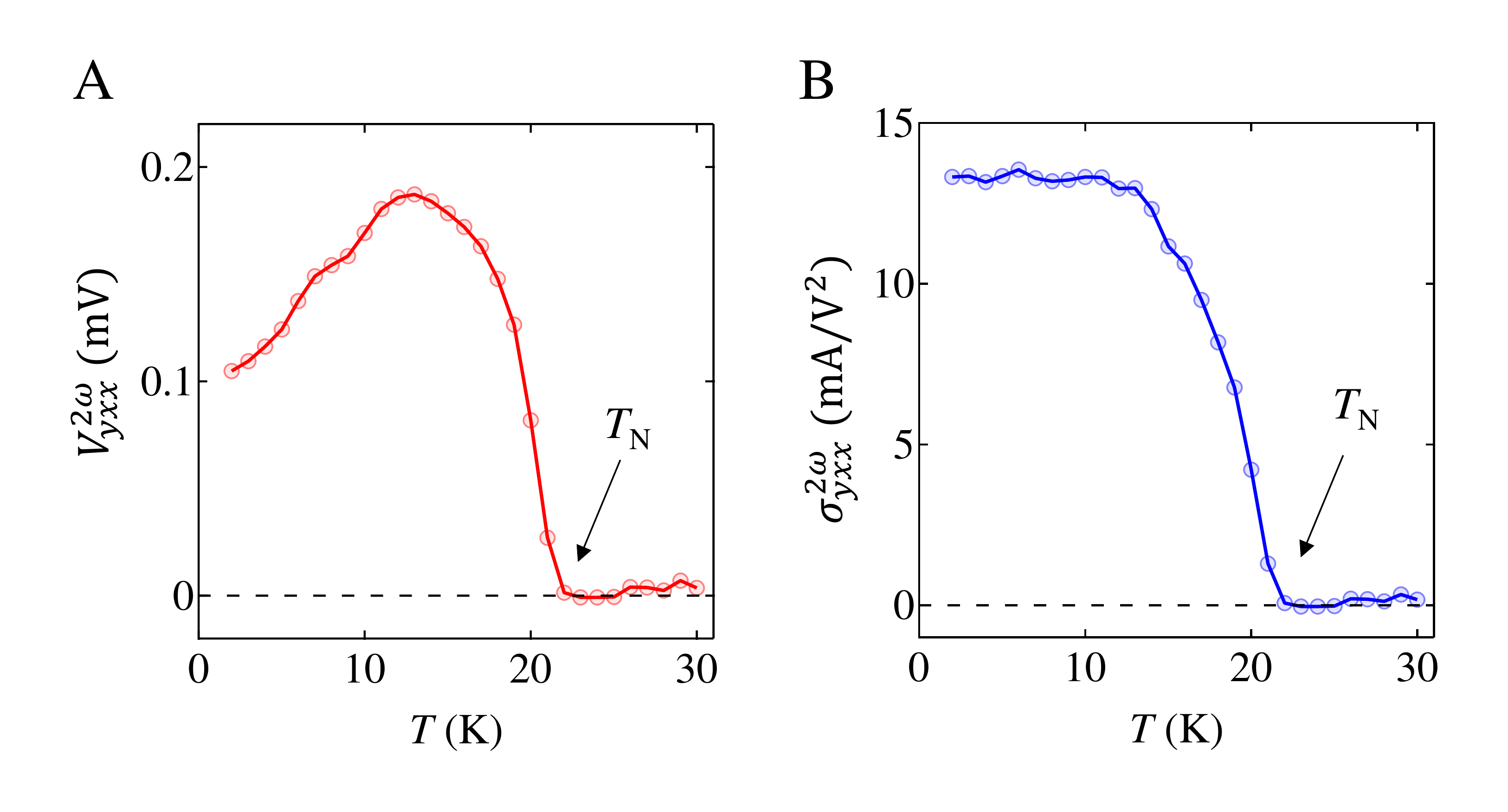}
\caption{\textbf{Additional temperature dependence data corresponding to Fig. 3A.} (A) The nonlinear Hall voltages $V_{yxx}^{2\omega}$ as a function of temperature. (B) The nonlinear Hall conductivity $\sigma _{yxx}^{2\omega}$ as a function of temperature. When temperature is lower than 15 K, $\sigma _{yxx}^{2\omega}$ is almost independent of temperature. 
}
\label{Additional_Temp_Fig3}
\end{figure*}

\clearpage
\begin{figure*}[h]
\centering
\includegraphics[width=16cm]{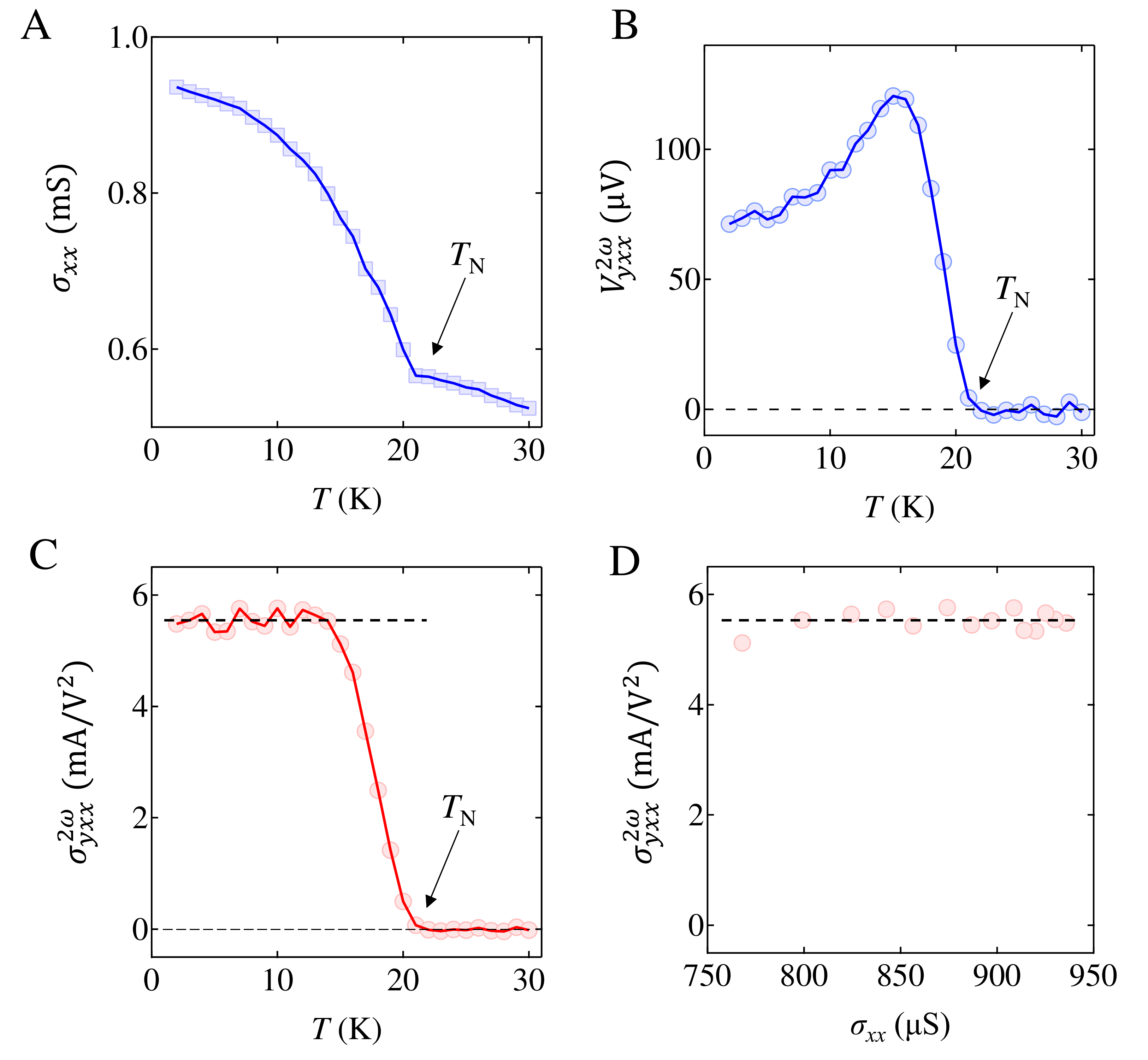}
\caption{\textbf{Additional temperature dependence data on another device (Device-BM21).} (A to C) The temperature dependence of the linear longitudinal conductivity $\sigma_{xx}$, nonlinear Hall voltage $V_{yxx}^{2\omega}$ and nonlinear Hall conductivity $\sigma _{yxx}^{2\omega}$. The nonlinear Hall signals $V_{yxx}^{2\omega}$ and $\sigma_{yxx}^{2\omega}$ decrease to zero when temperature is higher than Neel temperature ($T_N > 21$ K). (D) $\sigma _{yxx}^{2\omega}$ as a function of $\sigma_{xx}$. $\sigma _{yxx}^{2\omega}$ is independent of $\sigma_{xx}$. 
}
\label{Additional_Temp}
\end{figure*}

\clearpage
\begin{figure*}[h]
\centering
\includegraphics[width=16cm]{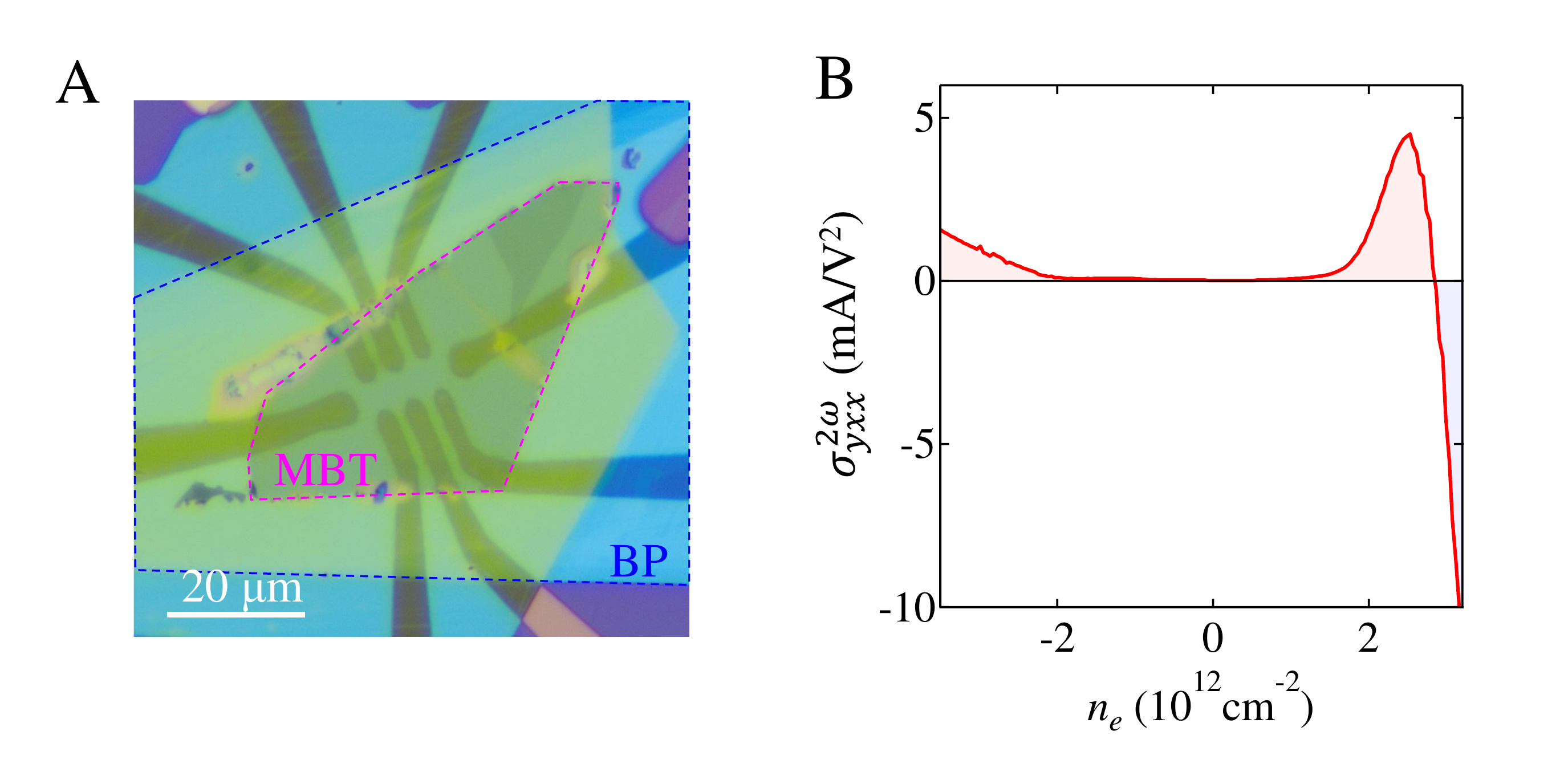}
\caption{\textbf{Additional data of the nonlinear Hall signals on a BP/4SL MnBi$_2$Te$_4$ device.} (A) The microscopic image of a BP/4SL MnBi$_2$Te$_4$ device. BP and 4SL MnBi$_2$Te$_4$ are outlined by blue and red dashed lines. (B) The nonlinear Hall conductivity $\sigma_{yxx}^{2\omega}$ as a function of carrier density $n_e$. The $\sigma_{yxx}^{2\omega}$ in the BP/4SL MnBi$_2$Te$_4$ device shows the same behavior as that in the BP/6SL MnBi$_2$Te$_4$ device (Fig. 4A).
}
\label{Additional_4SL}
\end{figure*}

\clearpage
\begin{figure*}[h]
\centering
\includegraphics[width=18cm]{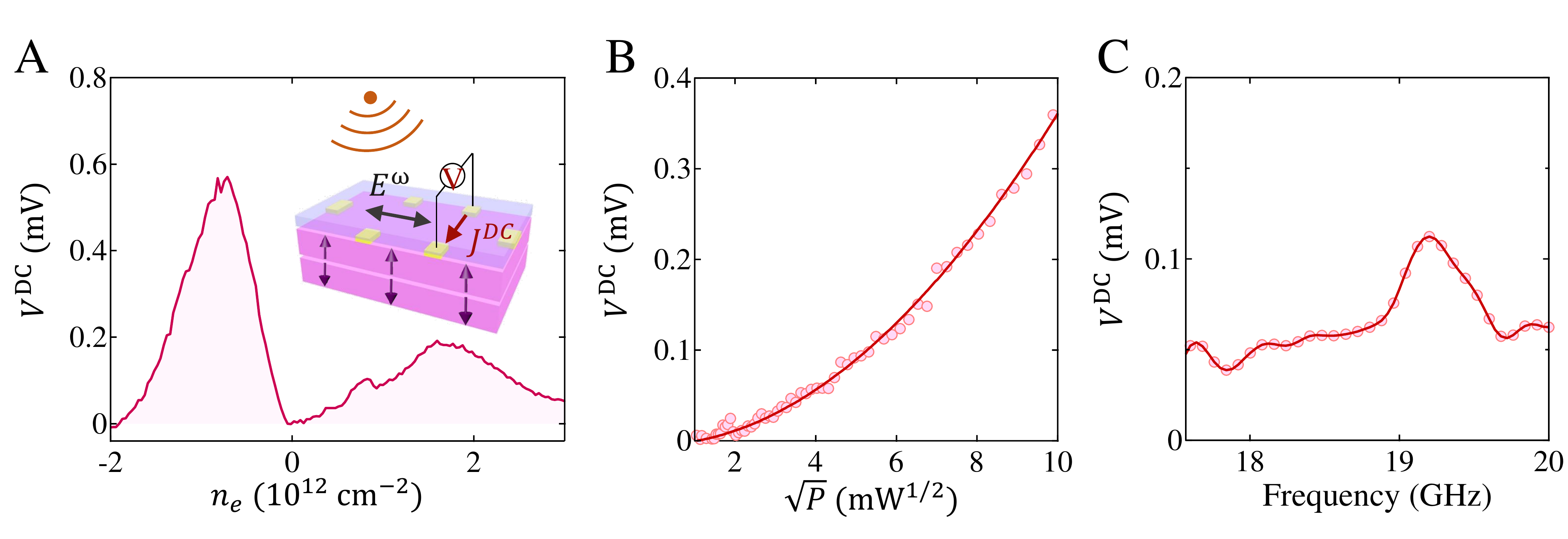}
\caption{\textbf{Microwave rectification measurements on a BP/6SL MnBi$_2$Te$_4$ device at 2 K.} (A) The DC signals $V^{\textrm{DC}}$ as a function of carrier density $n_e$ at 2.8 GHz. The inset shows the DC microwave rectification measurement circuit. The electrical field direction of the microwave is not well defined. Therefore, the recorded DC signals are only part of the the microwave generated signals. (B) The power dependence of the $V^{\textrm{DC}}$. The solid line is the fit of the $V^{\textrm{DC}}$. It scales quadratically with $\sqrt{P}$ ($\sqrt{P}\propto I$). (C) The $V^{\textrm{DC}}$ as a function of Frequency. The fluctuation is caused by the transmission losses of the RF signal.
}
\label{Additional_RF}
\end{figure*}

\clearpage
\begin{figure*}[h]
\centering
\includegraphics[width=18cm]{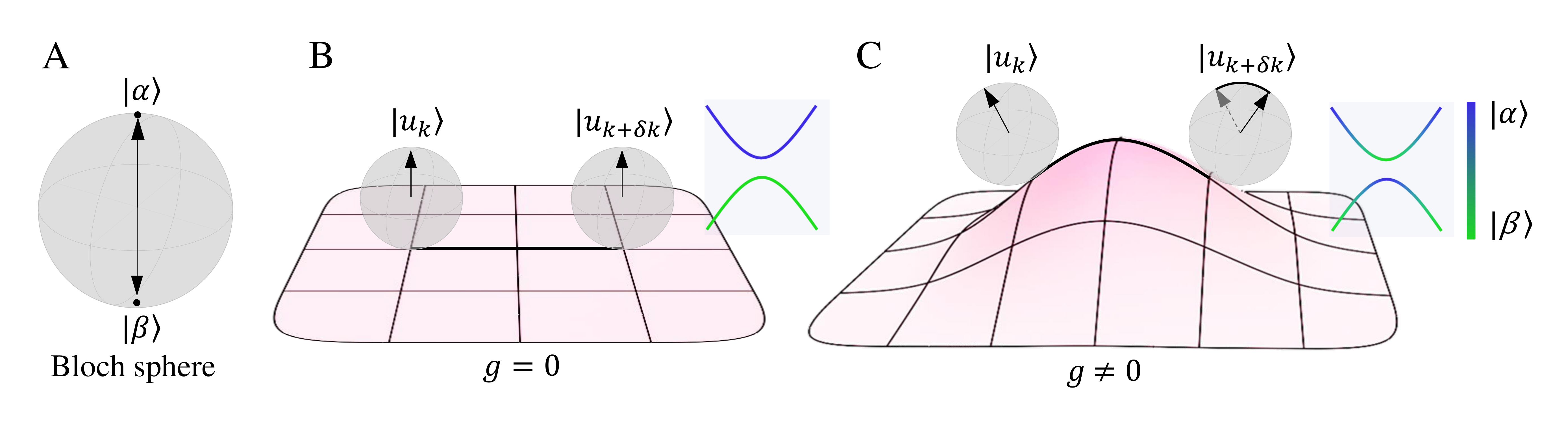}
\caption{\textbf{An intuitive picture of quantum metric.} (A) The Bloch sphere. The north and south poles of the Bloch sphere correspond to the the basis vectors of $\arrowvert\alpha\rangle$ and $\arrowvert\beta\rangle$, respectively. (B) A trivial case ($g=0$) without band inversion. When electrons momenta change from $k$ to $k+\delta k$, they occupy the same state ($\arrowvert\alpha\rangle$) and map on the same point on the Bloch sphere. Therefore, the distance  between the two states ($\arrowvert u_k\rangle$ and $\arrowvert u_{k+ \delta k}\rangle$) on the Bloch sphere is zero. (C) A nontrivial case ($g \neq 0$) in which conducting band and valance band go through a band inversion. In this case, each band becomes a $\mathbf{k}$-dependent superposition of $\arrowvert\alpha\rangle$ and $\arrowvert\beta\rangle$. The change of $\arrowvert u\rangle$ is dramatic near $\mathbf{k}=0$ where band inversion occurs but weak at large $\mathbf{k}$. When electrons momenta change from $k$ to $k+ \delta k$, the state changes from one to another. The nonzero distance between the two states is drawn on the Bloch sphere.
}
\label{Theory_QM}
\end{figure*}

\clearpage
\begin{figure*}[h]
\centering
\includegraphics[width=12cm]{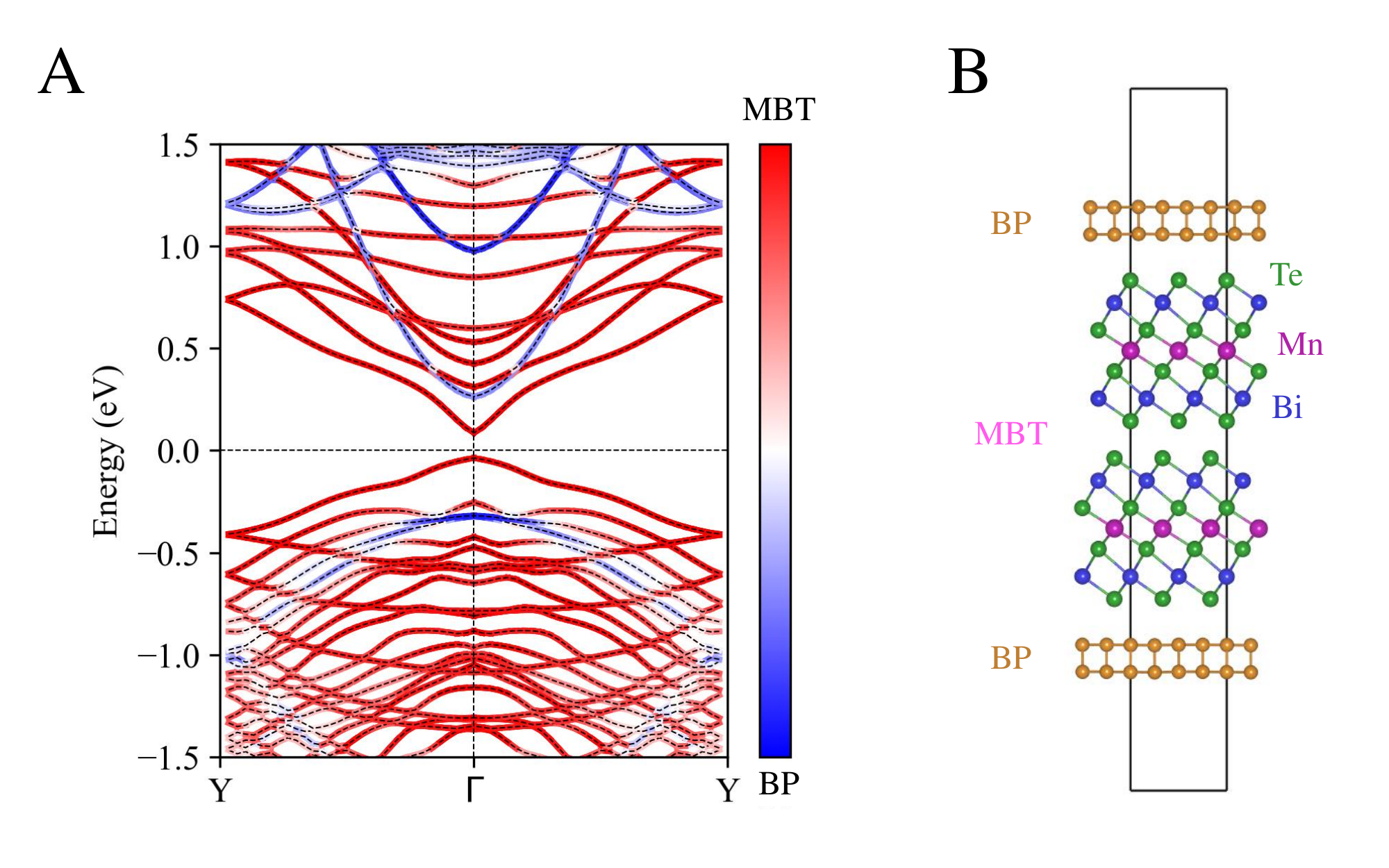}
\caption{\textbf{The DFT computed band structure of the BP/MnBi$_2$Te$_4$/BP heterostructure.} (A) The MnBi$_2$Te$_4$ (MBT) bands are marked in red, and the BP bands are shown in blue. As evident from the plot, MBT and BP share similar work functions, and as a result, the low-energy BP bands hybridize with MBT bands around the Fermi level of MBT, forming a straddling gap (type-I) arrangement. (B) The heterostructure considered here is created by a rectangular supercell of 2SL-MBT and a 1x2 supercell of BP. The armchair direction of the MBT is aligned along the zigzag direction of the BP. 
}
\label{DFT}
\end{figure*}

\clearpage
\begin{figure*}[h]
\centering
\includegraphics[width=14cm]{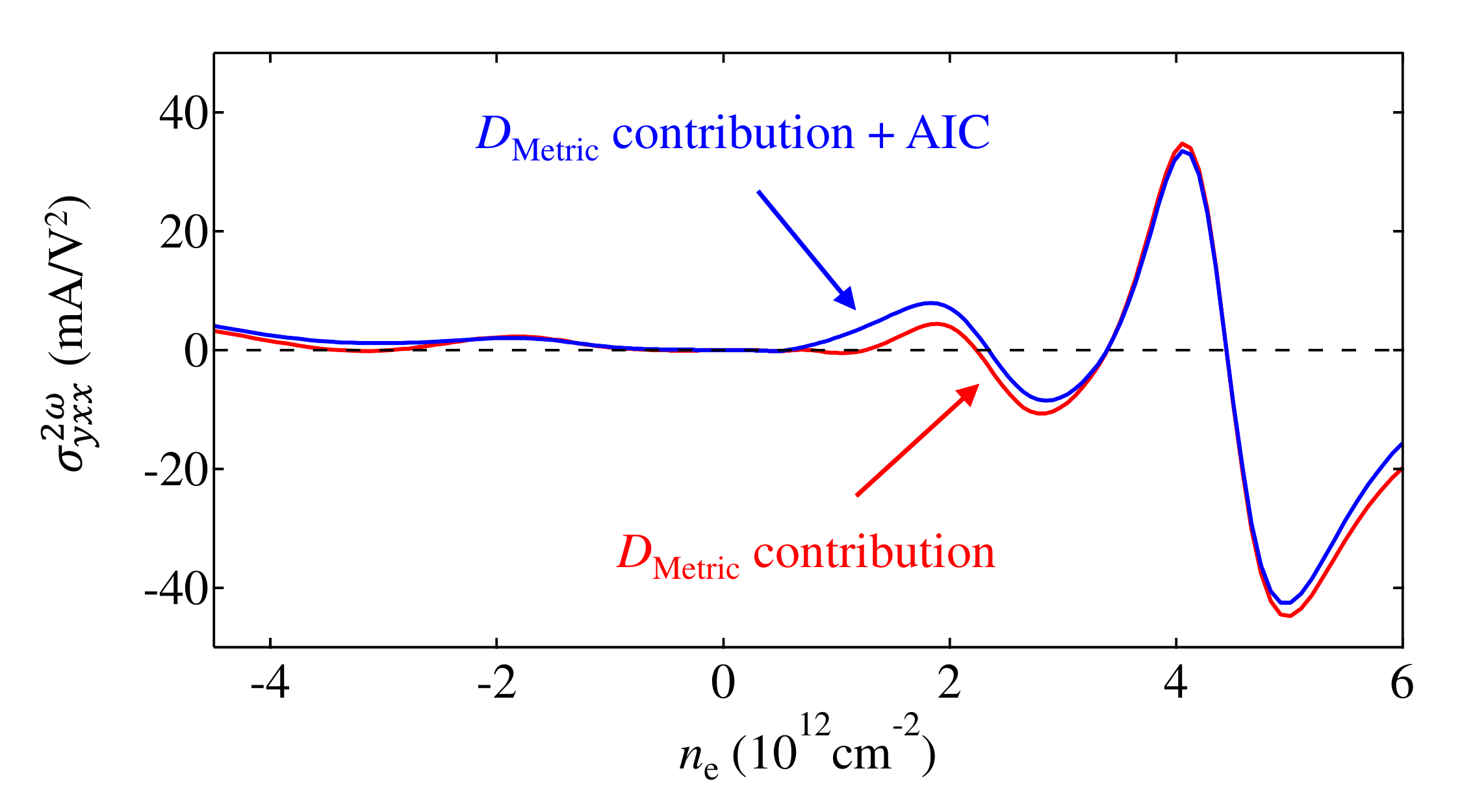}
\caption{\textbf{Quantum metric dipole dominated nonlinear Hall conductivity.} Two-band model (quantum metric dipole $D_{\textrm{Metric}}$ contribution) and multiband model ($D_{\textrm{Metric}} +$ additional inter band contributions AIC) calculated nonlinear Hall conductivity as a function of carrier density. The nonlinear Hall signals are dominated by the quantum metric dipole contribution.
}
\label{Theory_Twobands}
\end{figure*}

\clearpage
\begin{figure}[h]
    \centering
    \includegraphics[width=0.9\linewidth]{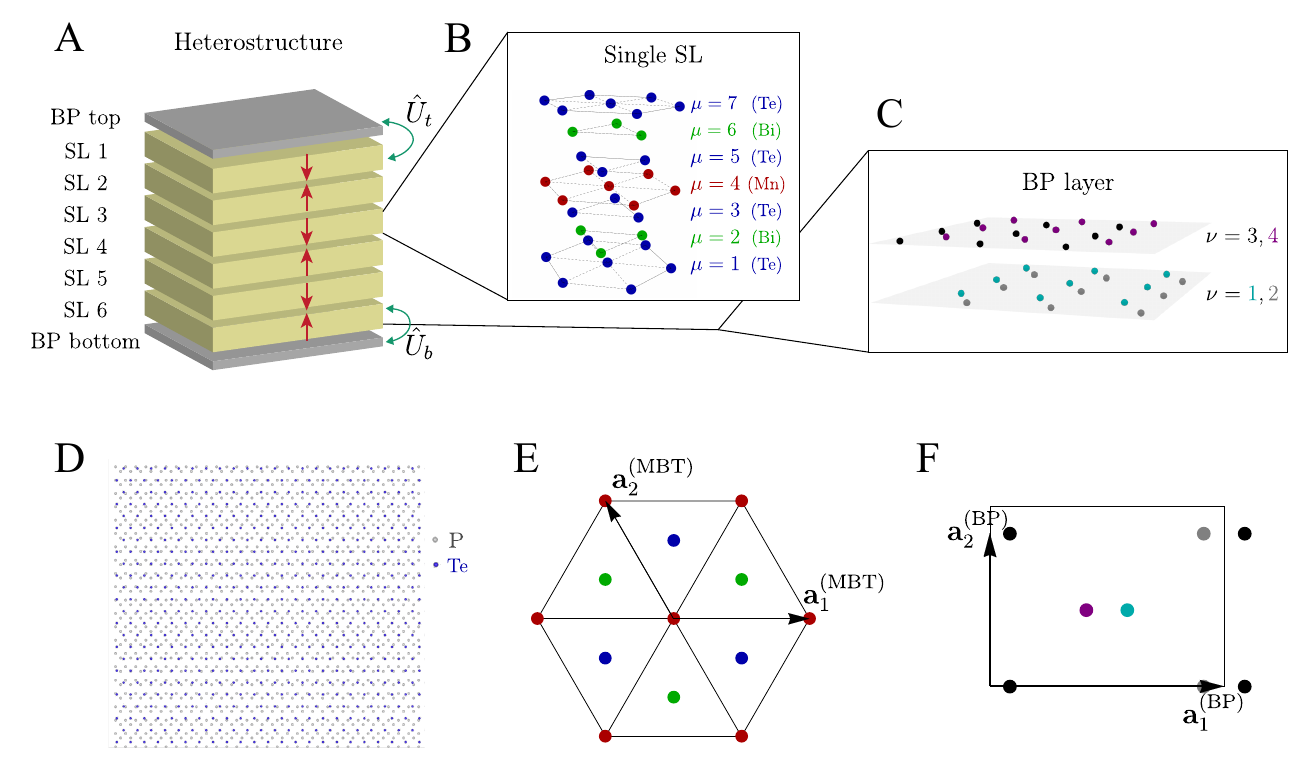}
\caption{\textbf{Real-space lattice structure of the MBT-BP heterostructure.} (A) Schematic drawing of the $\mathcal{PT}$ symmetric MBT-BP heterostructure consisting of 6SL MBT encapsulated by BP monolayer. Red arrows indicate AFM order of the MBT SLs. The direct electronic coupling between MBT and BP is dominated by hopping processes between BP top and SL 1 $(\hat{U}_t)$ and between BP bottom and SL 6 $(\hat{U}_b)$. (B) Lattice structure of a single MBT SL with seven atomic layers labeled by index $\mu$. (C) Lattice structure of a BP monolayer with four basis atoms labeled by index $\nu$. (D) Schematic top view on an incommensurate bilayer structure consisting of bottom layer of MBT  (blue) and top P layer of BP (grey) with different lattice constants. (E) Two-dimensional projection of MBT SL with primitive vectors $\mathbf{a}^{\text{MBT}}_1$ and $\mathbf{a}^{\text{MBT}}_2$. Color coding corresponds to that used in panel (B). (F) Two-dimensional projection of BP monolayer with primitive vectors $\mathbf{a}^{\text{BP}}_1$ and $\mathbf{a}^{\text{BP}}_2$. Color coding corresponds to that used in panel (C). }
\label{fig:SM_theory_1}
\end{figure}

\clearpage
\begin{figure}[h]
    \centering
\includegraphics[width=0.99\linewidth]{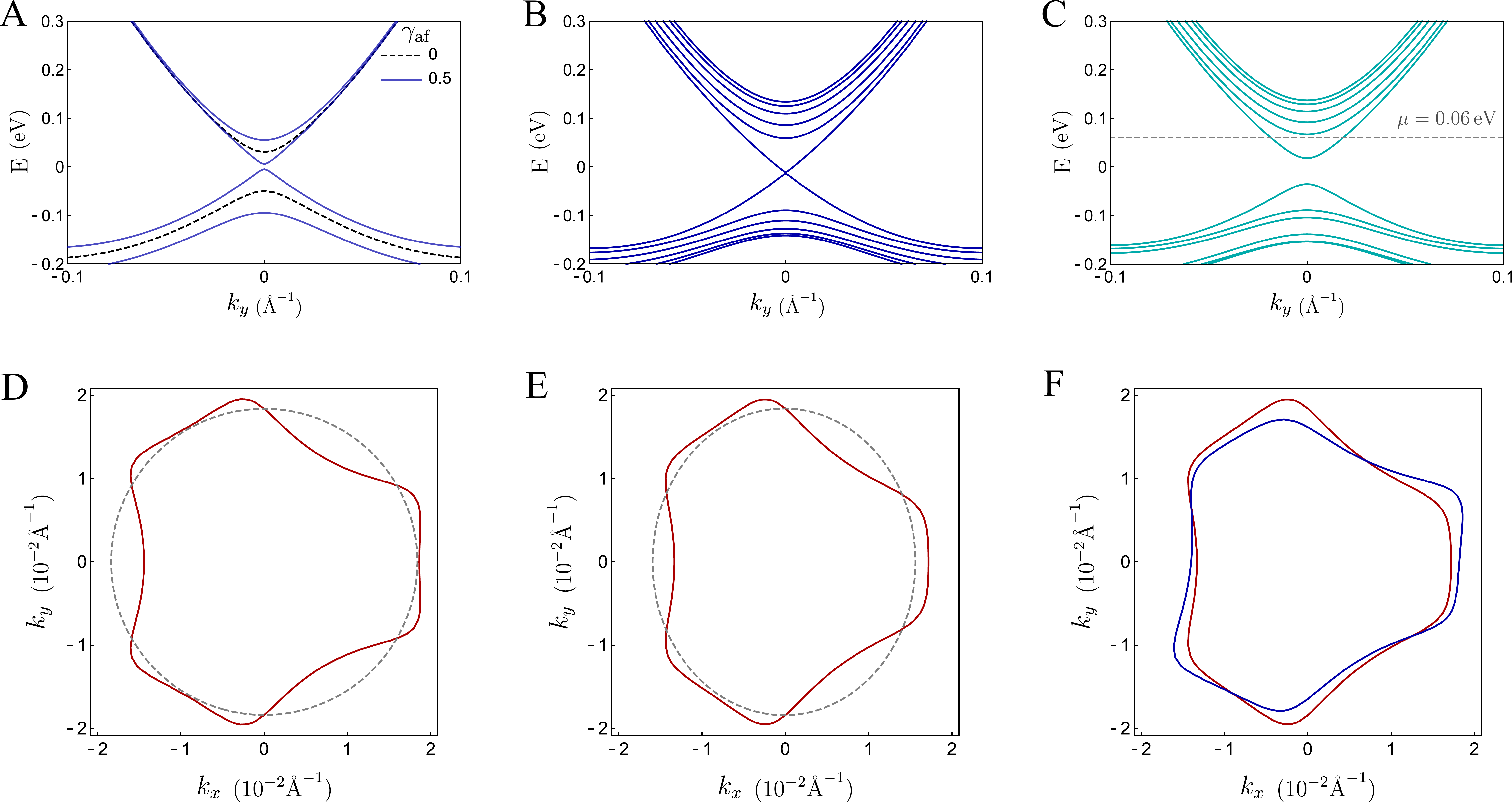}
\caption{\textbf{Energy band Fermi surface of MBT modeled by Eqs.(\ref{h_MBT}), (\ref{eq:newh_SL}) and (\ref{eq:newT}).} (A) A cut along $k_y$ of the bands of a single MBT SL in the paramagnetic state (dashed) and after the onset of the AFM order (solid line). The band gap at $\Gamma$ ($\mathbf{k}=0$) in the paramagnetic state originates from the hybridization of the surface states on the top and bottom of the SL and is reduced after the onset of magnetism. The hybridization of the surface states is suppressed by stacking multiple SLs along $\hat{\mathbf{z}}$, which increases the system's thickness. (B) The energy bands of a set of six SLs in the paramagnetic state ($\gamma_{af}=0$), where the surface Dirac cone at $\Gamma$ can be readily seen. (C) The onset of the AFM order open a gap in the surface Dirac cone. (D to F) The Fermi surface of 6SL MBT when the chemical potential is set to $\mu=0.06$~meV. The dashed (solid) line in (D) and (E) refer to the Fermi surface without (with) the warping term. The red curve in panel (F) is the same as in (E), while the blue curve is obtained by rotation the red by $2\pi/3$ around $\hat{z}$ to highlight the break of $C_{3z}$ symmetry induced by strain. We set the parameters of the model to the values listed in Table \ref{tab:kp_parameter_values}. Note that the choice of $k_x$ and $k_y$ axis is rotated by $\pi/2$ with respect to the main text. }
\label{fig:SM_theory_2}
\end{figure}

\clearpage
\begin{figure}[h]
    \centering
\includegraphics[width=0.9\linewidth]{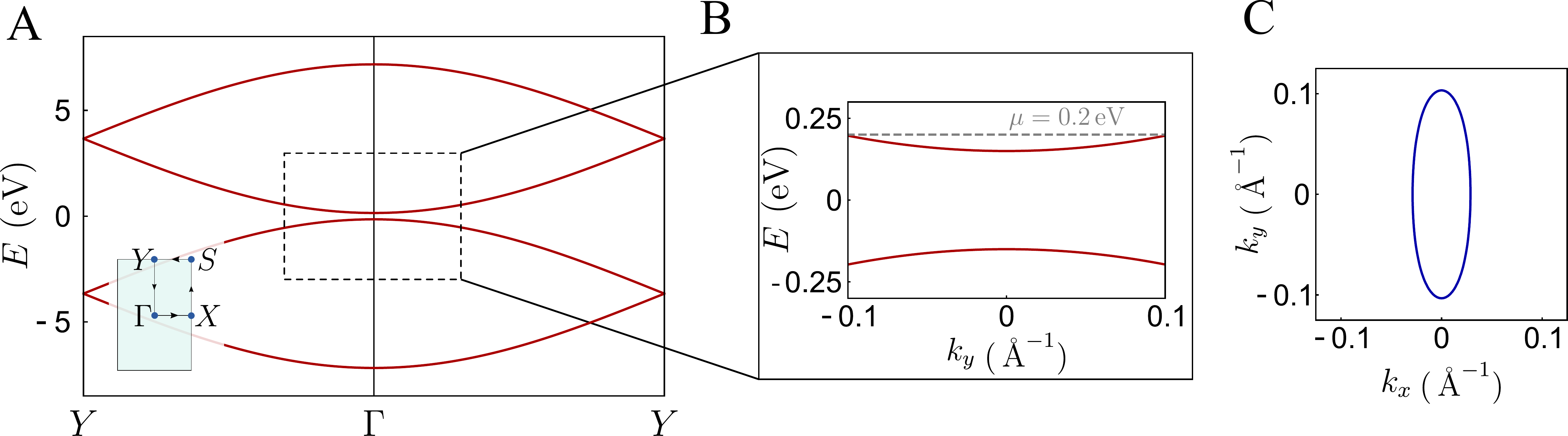}
\caption{\textbf{Energy band and Fermi surface of monolayer BP.} (A) Energy bands of BP monolayer modeled by Eq.(\ref{eq:hBP_sigma}) along a high-symmetry direction of the Brillouin zone shown in blue as a inset. (B) A zoom-in view of the bands around the $\Gamma$ point (e.g. $\mathbf{k}=0$). (C) The Fermi surface of BP where we set the chemical potential to be $\mu=0.2$ eV (dashed gray line in (B)). BP does not preserve 3-fold rotational symmetry around $\hat{z}$, which is reflected in the elliptical shape of its Fermi surface. Therefore, similar to the role of uniaxial strain, encapsulating the MBT SLs with BP monolayer leads to a breaking of $C_{3z}$ symmetry of the heterostructure. We set the parameters of the model to the values listed in Table \ref{tab:kp_parameter_values}.}
\label{fig:SM_theory_3}
\end{figure}

\clearpage
\begin{figure}[h]
    \centering
    \includegraphics[width=0.6\linewidth]{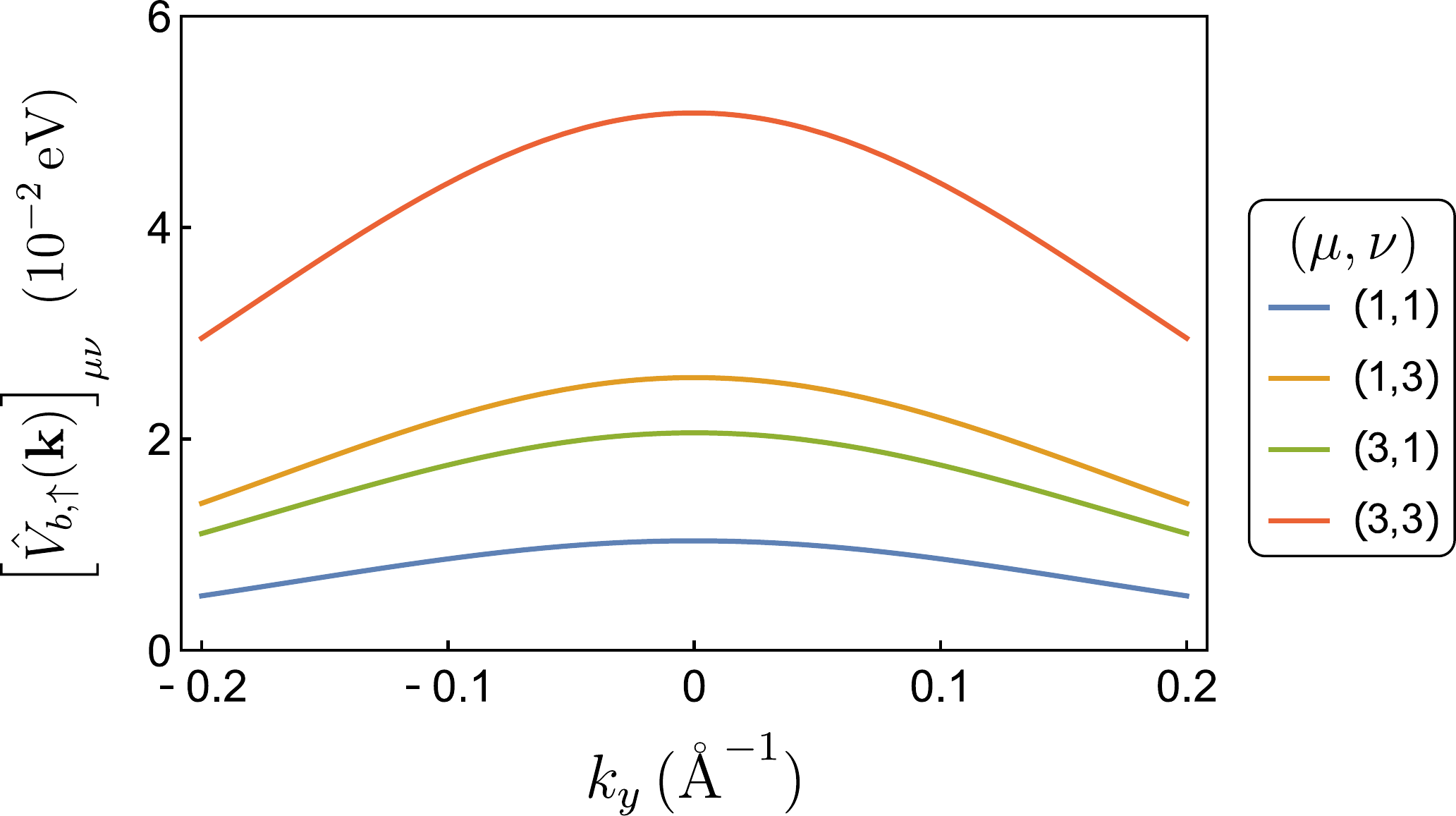}
\caption{\textbf{Matrix elements of $\hat{V}_{b,\uparrow}(\mathbf{k})$ along the $k_y$ axis.} We set the parameters of the model to the values listed in Table \ref{tab:kp_parameter_values}. }
    \label{fig:SM_theory_4}
\end{figure}

\normalsize
\clearpage
\begin{table}[h]
\begin{center}
\begin{tabular}{p{2.5cm}|p{1.3cm}|p{1.7cm}|p{2.5cm}|p{2.3cm}|p{2.3cm}|p{1.3cm}|p{1.7cm}}
\hline
Device type & Device number & Contact geometry & Angle between top/bottom BP and MBT & Angle between MBT and contacts & top/bottom BP thickness & $\sigma_{xx}^{\omega}$ ($\mu$S) & $\sigma_{yxx}^{2\omega}$ (mA/V$^2$)\\ \hline
      		& BMB1 & Hall bar & 0/0		  & 0 	  & 2L/2L         & 227	  & 8.93	\\ \cline{2-8}
      		& BMB2 & Hall bar & 0/-1	  & 0 	  & 2L/1L	      & 671	  & 6.62	\\ \cline{2-8} 
BP/MBT/BP 	& BMB3 & Hall bar & -11/-6	  & 1	  & 2L/$\sim$10nm  & 249	  & 10.25	\\ \cline{2-8} 
	  		& BMB4 & Hall bar & 53/15	  & 27	  & 1L/$\sim$15nm  & 585	  & 10.5	\\ \cline{2-8} 
	  		& BMB5 & Hall bar & -16/-17	  & 16	  & 2L/$\sim$10nm  & 1466  & 15.68	\\ \hline
	  		& BM1  & Circular & -5		  & 0	  & $\sim$12nm	  & 154	  & 13.97	\\ \cline{2-8} 
	  		& BM2  & Hall bar & -		  & -	  & -		  & 500	  & 6.24	\\ \cline{2-8} 
 	  		& BM3  & Hall bar & -30		  & 25	  & -		  & 141	  & 9.70	\\ \cline{2-8} 
      		& BM4  & Hall bar & -         & -     & -         & 119   & 19.91 \\ \cline{2-8} 
      		& BM5  & Hall bar & -         & -     & -         & 403   & 10.42 \\ \cline{2-8} 
      		& BM6  & Hall bar & -         & -     & -         & 206   & 9.06  \\ \cline{2-8} 
      		& BM7  & Hall bar & -5        & -2    & -         & 498   & 8.82  \\ \cline{2-8} 
      		& BM8  & Hall bar & -1        & 1     & -         & 608   & 10.26 \\ \cline{2-8} 
BP/MBT 		& BM9  & Hall bar & -         & -     & 4L        & 1070  & 8.37  \\ \cline{2-8} 
      		& BM10 & Hall bar & -         & -     & -         & 2890  & 12.22 \\ \cline{2-8} 
      		& BM11 & Hall bar & -         & -     & -         & 307   & 6.24  \\ \cline{2-8} 
     		& BM12 & Hall bar & -         & -     & 5-7L      & 398   & 7.82  \\ \cline{2-8} 
      		& BM13 & Hall bar & -         & -     & -         & 450   & 5.95  \\ \cline{2-8} 
      		& BM14 & Hall bar & -         & -     & -         & 943   & 8.75  \\ \cline{2-8} 
      		& BM15 & Hall bar & -         & -     & -         & 483   & 7.97  \\ \cline{2-8} 
      		& BM16 & Hall bar & -26       & -2    & $\sim$8nm  & 238   & 5.10  \\ \cline{2-8} 
      		& BM17 & Hall bar & -         & -     & -         & 357   & 4.60  \\ \cline{2-8} 
      		& BM18 & Hall bar & -35       & 54    & -         & 500   & 7.00  \\ \cline{2-8} 
      		& BM19 & Hall bar & 16        & 26    & -         & 357   & 8.70  \\ \cline{2-8}
      		& BM20 & Hall bar & -         & -	  & -         & 700   & 5.86  \\ \cline{2-8} 
      		& BM21 & Hall bar & -         & -     & -         & 930   & 5.69  \\ \hline
\end{tabular}
\end{center}
\caption{\textbf{Device summary of 5 BP/MnBi$_2$Te$_4$/BP hetrostructures (BP/MBT/BP) and 21 BP/MnBi$_2$Te$_4$ hetrostructures (BP/MBT)}}
\label{Device summary}
\end{table}

\normalsize
\clearpage
\begin{table}[h]
\begin{center}
\begin{tabular}{c|ccc}
\hline
\multirow{2}{*}{P basis atom}     & \multicolumn{3}{c}{$\text{Components (\AA)}$}                   \\ \cline{2-4} 
                      & \multicolumn{1}{c|}{$\hat{x}$} & \multicolumn{1}{c|}{$\hat{y}$} & $\hat{z}$ \\ \hline
$\boldsymbol{\rho}_1$ & \multicolumn{1}{c|}{2.71}      & \multicolumn{1}{c|}{1.65}      & 15.05     \\ \hline
$\boldsymbol{\rho}_2$ & \multicolumn{1}{c|}{4.2}       & \multicolumn{1}{c|}{0}         & 15.05     \\ \hline
$\boldsymbol{\rho}_3$ & \multicolumn{1}{c|}{0.41}      & \multicolumn{1}{c|}{0}         & 17.15     \\ \hline
$\boldsymbol{\rho}_4$ & \multicolumn{1}{c|}{1.9}       & \multicolumn{1}{c|}{1.65}      & 17.15     \\ \hline
\end{tabular}
\end{center}
\caption{\textbf{Cartesian components of the position relative to the origin of the unit cell of the four nonequivalent phosphorus atoms in the BP monolayer.}}
\label{tab:Ppositon}
\end{table}

\normalsize
\clearpage
\begin{table}[h]
    \centering
    \begin{tabular}{*{8}{c}}

    \multicolumn{2}{c}{MBT} \\
    \cline{1-2}
       $\gamma_0$ (eV) & $\gamma$ (eV \AA$^2$) & $m_0$ (eV) & $\beta_0$ (eV \AA$^2$) & $\beta_1$ (eV \AA$^3$) & $\alpha$ (eV \AA) & $t_1$ (eV) & $t_2$ (eV) \\
       $-0.01$ & $17$ & $0.04$ & $9.4$ & $8000$ & $3.2$ &  $-0.0533$ & $0.0463$  \\
    \hline
       $\lambda$ (eV) & $\alpha_3$ (eV \AA$^2$) & $\gamma_{af}$ & $m_1$ (eV) & $m_2$ (eV) & $\gamma_s$ & $s_1$ (eV \AA$^2$) & $s_2$ (eV \AA$^2$)    \\
       $0.0557$ & $0.0$ & $0.5$ & $0.05$ & $0.09$ & $1.0$ & $0.1$ & $0.1$  \\
    \hline
       $s_3$ (eV \AA$^2$)  & $s_4$ (eV \AA$^2$) & $s_5$ (eV \AA) & $s_6$ (eV \AA) & $s_7$ (eV \AA) & $s_8$ (eV \AA) & $s_9$ (eV) \\
       $50$  & $0.1$ & $0.1$ & $0.0$ & $0.1$ & $0.1$ & $0.0$\\ 
    \hline
\\    
    \multicolumn{2}{c}{BP} \\
    \cline{1-2}
    $\tilde{t}_1$ (eV) & $\tilde{t}_2$ (eV) \\
    $-1.7575$ & $3.665$ \\
    \hline
    \\
    
    \multicolumn{2}{c}{MBT-BP coupling} \\
    \cline{1-2}
    $A_\sigma/\Omega_{\text{BP}}$ (eV/\AA$^2$) & $A_\pi/\Omega_{\text{BP}}$ (eV/\AA$^2$) & $a_\sigma$ (\AA) & $a_\pi$ (\AA) & &  & & \\
    $0.34$ & $0.0$ & $2.0$ & $2.0$ & & & & \\
  
    \hline
    \end{tabular}
\caption{\textbf{Numerical $k \cdot p$ model parameter values that are used in the main text.} The only other parameters that enter the model are the basis positions of the atoms in the unit cell that enter via the parameters $s_{\mu \nu}$. We provide the basis location of the P atoms in Table~\ref{tab:Ppositon} and refer to standard literature for the basis positions of the atoms in the MBT SL. Note that the Bohr radii $a_\sigma$, $a_\pi$ enter the model via the fit parameters $\gamma_j$ ($j=1,\ldots, 6$). Finally, we note that $\Omega_{\text{BP}} = 14.5$\AA$^2$ such that $A_\sigma = 4.9$~eV.}
    \label{tab:kp_parameter_values}
\end{table}

\normalsize
\clearpage

\bigskip
\end{document}